\documentclass[draft]{agujournal2019}
\usepackage{url}
\usepackage{lineno}
\usepackage{soul}
\usepackage{longtable}
\usepackage{amsmath}
\usepackage[protrusion=true,expansion=false]{microtype}
\draftfalse

\journalname{Journal of Advances in Modeling Earth Systems (JAMES)}

\begin{document}

\title{ClimateBench v2.0: Probabilistic Climate Model Benchmarking}

\authors{Duncan Watson-Parris\affil{1,2}, Willa Tobin\affil{1}, Aytaç Paçal\affil{3,4}, Manuel Schlund\affil{5}, V. Balaji\affil{6}, Kevin Bowman\affil{7}, Chris Bretherton\affil{8}, Peter M. Caldwell\affil{9}, Will Chapman\affil{10}, William D. Collins\affil{11}, Gregory S. Elsaesser\affil{12,13}, Pierre Gentine\affil{12}, Helene Hewitt\affil{14}, Stephan Hoyer\affil{15}, Ralph Keeling\affil{1}, Nikolay Koldunov\affil{16}, David M. Lawrence\affil{17}, Christian Lessig\affil{18}, Daniel J. Lunt\affil{19}, J. David Neelin\affil{20}, Mike Pritchard\affil{21}, Sarah Purkey\affil{1}, Gavin Schmidt\affil{13}, Tapio Schneider\affil{22}, Michael Schulz\affil{23}, Tiffany Shaw\affil{24}, Isla R. Simpson\affil{17}, Graeme Stephens\affil{7}, Aneesh C. Subramanian\affil{10}, Joao Teixeira\affil{7}, Jessica Tierney\affil{25}, Andrew I.L. Williams\affil{1}, Laure Zanna\affil{26}, Veronika Eyring\affil{3,4}, and Rose Yu\affil{2}}

\affiliation{1}{Scripps Institution of Oceanography, University of California San Diego}
\affiliation{2}{Halicioglu Data Science Institute, University of California San Diego}
\affiliation{3}{University of Tübingen, connAIx Research School for Applied AI, Heilbronn, Germany}
\affiliation{4}{University of Tübingen, Tübingen AI Center, Tübingen, Germany}
\affiliation{5}{Deutsches Zentrum für Luft- und Raumfahrt e.V. (DLR), Institut für Physik der Atmosphäre, Oberpfaffenhofen, Germany}
\affiliation{6}{Schmidt Sciences}
\affiliation{7}{Jet Propulsion Laboratory, California Institute of Technology}
\affiliation{8}{Allen Institute for Artificial Intelligence}
\affiliation{9}{Lawrence Livermore National Laboratory}
\affiliation{10}{University of Colorado Boulder}
\affiliation{11}{Lawrence Berkeley National Laboratory and the University of East Anglia}  
\affiliation{12}{Columbia University}
\affiliation{13}{NASA Goddard Institute for Space Studies}
\affiliation{14}{Met Office, Exeter, UK}
\affiliation{15}{Google Research}
\affiliation{16}{Alfred Wegener Institute}
\affiliation{17}{National Center for Atmospheric Research}
\affiliation{18}{Otto von Guericke University Magdeburg}
\affiliation{19}{University of Bristol, UK}
\affiliation{20}{University of California, Los Angeles}
\affiliation{21}{NVIDIA Research}
\affiliation{22}{California Institute of Technology}
\affiliation{23}{Norwegian Meteorological Institute}
\affiliation{24}{University of Chicago}
\affiliation{25}{University of Arizona}
\affiliation{26}{New York University}

\correspondingauthor{Duncan Watson-Parris}{dwatsonparris@ucsd.edu}

\begin{keypoints}
\item A three-tier protocol evaluates physics-based, data-driven, and hybrid models against diagnostics informative for mid-century projections.
\item Ensembles are scored against held-out post-2015 observations, paleoclimate proxies, and perfect-model simulations using probabilistic tests.
\item All evaluation code, observational and perfect-model data are made publicly available  to drive measurable progress in climate projection.
\end{keypoints}

\begin{abstract}
We present ClimateBench v2, a standardized protocol for evaluating climate models on diagnostics expected to be informative for their skill in projecting mid-century regional temperature and precipitation changes. The protocol is designed to evaluate any physics-based, data-driven, or hybrid climate model on equal footing using a common set of observational and out-of-distribution tests. We define three tiers of evaluation. Tier I establishes physical credibility through entry-ticket tests of energy conservation, coupled (co-)variability, and basic forced responses. Tier II scores models against post-2015 observations of surface temperature, precipitation, radiative fluxes, sea ice, and key modes of variability using fair CRPS as the primary probabilistic score, complemented by distributional and ensemble-consistency diagnostics. Tier III tests out-of-distribution generalization through paleoclimate simulations spanning the Last Interglacial, Last Glacial Maximum, and Mid-Holocene, and through perfect-model experiments in which data-driven models must predict the future climate of existing Earth system models from historical data alone. We reserve all observational data after 2015 for testing, and submissions must include multiple ensemble members to enable probabilistic evaluation. This reservation exploits a new opportunity provided by the decade of observations accumulated since the end of the CMIP6 historical experiment, which constitutes an out-of-sample record of forced climate change (and internal variability) for the current generation of models, and we quantify, in an idealized setting, the information it carries about mid-century warming. We provide the evaluation code, observational reference datasets, and perfect-model training data as an open benchmark to drive measurable progress in climate projection across all modeling approaches.
\end{abstract}

\section*{Plain Language Summary}
Climate models are essential tools for understanding and projecting future climate change. However, many climate models exist, and they often produce different projections. ClimateBench v2 is a benchmark of standardized tests that compares how well different climate models capture aspects of the climate system thought to matter for projecting regional temperature and rainfall changes by 2050. The benchmark is designed to test three types of models: traditional physics-based Earth system models, new data-driven models, and hybrid approaches that combine physics and machine learning (ML). We evaluate models using observations after 2015 and by checking whether they can correctly simulate past climates different from today. By providing open access to all code and data, we aim to help researchers improve climate projections across all modeling approaches.

\section{Introduction}
Climate models are central to our understanding of past, present and future climate change and underpin the scientific basis for international climate policy~\cite{ipccClimateChange20212021}. Since the first coupled atmosphere--ocean simulations, successive generations of models have grown dramatically in scope and complexity, now additionally representing atmospheric chemistry, the carbon cycle, ice sheets, and dynamic vegetation \cite{randall100YearsEarth2019}. The degree of confidence we place in these models depends on whether we can quantify the uncertainty in their projections~\cite{allenQuantifyingUncertaintyForecasts2000,knuttiClimateModelProjection2017}. Climate change, however, is generally an out-of-distribution problem: we are projecting into conditions the Earth system has not experienced during the instrumental period, and unlike weather forecasting, the opportunities for timely verification of the predictions we actually care about are limited \cite{watson-parrisMachineLearningWeather2021}. This makes evaluating and scoring climate models both critically important and fundamentally difficult \cite{eyringTakingClimateModel2019}.

This question has taken on new urgency as recent improvements in machine learning and data-driven modelling have expanded both the opportunities for improving climate projections and the diversity of models that produce them: by supporting improved calibration of physical models; by enabling parameterizations learnt from high-resolution simulations or from observations; and by replacing certain model components, or even whole models, with machine-learned surrogates. Throughout this paper we use `climate model' to mean any model that, given a scenario of external forcings, produces the multi-decadal statistical properties of the Earth system (means, variability, and extremes) regardless of how it is constructed. Within this broad class we adopt the terminology of \emph{physical} models, built from discretized fluid dynamics with process-based parameterizations; \emph{data-driven} models, which learn their dynamics and/or parameterizations directly from simulation output or observations with no a priori commitment to physical structure (encompassing emulators and AI weather and climate models); and \emph{hybrid} models, which combine the two by embedding learned components, typically replacing individual parameterizations, within an otherwise physical dynamical core \cite{raspDeepLearningRepresent2018,kochkovNeuralGeneralCirculation2024,arcomanoHybridAtmosphericModel2023}. These classes differ not only in implementation but in the structural commitments they make: physical and hybrid models inherit standard assumptions such as the vertical column representation of subgrid-scale physics, while data-driven models make no such a priori commitments about scale interactions, which is both a potential strength and a source of new failure modes \cite{hakimDynamicalTestsDeep2024}.

Any benchmark aiming to span these diverse model classes must confront the fundamental realities that separate climate projection from weather forecasting or computer vision. Chiefly among these is that our primary quantity of interest, the multi-decadal forced response, is never directly observable on a useful timescale, meaning a benchmark cannot serve as a simple held-out test set. Furthermore, data-driven emulators easily achieve low error on present-day statistics by exploiting shortcut features or memorizing historical forcing correlations, mechanisms that inevitably break down under unseen emissions pathways. Crucially, climate projections are inherently probabilistic. Because the true object of evaluation is a distribution of outcomes shaped jointly by forced responses and internal variability, a single deterministic projection is simply not a meaningful metric on climate timescales. Alongside these challenges, however, a new opportunity has opened. The CMIP6 historical experiment ends in 2014, and the decade of well-observed, rapid climate change since (unavailable to the development and tuning of the current generation of physical models, and withheld from data-driven submissions by this protocol's train/test contract) constitutes the first community-wide out-of-sample record of the forced response available while the models it tests are still in active use. A central aim of this protocol is to exploit this record, and we quantify the substantial information it carries about mid-century warming in Section~\ref{sec:task}. 

ClimateBench v2 is therefore organized around four design principles: (i) physical admissibility as a necessary precondition (Tier I), (ii) probabilistic scoring against a strictly held-out observational record (Tier II), (iii) explicit out-of-distribution stress tests using paleoclimate states and perfect-model experiments (Tier III), and (iv) full transparency of the train/validation/test contract, with all observational data after 2015 reserved for testing and all evaluation code, reference datasets, and baselines openly released.

Thus, although this protocol is motivated by the rapid recent development of data-driven models, it is designed to place all classes of climate model on an equal footing. Section~\ref{sec:background} reviews the extensive body of work on climate model testing and evaluation that this effort builds upon. The rest of the paper defines the specific task we wish to benchmark (Section~\ref{sec:task}) and outlines three levels of tests against this task (Sections~\ref{sec:tier1}-\ref{sec:tier3}), before providing a discussion of the choices we made in doing so (Section~\ref{sec:discussion}) and concluding (Section~\ref{sec:conclusion}).

\section{Background}
\label{sec:background}

Substantial efforts have been made to evaluate successive generations of climate models against observations \cite{eyringStrategyProcessOriented2005}. These multi-model comparisons were made possible through the definition of common simulation protocols and forcings defined for example as part of the Coupled Model Intercomparison Project (CMIP), now in its seventh phase \cite<CMIP7,>{dunneEvolvingCMIP2025}. Over the last decades, evaluation tools such as ESMValTool \cite{eyringEarthSystemModel2020,lauerEarthSystemModel2020}, PCMDI Metrics Package \cite{glecklerPerformanceMetricsClimate2008}, and ILAMB \cite{collierInternationalLandBenchmarking2018} have enabled increasingly systematic and reproducible evaluation across variables, regions, timescales, and Earth system components. For most large-scale indicators of climate change, the latest generation CMIP6 models show improved performance compared to their CMIP5 predecessors, even though some important differences from observations remain, for example in regional precipitation patterns \cite{eyringHumanInfluenceClimate2021,bockQuantifyingProgressCMIP2020}.

A key aspiration emerging from coordinated CMIP activities has been to move beyond post-hoc evaluation toward systematic, open, and rapid performance assessment, with community tools applied to model output as soon as it is published to the ESGF \cite{eyringImprovedRoutineEvaluation2016}. This has now been realized through the CMIP7 Rapid Evaluation Framework \cite<REF,>{hoffmannRapidEvaluationFramework2025}, a community-driven, containerized platform that automatically executes standardized evaluations as soon as model data are published, giving developers immediate feedback on model biases. Yet these efforts focus on evaluating the historical simulation against Earth observations rather than on performance-based weighting of climate projections.

The prevailing paradigm of `model democracy', in which each model is treated as equally plausible, persists despite growing evidence that equal weighting is suboptimal \cite{knuttiClimateModelProjection2017,eyringTakingClimateModel2019}. \citeA{glecklerPerformanceMetricsClimate2008} and \citeA{reichlerHowWellDo2008} showed that model rankings vary considerably across variables, cautioning against any single index of performance, and \citeA{knuttiChallengesCombiningProjections2010} demonstrated that present-day skill (at the time of writing) related only weakly to the magnitude of future projections; we note an important shift in this assessment given present observations (Section~\ref{sec:task}). More recent model weighting schemes that account for both performance and inter-model dependence have shown promise for specific applications \cite{knuttiClimateModelProjection2017,merrifieldClimateModelSelection2023}, and IPCC AR6 applied observational constraints to assessed ranges of global surface temperature, ocean heat uptake, and sea-level rise \cite{leeFutureGlobalClimate2021}, but developing robust constraints for precipitation, atmospheric circulation, and climate extremes remains a challenge and there remains no consensus on how observations of the present climate can be used to constrain projections of the future more broadly.

Despite the extremely large number of Earth system measurements available, especially in recent times, observations of long-term statistical changes due to anthropogenic forcing are much sparser, leaving few opportunities for out-of-sample tests. Furthermore, the substantial role of internal variability in the historical evolution of the climate system makes it difficult to discern which aspects of long-term change reflect anthropogenically forced signals that these models should reproduce, and which reflect internal variability that they should reproduce in their statistical properties but not the exact time evolution. The ForcesMIP project provided a recent comparison of methods to do this, but none are perfect and they each have their strengths and weaknesses \cite{willsForcedComponentEstimation2026}. Single Model Initial-Condition Large Ensembles (SMILEs) have become an important tool for addressing this challenge, providing the multi-member realizations needed to separate forced response from internal variability within a given model \cite{deserInsightsEarthSystem2020,maherLargeEnsembleClimate2021}, although they cannot by themselves resolve structural uncertainty across models.

Observational evidence for constraining projections falls into two complementary classes. Higher-frequency fluctuations such as the seasonal cycle, interannual variability, and transient responses to volcanic eruptions offer large signal-to-noise ratios~\cite{Hall19z, Williamson21a, brientReducingUncertaintiesClimate2020, wenzelProjectedLandPhotosynthesis2016, wenzelEmergentConstraintsClimate2014, schlundConstrainingUncertaintyProjected2020}, and via fluctuation-dissipation arguments and model-based evidence can be used to constrain projections of the future \cite<e.g.,>{Schneider21b}. Their robustness across model generations remains debated though  \cite{schlundConstrainingUncertaintyProjected2020,nowackOpinionWhyAll2025}. Long-term trends, by contrast, provide a more direct mapping to the projected changes of ultimate interest but with much lower signal-to-noise relative to internal variability and forcing uncertainty \cite{tokarskaPastWarmingTrend2020}. Paleoclimate proxy reconstructions extend the trend-based evidence to climate states with substantially different CO$_2$ concentrations \cite{tierneyClimatesInformOur2020} and have been used to narrow the upper bound of ECS \cite{cooperLastGlacialMaximum2024,cooperPaleoclimatePatternEffects2026}, with some evidence that models with very high ($>5$~$^\circ$C) or very low ($<2$~$^\circ$C) ECS produce paleoclimate simulations inconsistent with the proxy record \cite{luntPaleoclimateDataProvide2024}.

It should be noted that even `physical models' are not purely physical: they require many semi-empirical relationships fitted to limited observations, and as their complexity has grown they are increasingly at risk of being overfitted to the present climate. Moreover, the conceptual frameworks of many parameterizations were developed decades ago for resolutions of O(100~km)--O(1000~km), and rely on assumptions such as scale separation between resolved and unresolved scales that become inadequate at today's O(10~km) resolutions \cite{arakawa2004cumulus}, where processes such as convection and turbulence are only partially resolved \cite{schneiderOpinionOptimizingClimate2024,shawOtherClimateCrisis2025}. The same tension arises in the ocean, where mesoscale eddies are only partially resolved at the resolutions typical of current ESMs, and learned closures have been developed both for mesoscale eddy fluxes \cite{zannaDataDrivenEquationDiscovery2020,guillauminStochasticDeepLearning2021,perezhoginGenerativeDataDriven2023} and for vertical mixing in the surface boundary layer \cite{saneParameterizingVerticalMixing2023}. Systematic benchmarking of such closures has already shown that offline skill is a poor predictor of online performance, and that learned closures may fail to generalize under distribution shift \cite{rossBenchmarkingMachineLearning2023}, a finding that directly motivates the online, out-of-distribution emphasis of the present protocol \cite{eyringPushingFrontiersClimate2024}. The recently proposed AIMIP protocol \cite{AIMIP} provides a useful intermediate step, mirroring the trajectory of physical models from AMIP to CMIP by defining matched prescribed-SST experiments and data standards that enable direct comparison of data-driven, hybrid, and physical atmospheric models on a common footing. ClimateBench v2 extends this approach to the fully coupled Earth system; future coupled data-driven and hybrid models will need to adhere to analogous experimental and data standards to be admissible for benchmarking and CMIP submission. Benchmarks of this kind will be essential to guide and assess progress towards more accurate and trustworthy simulations across all model classes \cite{reichlerHowWellDo2008}.

ClimateBench~v1 \cite{watson-parrisClimateBenchV10Benchmark2022} was designed to evaluate climate model emulators: statistical or machine-learning models trained to reproduce the input--output relationship of a full-complexity Earth system model at a fraction of the computational cost. For that class of model, evaluation is relatively straightforward; the emulator can be tested against held-out simulations from the ESM it was trained on. Emulators of this kind have since proven remarkably effective at reproducing global and regional patterns of temperature and precipitation response to forcing scenarios \cite{bouabidFaIRGPBayesianEnergy2024,mathisonRapidapplicationEmissionstoimpactsTool2025,tebaldiEmulatorsClimateModel2025}, but are, by construction, limited to interpolating within the behavior of the parent model and cannot improve upon its biases or extend beyond its range of simulated climates. For models that generate their own climates -- whether physical ESMs, data-driven models of the atmosphere \cite{watt-meyerACE2AccuratelyLearning2025} or ocean \cite{dheeshjithSamudraAIGlobal2025}, coupled data-driven systems combining the two \cite{cresswell-clayDeepLearningEarth2025,duncanSamudrACEFastAccurate2026}, or hybrid approaches that embed learned components within a physical dynamical core such as NeuralGCM \cite{kochkovNeuralGeneralCirculation2024,arcomanoHybridAtmosphericModel2023} -- the emulation framing is insufficient, and we need to define a suitable observational ground truth against which all models can be compared on equal terms.

\section{Task definition}
\label{sec:task}

We now define this prediction task more precisely. Because climate projections are based on a complex interplay of physical, chemical and even biological processes and have a broad set of uses, it is difficult to define a single set of success criteria for them. In order to keep the protocol manageable, we define a fairly narrow and specific use-case for this objective: ``\textbf{Given a plausible future emissions pathway, what will regional temperature and precipitation changes be in 2050 relative to 1995-2014?}''. By framing the task around 2050, we minimize the effects of current ocean conditions and set aside concerns about deep-ocean equilibration, which will take much longer. It is also the period at which the model uncertainty dominates over both internal variability and scenario uncertainty in future projections for many variables and regions \cite{hawkinsPotentialNarrowUncertainty2009}. By using a relatively recent baseline period we also minimize uncertainties in the observational comparisons (as compared to e.g. 1850).

This framing also reflects an asymmetry in what the observational record can constrain: the equilibrium climate sensitivity is considerably harder to constrain than the transient climate response. Observed warming trends provide a powerful constraint on multi-decadal warming but leave the long-term equilibrium response highly uncertain \cite{allenQuantifyingUncertaintyForecasts2000,knuttiClimateModelProjection2017,bauerLearningClimateSensitivity2025,tokarskaPastWarmingTrend2020}, and recent work has shown that targeted model modifications can substantially shift ECS while leaving TCR essentially unchanged \cite{duffyHighECSCESM22026}. Taken together with the policy relevance of multi-decadal warming, this asymmetry motivates focusing the benchmark on the transient, decadal-to-mid-century response rather than the equilibrium response \cite{knuttiBeyondEquilibriumClimate2017,sherwoodAssessmentEarthsClimate2020}.

So, we define a climate model scoring framework based on observed changes in the recent past and over three paleoclimate time slices spanning the last 127,000 years: the Last Interglacial (127 ka), the Last Glacial Maximum (21 ka), and the Mid-Holocene (6 ka). The premise is that skill at reproducing the climate of these periods is informative about skill at predicting mid-century change; this is a premise the benchmark is designed to test rather than to assume, and Section~\ref{sec:task} shows that the reserved observational record already carries quantifiable information about the mid-century target. The diagnostics scored here are necessary proxies for skill at the 2050 target, not a direct measurement of it, since the 2050 change is by construction unobservable at the time of evaluation. The post-2015 observational window in particular is short and dominated by a single decade of internal variability, so consistency with it is best read as a necessary condition for trustworthy mid-century projection rather than a sufficient one.

Because of the wide range of possible AI approaches and the rapidly evolving nature of climate modelling, submitted models may not conform to the implicit assumptions we hold about how `traditional' climate models behave. Regardless of its underlying architecture, any candidate model must successfully capture the forced response to changing greenhouse gases, handle the complex adjustments of aerosol precursor emissions, and accurately reproduce the statistical properties of internal variability. This underlying internal variability sets the baseline envelope of outcomes for probabilistic projections. Because climate prediction is fundamentally an out-of-distribution problem lacking an immediate verification target, we must instead rely on strict adherence to physical laws as our primary surrogate for trustworthiness.

The reservation of the post-2015 record exploits an opportunity that was not available to previous generations of models. The CMIP6 historical experiment ends in 2014, and the physical models of that generation were developed, calibrated, and tuned against observations available before that date; our benchmark protocol requires that data-driven submissions restrict their training to the same period (Section~\ref{sec:tier2}). The decade of observations accumulated since is the best-observed period in history, spanning nearly a third of the total warming to date, and therefore constitutes an acceptably out-of-sample record of the forced response for an entire generation of models. Out-of-sample evaluations of climate projections have previously been possible only in retrospect, decades after the projections were made and long after the models themselves had been superseded \cite{hausfatherEvaluatingPerformancePast2020}. The present moment is different: this record has accumulated while the models of the CMIP6 era are still in active use, and just as a new class of data-driven models, for which train/test separation is both essential and natural, emerges.

\begin{figure}[htbp]
    \centering
    \includegraphics[width=\linewidth]{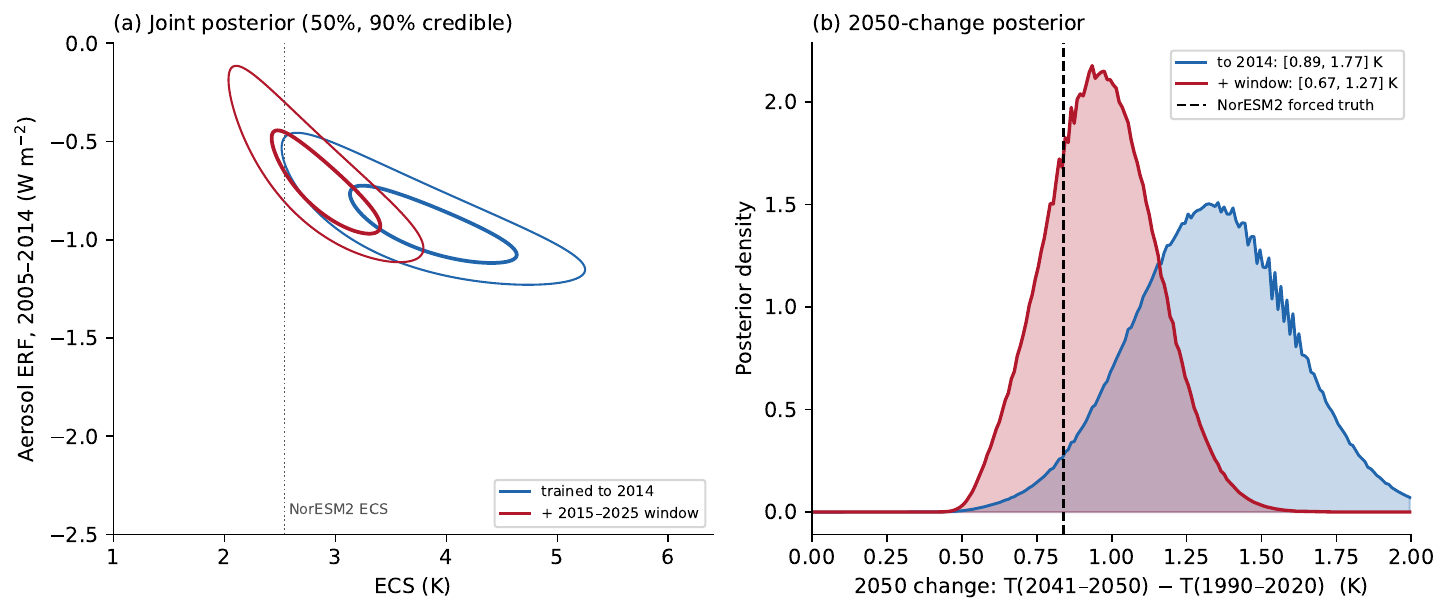}
    \caption{Quantifying the information the reserved test period carries about the benchmark target, following \citeA{watson-parrisIntegratingTopDown2025}. (a) Joint posterior over equilibrium climate sensitivity (ECS) and aerosol forcing amplitude, conditioned on the NorESM2-LM \cite{selandOverviewNorESM22020} global-mean temperature record to 2014 only (blue) and additionally on the 2015--2025 test window (red); contours enclose 50\% and 90\% of the posterior mass. (b) The corresponding posteriors for the benchmark target, the 2050 change $T(2041$--$2050) - T(1990$--$2020)$: adding the test window reduces the posterior variance by ${\sim}50\%$ (42--54\% across the three realizations of internal variability) and recenters the distribution on the true forced response of NorESM2-LM (dashed). Full methods are given in~\ref{app:testwindow}.}
    \label{fig:testwindow}
\end{figure}

Moreover, the information this record carries about the benchmark target can be quantified, at least in an idealized setting. Following the approach of \citeA{watson-parrisIntegratingTopDown2025}, we represent the climate system with a probabilistic version of the FaIR impulse-response model \cite{leachFaIRv200GeneralizedImpulse2021,bouabidFaIRGPBayesianEnergy2024}, treat the equilibrium climate sensitivity and the amplitude of the aerosol forcing as uncertain, and perform a Bayesian update of these parameters against the global-mean temperature record simulated by a full ESM (NorESM2-LM; \citeA{selandOverviewNorESM22020}) under SSP2-4.5. Conditioned on the record to 2014 alone, the posterior exhibits the well-known degeneracy between climate sensitivity and aerosol forcing \cite{kiehlTwentiethCenturyClimate2007}, and the implied 2050 warming remains spread over almost a kelvin (Figure~\ref{fig:testwindow}). Adding the 2015--2025 test window breaks this degeneracy: the posterior variance of the 2050 change is reduced by roughly half, and the posterior recenters on the underlying ESM's true forced response. Notably, the information resides almost entirely in the \emph{realized warming since 2014} (the mean level of the window relative to the training period, an integrated and internal-variability-robust measure of the recent warming rate; \citeNP{watson-parrisIntegratingTopDown2025}), rather than in the trend within the window, which at this length is dominated by internal variability. This is an idealized, global-mean demonstration, with a single ESM as truth and simplified uncertainties, and constraining regional changes will certainly be harder; on the other hand, present-day regional and seasonal observations carry additional information that is not exploited here. Global-mean temperature is in any case the natural first target, since regional impacts are commonly assessed as functions of global warming level \cite{leeFutureGlobalClimate2021}.

That the recent warming record carries information about future warming is not itself new: emergent-constraint analyses have used observed trends to narrow the CMIP6 projection range, finding that models with the strongest recent warming are least consistent with observations \cite{tokarskaPastWarmingTrend2020}. What the protocol adds is to treat that record as a formally held-out test set rather than as a post-hoc constraint applied across an ensemble, and Fig.~\ref{fig:testwindow} quantifies, in an idealized single-model setting, how much of the mid-century uncertainty the reserved window resolves. We are careful, however, to distinguish this from a stronger claim that the demonstration does not support: that probabilistic scores of free-running models against the reserved record rank those models by their mid-century accuracy. The information in Fig.~\ref{fig:testwindow} resides in a single global scalar, the realized warming level, and a single realized value of a scalar cannot reliably rank models (Section~\ref{sec:scoring}). The protocol therefore uses the realized warming level as a \emph{falsification test}, reported prominently for every submission, of whether a model is consistent with the observed forced response (Section~\ref{sec:trends}); the Tier~II scores measure a model's \emph{adequacy} against the held-out record; and the link between those scores and skill at the 2050 target remains a working hypothesis, which the benchmark is designed to test as observations accumulate rather than one we assume.

This leads us to define three tiers of metrics of increasing difficulty against which models will be measured. First, we enumerate a set of physical laws which all climate models must satisfy to be considered. Some of these laws may be built into physics-based climate models but are far from guaranteed in data-driven models. Second is the ability to reproduce observed climate with particular focus on features which are informative about response to forcing like diurnal and seasonal cycles or historical trends \cite{simpsonConfrontingEarthSystem2025}. Our final category focuses on the ability to capture future climate. Paleo records are particularly useful for ensuring appropriate sensitivity to a variety of external forcings. We also include ``perfect model experiments'', which unlike the previous two categories, are only relevant for ML-based models since climate change is inherently an emergent property of physics-based models (as opposed to a target). Finding a sufficient condition for trusting climate model projections is almost impossible because the future is unknowable, but an ML model that can correctly predict the future of several different physics-based climate models when trained on their current-climate data would be a compelling tool.

The experiments described below follow the protocols defined for the CMIP6 ensemble \cite{eyringOverviewCoupledModel2016}, and many of the required simulations (piControl, historical, abrupt-4xCO2, amip, and the PMIP4 paleoclimate experiments) already exist for many physics-based models. Where those simulations are available, they may be submitted directly; no new runs are required. For data-driven or hybrid models without an existing CMIP6 submission, the same experimental designs should be followed as closely as the model architecture allows. All simulations are driven by emissions and concentrations as specified by the input4MIPs datasets used for CMIP6 \cite{durackStandardizedDataSets2018}, with extensions to the present day following the approach adopted by the CMIP6 historical extensions. This ensures consistency across model types and allows scores to be compared on equal terms. A complete list of required and extended simulations, with ensemble requirements and estimated computational cost, is provided in Table A1. Throughout the protocol, tests, simulations, and output variables are designated as either \emph{required} or \emph{extended}. Required elements constitute the minimum conforming submission, and all applicable required Tier I tests must be passed for a model to be scored. Because the protocol admits architectures that do not represent every component of the climate system, a required test may additionally be recorded as \emph{not applicable by architecture} when a submission does not produce the state it diagnoses (for example, a model that predicts no winds or geopotential, or that has no ocean component). Such a test neither gates entry nor counts against the model, and the scorecard reports which required tests were applicable to each submission. The designation must be declared in the model card (Section~\ref{sec:discussion}), and cannot be claimed for a quantity the model does produce. Extended elements are optional, typically because they demand additional simulations or non-standard output; they do not gate entry, but their results are reported as additional entries in the scorecard alongside the required scores.

\section{Tier I: Entry Ticket}
\label{sec:tier1}
Before a model can be assessed in terms of ability to reproduce observations, it is important to perform `sanity checks' to ensure its fundamental viability. This necessity stems from the nature of climate projections as a strictly out-of-distribution problem; unlike weather forecasting, we lack benchmarks for mid-century projections on practically useful timescales. Consequently, we must rely on adherence to physical laws as a surrogate for the unavailable prediction benchmark. Validating that a model respects these fundamental constraints establishes a necessary, but not sufficient, floor of physical plausibility. This guards against gross failures and overfitting, while leaving the deeper question of whether a model is right for the right reasons to later tiers and to ongoing process-level evaluation. While some of these tests apply to all models, others will evolve as we learn more about the specific failure modes of data-driven architectures.

Table~\ref{tab:tier1} summarizes the Tier~I tests, the experiments they require, and their pass criteria. Except where noted, all thresholds are applied to the ensemble mean; the required tests must all be passed for a model to proceed to scoring in Tiers~II and III, except those recorded as not applicable by architecture (Section~\ref{sec:task}).

\begin{table}[ht]
\caption{Summary of Tier~I entry-ticket tests. All criteria are pass/fail. A required test is recorded as not applicable by architecture for a submission that does not produce the state it diagnoses (Section~\ref{sec:task}). Experiments follow the CMIP6/PMIP4 protocols (Table~\ref{tab:simulations}); precise definitions and formulae are given in ~\ref{app:tier1}.}
\label{tab:tier1}
\small
\begin{tabular}{p{3.6cm}p{2.9cm}p{5.0cm}l}
\hline
Test & Experiment(s) & Pass criterion & Status \\
\hline
Energy balance & piControl & $|\mu(N)| < 0.5$~W~m$^{-2}$; drift $|\delta| < 0.02$~W~m$^{-2}$~decade$^{-1}$ & Required \\
Water budget closure & piControl & $|\langle P\rangle-\langle E\rangle| < 0.05$~mm~day$^{-1}$ & Required \\
Atmospheric energy closure & piControl & $|\langle Q_\text{rad}\rangle-(\langle LP\rangle+\langle\text{SHF}\rangle)| < 2$~W~m$^{-2}$ & Required \\
Clear-sky LW feedback & abrupt-4xCO2 & forced global-mean $\Delta\text{rlutcs}/\Delta T_s$ within $\pm25\%$ of $1.9$~W~m$^{-2}$~K$^{-1}$ & Required \\
Geostrophic balance & historical (daily) & $\rho(u,u_g)$ at 850~hPa, $30$--$60^\circ$, $> 0.9$ & Required \\
Precipitation--buoyancy & historical & $P'$--MSE$'$ slope (1983--2014) within $\pm30\%$ of GPCP/ERA5 & Required \\
SST-patch feedback contrast & amip, patch-EP, patch-WP & $\Delta\lambda > 0.5$~W~m$^{-2}$~K$^{-1}$ & Extended \\
Fast CO$_2$ forcing & amip-4xCO2 & ERF $\in [6.5, 9.0]$~W~m$^{-2}$ & Required \\
ENSO amplitude, spectrum, teleconnections & piControl ($\geq$100~yr) & $\sigma_\text{N34}\in[0.5,1.4]$~K; interannual (2--7~yr) spectral peak; teleconnection pattern correlation $>0.7$ & Required \\
MJO propagation & piControl (daily) & eastward/westward power ratio $> 1.5$ & Extended \\
Land--ocean warming contrast & abrupt-4xCO2 & ratio $\in [1.2, 1.6]$ & Required \\
Arctic amplification & abrupt-4xCO2 & $(\Delta T_{>66.5^\circ\text{N}})/\Delta T_\text{global} \geq 1.5$ & Required \\
ECS (Gregory regression) & abrupt-4xCO2 & $\in [1, 7]$~K (recorded as metadata) & Required \\
Aerosol forcing & hist-aer & ERF(2015) $\in [-2.0, -0.5]$~W~m$^{-2}$; $\Delta T$(2015) $< 0$ & Required \\
Meridional heat transport & piControl & ocean peak 1.5--2.0~PW at 15--20$^\circ$; atmosphere peak 4--5~PW at $41\pm2.5^\circ$N & Required \\
ITCZ--EFE relationship & historical & ITCZ lagged 2 months behind cross-equatorial transport; $|\text{slope}|$ within $\pm30\%$ of ${\sim}3^\circ$~PW$^{-1}$; $|r| > 0.85$ & Required \\
\hline
\end{tabular}
\end{table}

\begin{figure}[htbp]
    \centering
    \includegraphics[width=\linewidth]{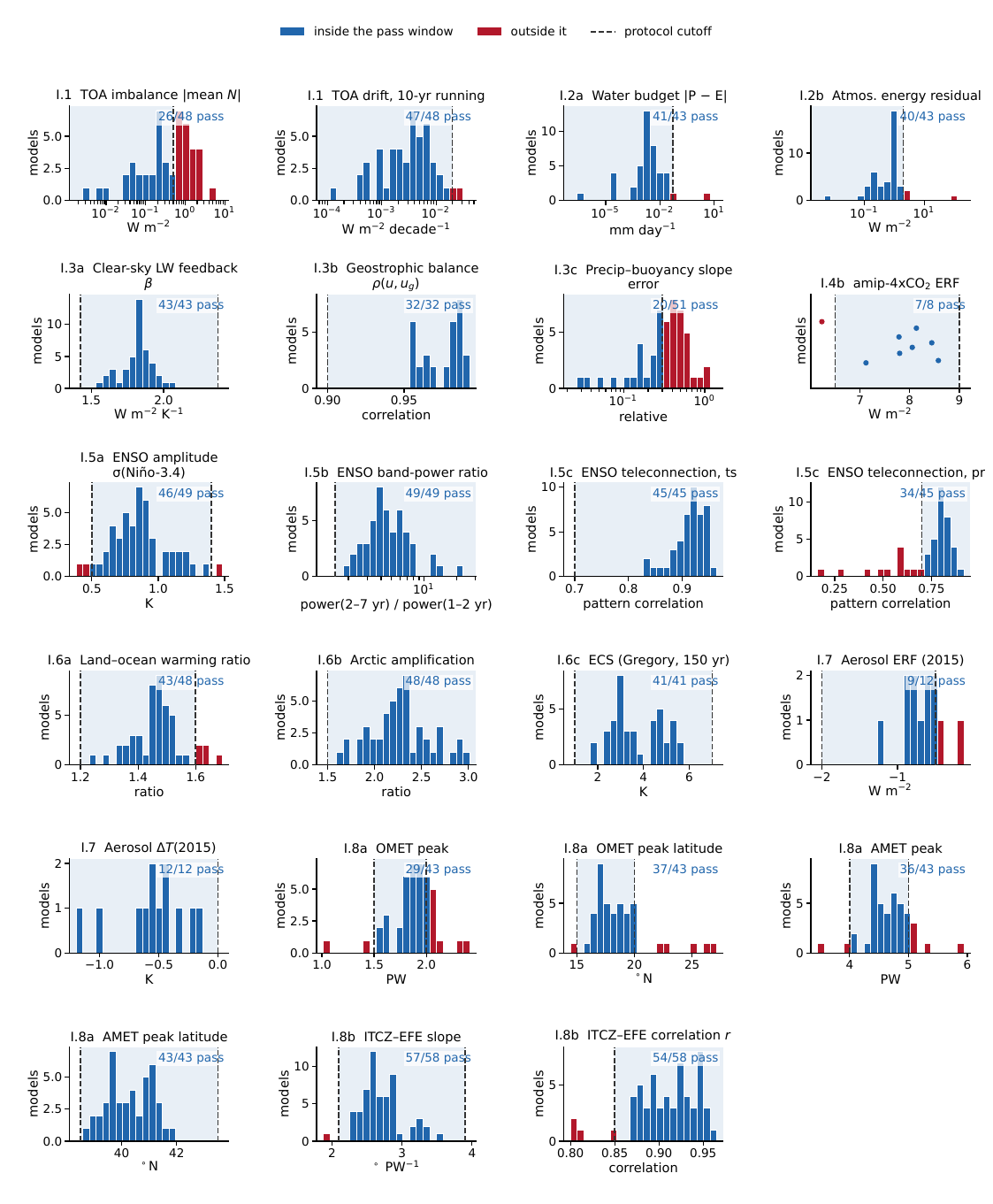}
    \caption{Distribution of each Tier~I gate statistic across the 60 CMIP6 models with required Tier~I output; shading and dashed lines mark the protocol pass window (Table~\ref{tab:tier1}). Models inside the pass window are shown in blue and those outside it in red, and each panel gives the number of models passing out of those for which the test could be evaluated. The fast CO$_2$ forcing test (I.4b), available for only 8 models, is shown as individual points. Panel labels (I.1--I.8b) follow the row order of Table~\ref{tab:tier1}; extended tests are not shown.}
    \label{fig:tier1hist}
\end{figure}

Several required tests rely on experiments from the CFMIP and DAMIP programmes that only a subset of CMIP6 models performed: of the 60 CMIP6 models for which we could evaluate required Tier~I output, 8 provide amip-4xCO2 (fast CO$_2$ forcing) and 12 provide hist-aer (aerosol forcing), and only 7 provide every required input. The CMIP6 results in Figure~\ref{fig:tier1hist} therefore report each test over the models for which it could be evaluated rather than as a joint pass rate, and for new submissions alternative routes to the same quantities (e.g., RFMIP-style fixed-SST forcing diagnostics; \citeNP{smithEffectiveRadiativeForcing2020}) may be accepted. Figure~\ref{fig:tier1hist} shows the distribution of each Tier~I statistic across the CMIP6 ensemble together with its pass window. A CMIP6 model failing a gate does not by itself show that the gate is miscalibrated: the thresholds are set from physical reasoning and observations rather than tuned to the CMIP6 ensemble, and many of the failures in Figure~\ref{fig:tier1hist}, notably in the piControl TOA imbalance and the aerosol forcing, reflect known deficiencies of current models, while for others, such as the precipitation--buoyancy relationship and the peak ocean heat transport, it remains open how much reflects model error and how much the tightness of the threshold. We will monitor pass rates as submissions accumulate and revise the gates as evidence emerges, with each change made at a protocol version increment and its rationale published (Section~\ref{sec:discussion}).

\subsection{Energy Balance Constraint}
The most critical of these tests involves the global energy budget. Climate change is first and foremost a perturbation of Earth's energy balance: increased greenhouse gas concentrations and anthropogenic aerosols modulate radiative energy fluxes, eventually driving the system toward a new equilibrium where absorbed shortwave radiation matches outgoing longwave radiation. It follows that a trustworthy model must accurately capture these radiative balances. For a model to establish a pre-industrial steady state that does not exhibit runaway warming or cooling independent of forcing, it must close the energy balance at the top of the atmosphere. While strict energy conservation to machine precision is the ideal, we recognize that not all models, whether physical or data-driven, achieve this due to numerical complexities.

We test the stability of the simulated climate system through two related quantities computed over the last 100 years of the piControl simulation: the long-term mean of the net top-of-atmosphere radiative imbalance, and the drift of its decadal running mean (Table~\ref{tab:tier1}; precise definitions in~\ref{app:tier1}). Minimizing both is desirable because it cleanly isolates the anthropogenic perturbation of the energy balance from numerical drift or residual spin-up in components with high thermal inertia, such as the deep ocean. We recognize that a 500-year control run is a substantial request, particularly for expensive or novel architectures; the mean-imbalance and drift criteria can be meaningfully evaluated on shorter control runs (of order 100 years), albeit with reduced power to detect slow drifts, and such submissions will be accepted with the shorter record length noted in the scorecard.

\subsection{Budget Closures and Physical Covariances}
Beyond unforced energy stability, models must respect a set of well-established physical relationships among state variables. We separate these into closure constraints, where a residual is computed and bounded, and diagnostic covariances, where a regression or correlation statistic is computed and compared against observations.

The \textit{closure constraints} require the global-mean water budget and the global-mean atmospheric energy budget to close in the piControl simulation, to within tolerances set respectively by the precision typical of current ESMs and by observational uncertainty in the atmospheric radiative cooling rate \cite<e.g.>{stephensUpdateEarthsEnergy2012}; the two tolerances are energetically consistent (a surface water imbalance of $0.05$~mm~day$^{-1}$ corresponds to ${\approx}1.45$~W~m$^{-2}$ of latent heat flux). The \textit{diagnostic covariances} require three well-established relationships to be reproduced quantitatively: the clear-sky longwave feedback in the forced response to CO$_2$ \cite{kollEarthsOutgoingLongwave2018,zhangLinearityOutgoingLongwave2020} and, in the historical simulation, midlatitude geostrophic balance and the tropical precipitation--buoyancy relationship \cite{chouEvaluatingRichGetRicherMechanism2009}. Finally, because covariances alone can be reproduced by spurious correlations, two \textit{causal response tests} probe the causal structure of the response through idealized experiments: paired tropical-Pacific SST-patch experiments following the GFMIP protocol (an extended diagnostic, as they rely on prescribed-SST integrations), whose east--west contrast in global radiative response per unit local warming is robustly positive in CMIP6 models and observational reconstructions \cite{williamsDependenceFastChanges2023, andrewsDependenceGlobalCloud2018}, and the amip-4xCO2 experiment, whose diagnosed effective radiative forcing must fall within the assessed likely range \cite{forsterEarthsEnergyBudget2021}. Precise definitions, formulae, and thresholds for each of these tests are given in \ref{app:tier1}.

\subsection{Coupled Variability}
The El Niño–Southern Oscillation (ENSO) is the dominant mode of interannual climate variability and the Madden–Julian Oscillation (MJO) is the dominant mode of tropical intraseasonal variability; both arise from coupled ocean–atmosphere or atmosphere–convection interactions that cannot be straightforwardly tuned for. Robustly evaluating their statistics requires multi-decadal records that far exceed the post-2015 test window of Tier II. Nevertheless, a model that claims to represent the coupled climate system should produce recognizable ENSO- and MJO-like variability as a basic indicator of credible coupled dynamics (see \citeA{plantonAnnexIVModes2021}).

For models producing their own sea-surface temperatures, we require the Niño-3.4 index, computed over at least 100 years of piControl, to exhibit a realistic amplitude of variability, a spectral peak in the interannual (2--7~yr) band, and tropical teleconnections of the correct sign and approximate magnitude (the precise definitions and thresholds are given in~\ref{app:enso}). We restrict the teleconnection check to tropical signatures, where signal-to-noise is high; extratropical teleconnections (e.g., the Pacific--North American, or PNA, pattern) are poorly constrained even in long ensembles \cite{deserHowWellWe2017} and are not used as pass/fail criteria. As an extended diagnostic, we additionally require coherent eastward-propagating intraseasonal variability, measured by the ratio of eastward- to westward-propagating power in the Wheeler--Kiladis spectrum of equatorial precipitation. This is intentionally a weaker test than the full CLIVAR (Climate and Ocean: Variability, Predictability and Change) MJO Working Group metrics \cite{waliserMJOSimulationDiagnostics2009}, designed to catch models with no coherent MJO-like signal at all. These are intended as sanity checks on the model's ability to generate realistic coupled and intraseasonal variability, not as detailed assessments of ENSO or MJO dynamics, which we defer to future work; the precise criteria are given in \ref{app:tier1}.

\subsection{Basic Forced Responses}
The abrupt-4xCO2 experiment provides a standardized perturbation from which several well-understood forced responses can be evaluated. Three robust features of the greenhouse warming response serve as entry-ticket checks, all evaluated in the annual mean over the last 50 years of the abrupt-4xCO2 experiment, by which point these response ratios are quasi-stationary even though the absolute warming has not fully equilibrated. First, the land--ocean warming contrast: because land surfaces have lower effective heat capacity and reduced evaporative damping relative to the ocean, they warm faster and more strongly under increased CO$_2$ forcing. Models must produce a land--ocean warming ratio greater than one, consistent with the observed and theoretically expected range of approximately 1.2--1.6 \cite{joshiMechanismsLandSea2008,suttonLandSeaWarming2007,byrneLandOceanWarming2013}. Second, Arctic amplification: the combination of sea-ice albedo feedback, lapse-rate feedback, and increased poleward energy transport produces enhanced warming at high northern latitudes that is among the most robust features of greenhouse forcing across the model hierarchy \cite{pithanArcticAmplificationDominated2014}. We require that the Arctic (north of 66.5$^\circ$N) warms at least 1.5 times the global mean. The seasonal structure of this amplification, which is strongly winter-enhanced, is not enforced at Tier I but is evaluated as part of the Tier II climatological diagnostics. Third, the equilibrium climate sensitivity (ECS) is diagnosed from a Gregory regression of the first 150 years of the abrupt-4xCO2 simulation \cite{gregoryNewMethodDiagnosing2004} and is required to fall within the broad envelope 1~K $\leq$ ECS $\leq$ 7~K. This is a deliberately wide criterion, well outside the IPCC AR6 very likely range of 2.0--5.0~K \cite{forsterEarthsEnergyBudget2021}, intended only to rule out structurally broken responses. The diagnosed value is recorded as metadata for each submission, providing essential context for interpreting a model's transient warming trajectory and allowing systematic comparison across architectures.

\subsection{Aerosol Forcing}
Anthropogenic aerosols represent the largest source of uncertainty in historical radiative forcing, and their net effect is a critical determinant of the transient climate response. As an entry-ticket test, we include an idealized aerosol-only historical simulation (hist-aer), following the DAMIP protocol \cite{gillettDetectionAttributionModel2016}, in which models are driven by time-varying anthropogenic aerosol emissions with all other forcings held at pre-industrial levels. Because no direct observational analog exists for an aerosol-only world, this experiment is evaluated as a pass/fail consistency check rather than scored against observations. Specifically, we require that the net, present-day (2015) aerosol effective radiative forcing (ERF) diagnosed from the simulation (\ref{app:aerosolerf}) falls within the assessed very-likely range of $-2.0$ to $-0.5$~W~m$^{-2}$ \cite{forsterEarthsEnergyBudget2021}, and that the global-mean surface temperature response (also at 2015) is a net cooling relative to pre-industrial. Models that produce aerosol warming or an ERF outside the physically plausible range fail the test. Because the hist-aer experiment is short, widely run, and already available for many CMIP6 models, this check is required; it provides a valuable safeguard against models that achieve realistic historical warming through unphysical compensating errors in aerosol and greenhouse gas forcing.

\subsection{Coupled Diagnostics}
The tests above evaluate individual physical constraints largely in isolation. The following diagnostics assess whether the model produces the correct large-scale interactions between the atmosphere and ocean, as expected from well-established theory.

\subsubsection{Meridional Heat Transport Partitioning}
The climate system transports energy poleward through two fluids whose relative contributions are governed by fundamental differences in mass transport and vertical stratification \cite{marshallMeanClimateVariability2007}. In the tropics, the ocean dominates meridional heat transport: although the mass overturning rates of the atmosphere and ocean are comparable at low latitudes, the much sharper vertical temperature contrast across the oceanic thermocline relative to the vertical moist static energy contrast means the ocean carries more energy per unit mass transported. In the extratropics, the atmosphere takes over: baroclinic eddies drive an atmospheric energy flux roughly five times larger than the oceanic contribution, making the atmosphere the dominant poleward heat carrier poleward of approximately 30$^\circ$. This crossover is a robust feature of the Earth system that emerges from fluid-dynamical first principles and is largely insensitive to continental geometry \cite{marshallMeanClimateVariability2007}. We require that models reproduce it: the implied oceanic and atmospheric heat transports, diagnosed from TOA and surface energy fluxes in the piControl simulation, must peak in the tropics and mid-latitudes respectively, at magnitudes consistent with estimates from ECCO and ERA5 (Table~\ref{tab:tier1}; \ref{app:tier1}). A model that achieves realistic surface temperatures while placing the atmospheric transport peak in the tropics has fundamental errors in its energy budget that will compromise forced projections.

\subsubsection{ITCZ--Energy Flux Equator Relationship}
Theoretical and modeling work has established that the latitude of the ITCZ is governed by the atmospheric energy budget: the zonal-mean precipitation maximum tracks the energy flux equator (EFE), defined as the latitude where the vertically integrated meridional atmospheric energy transport vanishes \cite{kangResponseITCZExtratropical2008,kangTropicalResponseExtratropical2009,schneiderMigrationsDynamicsIntertropical2014}. When a forcing asymmetry cools one hemisphere relative to the other, cross-equatorial energy transport increases and the ITCZ shifts toward the warmer hemisphere at a rate of approximately 3$^\circ$ per petawatt of cross-equatorial flux \cite{donohoeSeasonalCycleAtmospheric2013}. This relationship holds on the real Earth: it governs the seasonal migration of tropical rainfall and the observed ITCZ response to hemispheric aerosol forcing. We evaluate it from the co-variation of the precipitation-maximum latitude and the cross-equatorial atmospheric energy transport (which sets the EFE) over the climatological seasonal cycle of the historical simulation, with the ITCZ latitude lagged two months behind the transport to allow for its delayed response \cite{donohoeRelationshipITCZLocation2013}, requiring a regression slope consistent with the observed value and a high seasonal correlation (Table~\ref{tab:tier1}); the procedure is specified in \ref{app:tier1}.

\section{Tier II: Evaluation and Benchmarking}
\label{sec:tier2}

The core of the ClimateBench v2 protocol is a set of observational metrics against which models are scored on their ability to reproduce the observed climate and its recent changes. We focus evaluation on the quantities that are the direct objects of climate projection such as temperature, precipitation, radiative fluxes, and their statistics, rather than on dynamical intermediaries such as jet position or Hadley cell strength. This is a deliberate choice. While regional climate change is physically mediated by circulation changes, the scope of this benchmark encompasses model architectures that do not resolve atmospheric dynamics in any traditional sense (for example generative emulators such as cBottle; \citeNP{brenowitzClimateBottleGenerative2025}), and we wish to evaluate all models on equal footing. The inclusion criterion for a variable is therefore that it is either a direct target of prediction (e.g. regional temperature or precipitation) or that there is demonstrated observational evidence that correctly simulating that variable improves predictive skill for those targets (e.g. cloud-radiative properties as emergent constraints on climate sensitivity). We note that the dynamical evaluation of ML atmospheric models is the focus of the complementary AIMIP protocol, and we defer to that effort for detailed assessment of circulation fidelity (see Discussion).

\begin{figure}[htbp]
    \centering
    \includegraphics[width=\linewidth]{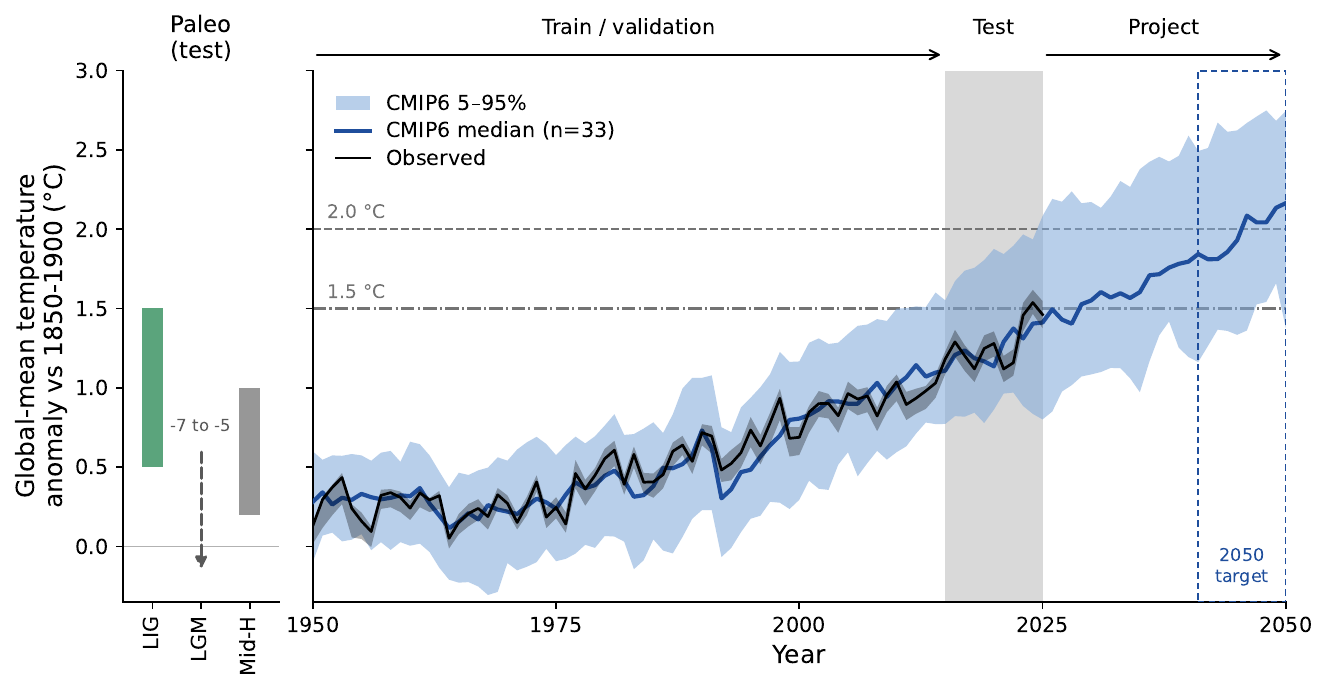}
    \caption{The ClimateBench v2 train/validation/test split. Right: annual global-mean
surface temperature anomalies relative to 1850--1900. The blue line and band show the
median and 5--95\% range across 33 CMIP6 models (\texttt{historical} to 2014 spliced to
SSP2-4.5 thereafter, variant \texttt{r1i1p1f1}, area-weighted global-mean \texttt{tas}). The black line and band show the median and range of three
observational products (HadCRUT5, GISTEMP v4, Berkeley Earth). Observational products
blend land air and sea surface temperatures whereas the model diagnostic is surface air
temperature; no blending correction is applied for this illustration. The grey band marks the reserved post-2015 test window and the dashed box the 2041--2050 benchmark target decade (scored relative to 1995--2014; see Section~\ref{sec:task}). Left: assessed global-mean temperature anomalies for the three Tier~III paleoclimate periods, also relative to
1850--1900, from IPCC AR6 WG1 Chapter~2; the Last Glacial Maximum lies below the plotted
range and is indicated by an arrow.}
    \label{fig:testsplit}
\end{figure}

Given the (relatively) limited amount of climate change observed in the recent past, a key consideration for evaluation is the test period. An important best practice in data driven modelling is to reserve some data (unused during training or validation) for independent model testing. As noted above, this consideration should also apply to any model which has tunable parameters, including traditional models. We propose to reserve all observational data since 2015 for testing (Figure~\ref{fig:testsplit}). While relatively short, this period has the best available data, has seen rapid warming (nearly a third of warming to date), includes some representation of high frequency climate variability, such as ENSO, and already carries substantial information about the mid-century response (Section~\ref{sec:task}, Figure~\ref{fig:testwindow}). The period also coincides with the AIMIP protocol for atmospheric data driven models, and the end of the CMIP6 historical simulations. To provide a sense of the impacts of internal variability on the comparisons above, each model should include at least three ensemble members, and ideally ten; the statistical power that larger ensembles provide for the consistency tests is discussed in Section~\ref{sec:scoring}.

All variables are evaluated over a set of updated historical simulations with time-varying emissions that reflect the recent past. We anticipate updating these annually as new emissions estimates and observations become available, consistent with ongoing efforts within the CMIP community and providing a continuously improving benchmark. The CMIP6-style extensions used for the current Tier~II tests omit two forcing changes inside the test window: the 2020 International Maritime Organization reduction in the sulfur content of marine fuel, and the 2022 Hunga Tonga--Hunga Ha'apai injection of water vapour into the stratosphere. Part of the observed post-2015 anomaly may therefore reflect forcing absent from simulations run with prescribed forcing, and would be scored as model error; this bears most on physical models, whereas a data-driven model could absorb it from data. Both will be represented in the CMIP7 forcing datasets, and the protocol will continue to use the latest available forcing at each update. To balance rigor with accessibility, we distinguish between a required set of core variables that all submissions must provide, and an extended set that is optional but encouraged, whose scores are reported as additional scorecard entries. The full list of required output variables, including CMIP short names, temporal resolution, and tier applicability, is provided in Table A2. Table~\ref{tab:tier2_variables} summarizes the variables, observational references, and temporal resolution used for Tier~II evaluation, with each designated as core (required for all submissions) or extended (optional).

\begin{table}[ht]
\caption{Tier II evaluation variables and observational datasets.}
\label{tab:tier2_variables}
\small
\begin{tabular}{p{4.5cm}lp{5cm}l}
\hline
Variable & Temporal Resolution & Observational Datasets & Status \\
\hline
2m temperature (mean, trends) & Seasonal; annual & NASA GISS \cite{lenssenImprovementsGISTEMP2019}, Berkeley Earth \cite{rohdeBerkeleyEarthTemperature2020}, HadCRUT \cite{moriceUpdatedAssessmentNear2021}, NOAA GlobalTemp \cite{huangImprovementsNOAAGlobalTemp2020} & Core \\
2m temperature (extremes, daily PDF) & Daily & Berkeley Earth (daily), HadGHCND \cite{caesarLargescaleChangesObserved2006} & Core \\
Surface temperature & Seasonal; annual & HadSST, HadISST (ocean) \cite{raynerGlobalAnalysesSea2003}; CRU TS (land) \cite{harrisVersion4CRU2020} & Core \\
Precipitation (mean, trends) & Seasonal; annual & GPCP \cite{adlerGlobalPrecipitationClimatology2018}, IMERG \cite{huffmanIntegratedMultisatelliteRetrievals2020}, MSWEP \cite{beckMSWEPV2Global2019} & Core \\
Precipitation (intensity PDF, extremes) & Daily & IMERG (V07 Final, daily) \cite{huffmanGPMIMERGFinal2023}, MSWEP & Core \\
TOA radiative fluxes (LW, SW; clear-sky and all-sky) and cloud radiative effect & Seasonal; annual & CERES EBAF \cite{loebCloudsEarthRadiant2018} & Core \\
Sea ice extent and trends & Monthly; seasonal & NSIDC Sea Ice Index \cite{fettererSeaIceIndex2017}, OSI SAF \cite{lavergneVersion2EUMETSAT2019}, HadISST \cite{raynerGlobalAnalysesSea2003} & Core \\
Ocean heat content (0-100m, 0--2000m) & Annual & Argo (2005--present), \cite{chengImprovedEstimatesOcean2017,levitusWorldOceanHeat2012} & Extended \\
Surface fluxes (latent, sensible, LW, SW) & Seasonal; annual & CERES SYN (satellite-derived) \cite{rutanCERESSynopticProduct2015}; FLUXNET \cite{pastorelloFLUXNET2015DatasetONEFlux2020}, OceanSITES buoys (in situ) \cite{sendOceanSITES2010} & Extended \\
Cloud properties (LWP, cloud fraction, cloud-top temperature and pressure) & Seasonal; annual & CERES, MODIS (Aqua MYD08\_M3 C6.1) \cite{platnickMODISCloudOptical2017,platnickMODISAtmosphereL32017}, CloudSat/CALIPSO \cite{stephensCloudSatMissionATrain2002}, ISCCP \cite{youngInternationalSatelliteCloud2018}, MAC-LWP \cite{elsaesserMultisensorAdvancedClimatology2017} & Extended \\
Vertically integrated water vapour & Seasonal; annual & RSS \cite{mearsConstructionUncertaintyEstimation2018}, ERA5 (as reference) \cite{hersbachERA5GlobalReanalysis2020}, radiosondes \cite{durreEnhancedComprehensiveExtraction2018} & Extended \\
\hline
\multicolumn{4}{l}{LW = longwave; SW = shortwave; LWP = liquid water path}
\end{tabular}
\end{table}

Ocean heat content is not yet included in the example scorecard (Fig.~\ref{fig:benchmark}): few CMIP6 models publish \texttt{thetao} for more than one historical and SSP2-4.5 member (most provide one to three, and only five provide five or more), so neither the fair CRPS nor the leave-one-out reference score (Section~\ref{sec:baselines}) is well defined for it. We intend to score it against an Argo-constrained reference such as IAP \cite{chengImprovedEstimatesOcean2017} once sufficient ensembles are available.

Surface flux observations require particular care as an evaluation reference. Observation-based synthesis products (CERES SYN, OAFlux, HOAPS) do not close the global energy budget, with residuals of $\mathcal{O}(10)$~W~m$^{-2}$ arising from known biases in latent heat flux over the tropical warm pool and inconsistencies in the SST products used across individual flux components \cite{yuGlobalAirSea2019,croninAirSeaFluxes2019}. Reanalyses achieve closure by construction rather than through improved physical fidelity, and disagree with in situ measurements and with each other by 5--15~W~m$^{-2}$ regionally~\cite{valdiviesoAssessmentAirSea2017}, which would silently propagate the reference reanalysis's own surface-flux parameterization biases into the benchmark. We therefore evaluate surface fluxes against the observational products in Table~\ref{tab:tier2_variables} using spatial pattern and seasonal cycle statistics (anomalies relative to the local climatological mean) rather than absolute magnitudes, which are much better constrained observationally. Point-wise diurnal and seasonal-cycle evaluation at the in situ networks (Ameriflux, OceanSITES, BSRN) is retained with site-level observational uncertainty incorporated as an additive term to the internal-variability spread in the ensemble consistency test. 

\subsection{Scoring Metrics}
\label{sec:scoring}

All Tier~II quantities are scored with the same proper scoring rule, the continuous ranked probability score (CRPS). A proper score identifies the better model only in expectation, so a single realized value cannot rank models on its own; the quantities differ mainly in the axis along which we accumulate the many (semi-)independent verification instances that estimate the expectation. For \emph{time-resolved} metrics, such as monthly and annual anomaly series across the post-2015 window, that axis is time. Because our candidate models are completely free-running, no individual ensemble member is expected to match the exact temporal phase of observed internal variability. Instead, the CRPS evaluates whether the model-derived distribution at any given time step remains consistent with the real-world realization, effectively rewarding an accurate forced trajectory combined with a well-calibrated representation of internal variability. For these time-resolved metrics every series (each ensemble member and the reference) is first expressed as an anomaly about its own 1985--2014 climatology, so that mean-state bias is not scored; where the reference record begins later, the baseline is the overlapping part of that period (2001--2014 for CERES-EBAF, which begins in March 2000, and 2003--2014 for MODIS Aqua). The aggregated diagnostics below, by contrast, retain the reference dataset's climatology, so that mean-state bias is scored there. For an ensemble $x_1,\ldots,x_M$ and observation $y$, we use the finite-ensemble (`fair') estimator \cite{ferroFairScoresEnsemble2014},
\begin{equation}
    \operatorname{CRPS}_\text{fair} = \frac{1}{M}\sum_{m=1}^{M}|x_m-y| - \frac{1}{2M(M-1)}\sum_{m\neq n}|x_m-x_n|,
    \label{eq:fair_crps}
\end{equation}
which removes the systematic penalty that the standard empirical estimator imposes on smaller ensembles, so that submissions with different ensemble sizes are scored on an equal footing. For a deterministic baseline ($M=1$), fair CRPS is undefined. Observational uncertainty enters by averaging Equation~\ref{eq:fair_crps} over a common set of draws from the reference-product distribution and stated retrieval uncertainties, using identical draws for the submission and its baselines. Scores are averaged over the many time steps of the window. Temporal autocorrelation does not bias this mean but reduces the effective number of independent verifications, so we report moving-block-bootstrap confidence intervals (resampling years, ensemble members, and observational products) alongside every scorecard entry, and treat score differences smaller than those intervals as indistinguishable since a model can score better or worse on a realization that happens to have internal variability aligned or misaligned with observations.

For \emph{aggregated} diagnostics such as climatological means, spatial patterns, multi-decadal trends, or seasonal teleconnection amplitudes the same principle applies along a different axis. For a spatial field that axis is space: we project both model and observations onto the leading empirical orthogonal functions (EOFs) of the reference dataset, defined over the pre-2015 record and fixed in advance, standardize each retained coefficient by its pre-2015 observational standard deviation, score it with the fair CRPS of Equation~\ref{eq:fair_crps}, and take the equal-weight mean over coefficients as the variable-level score. Because the coefficients are spatially correlated, the block bootstrap resamples spatial blocks and discounts them to an effective sample size, widening the reported interval accordingly. The limiting case is a diagnostic that really is a single global number, such as a global-mean trend or warming level, for which this axis collapses to one observation: here no score can reliably rank models, because a better model will sometimes score worse through internal-variability luck alone, so we report such quantities with their (wide) bootstrap intervals and only use them for the  consistency check below, omitting them from headline ranks.

Alongside the score, and for these low-dimensional diagnostics especially, we retain an \emph{ensemble-consistency test} as a calibration check rather than as a score. It evaluates the null hypothesis that the observed coefficient vector is a plausible draw from the model's distribution, constructed from the active ensemble and supplemented, where necessary, with non-overlapping segments of the long pre-industrial control run. Its statistic is the squared Mahalanobis distance between the observed coefficients and the model distribution; with known mean and covariance this would follow a $\chi^2$ distribution with degrees of freedom equal to the number of retained modes, but because both moments must instead be estimated from finite, autocorrelated samples we calibrate the null empirically, using held-out control segments as pseudo-observations. The truncation limit is set per variable by a pre-registered variance-explained criterion, observational uncertainty enters as an additive covariance term with a two-sided test; its outcome is reported as a warning when the calibrated $p$-value falls below 0.05. We do not use this in scoring because a hypothesis-test statistic is not a cardinal measure of quality, and because a score built on it could anyway be gamed by deliberate overdispersion. Two members suffice to form the fair CRPS, but the discriminating power of both the aggregated score and this consistency diagnostic grows with the ensemble size and with the length of the control run used to sample internal variability (for reference, a purely rank-based check with $N$ members can attain a two-sided $p$-value no smaller than $2/(N+1)$, i.e., $0.5$ at $N=3$), which is why $\mathcal{O}(10)$ members are requested. For a single-member equilibrium paleoclimate experiment, pseudo-members and pseudo-observations are instead drawn from non-overlapping blocks of the equilibrated portion of that experiment, using blocks matching the length of the scored climatology, which preserves climate-state-dependent internal variability.

\subsubsection{Modes of variability}

In addition to mean-state and trend metrics, models are evaluated on their representation of key modes of variability, including the diurnal and seasonal cycle (amplitude and phase of temperature and precipitation). Seasonal cloud-radiative feedbacks, the covariance between SST and cloud radiative effect on seasonal timescales, are included as an emergent constraint on equilibrium climate sensitivity. These variability metrics are computed over the full historical period rather than the post-2015 test window, given the need for longer records to robustly estimate them. Because that period may have been used in the training of data-driven submissions, these entries are labelled in the scorecard as \emph{in-sample adequacy diagnostics}, visually distinguished from the held-out post-2015 entries. Longer modes of variability, for which we have physical understanding but lack long observational baselines, are evaluated separately in Tier I rather than here.

The daily-resolution extremes and distributional metrics designated as core in Table~\ref{tab:tier2_variables} are defined as follows. For temperature and precipitation extremes we adopt a subset of the ETCCDI indices \cite{zhangIndicesMonitoringChanges2011}: the annual maximum and minimum daily temperatures (TXx, TNn), the frequency of warm days (TX90p), and warm-spell duration for temperature; and the annual maximum 1-day and 5-day precipitation (Rx1day, Rx5day), the fraction of precipitation from very wet days (R95pTOT), and consecutive dry days for precipitation. Each index reduces to a climatological scalar (and trend) per region, and is therefore scored through the same aggregated-diagnostic framework (fair CRPS with the consistency flag) as the other aggregated diagnostics; no additional machinery is required. The shape of the daily distributions themselves is evaluated with the Perkins skill score \cite{perkinsEvaluationAR4Climate2007}, the overlap between the modelled and observed empirical probability density functions of daily temperature and wet-day precipitation intensity, pooled by season and region and reported as a skill score in the scorecard alongside the other metrics (Section~\ref{sec:aggregation}). To avoid conflating the secular warming trend with the distribution shape, we construct these PDFs from anomalies relative to a moving climatological baseline, so the score reflects distribution shape rather than non-stationarity in the mean. Because extreme-value indices are scale-dependent, computing them at each model's native resolution would make them different estimands across models of differing resolution. All extremes and distributional metrics are therefore computed after conservative regridding of both model and observational daily fields to a fixed common evaluation grid of approximately $1^\circ$, chosen to match the effective resolution of the gridded daily observational products; this sacrifices some tail amplitude to interpolation smoothing, but does so identically for every submission and for the reference. For the Perkins score, wet days are defined by a fixed $1$~mm~day$^{-1}$ threshold and the PDFs use a common, pre-registered bin width per variable. These metrics are computed over the full historical record rather than the test window alone, since stable estimation of the tails requires long records, and are accordingly labelled as in-sample diagnostics in the scorecard.

The seasonal cycle represents a strictly forced response of the climate system to varying insolation, governed by many of the same physical mechanisms that determine the long-term response to anthropogenic forcing. Because seasonal variations in variables such as temperature, sea-ice extent, and carbon fluxes often exceed the magnitude of projected climate changes over the coming decades, they offer a high signal-to-noise benchmark for model fidelity \cite{Schneider21b}. Consequently, we define specific seasonal metrics, some of which serve as proxies for long-term climate sensitivity. These include the mean annual temperature range over land, which tests the model's thermal damping and feedback strength, and the seasonal correlation between sea surface temperature and low-cloud cover, which probes the cloud-radiative feedbacks that continue to dominate uncertainty in future climate projections. (The seasonal amplitude of land carbon uptake, a demonstrated constraint on the future CO$_2$ fertilization effect \cite{wenzelProjectedLandPhotosynthesis2016}, is deferred to the future carbon-cycle extension of the benchmark discussed in Section~\ref{sec:conclusion}, since the present protocol does not request carbon-flux output.) By validating the amplitude and phase of these forced variations, we build confidence in the model's ability to simulate the unobserved response to future forcing.

The diurnal cycle provides an analogous forced, high signal-to-noise test at the shortest resolved timescale. Like the seasonal cycle, it is a strictly periodic response to insolation, but it probes the fast physics, such as boundary-layer development, convective triggering, and cloud formation, that underlie many persistent model biases. We evaluate the amplitude and phase of the diurnal cycle of precipitation, which successive generations of physical models have simulated poorly (typically triggering continental convection several hours too early), and of the cloud radiative effect, which tests the coupling between surface heating, convection, and cloud properties. Merged satellite products provide the observational reference at sub-daily resolution (IMERG at half-hourly and CERES SYN at 3-hourly sampling), and passive microwave retrievals of column water vapour enable approximately 3-hourly sampling of the humidity field. Diurnal metrics are computed as the amplitude and phase of the first diurnal harmonic in local solar time, averaged seasonally over the historical period, and scored as aggregated diagnostics (fair CRPS with the consistency flag). Because they require 3-hourly output (Table~\ref{tab:variables}) that not all architectures produce, the diurnal metrics are designated as extended.

\subsection{Historical trends and events}
\label{sec:trends}

A model's ability to reproduce the observed evolution of global warming is a direct test of whether it captures the net effect of greenhouse gas forcing, aerosol cooling, and climate feedbacks over the instrumental period. Rather than evaluating the transient climate response (TCR), which is a theoretical quantity for which the true value is not directly observable, we evaluate the realized warming itself against observational products. For the short test period the primary scalar diagnostic is the realized warming level, the mean global-mean surface temperature anomaly over all complete years from 2015 onward relative to the fixed 1985--2014 pre-test baseline, rather than the ordinary-least-squares trend within the window: over a single decade the trend is dominated by internal variability, whereas the mean level is less sensitive to it and is the same integrated measure of the recent warming rate used in the Section~\ref{sec:task} demonstration. Over the full historical period (1950--present), where the record is long enough that trend uncertainty is acceptable, we additionally compare the modelled global-mean surface temperature trend, and report the test-period trend as a secondary diagnostic. These scalar diagnostics are compared against NASA GISS, Berkeley Earth, and HadCRUT estimates, with the observational uncertainty taken from the spread across (and stated uncertainties of) the three products. Because each is a single global number, it is not used to rank models (Section~\ref{sec:scoring}); instead the realized warming level serves as a falsification test of the forced response, and its calibrated ensemble-consistency $p$-value, together with the bootstrap interval of the fair CRPS, is reported prominently at the head of every scorecard, with models inconsistent with the observed warming flagged. Because those products blend land air and sea surface temperatures whereas model \texttt{tas} is a surface air temperature diagnostic, we correct the observations to a surface air temperature basis for the simplest fair comparison \cite{cowtanRobustComparisonClimate2015,richardsonReconciledClimateResponse2016}. The annual global-mean temperature anomaly series itself is scored with the fair CRPS. Each model's reported equilibrium climate sensitivity and TCR (where available) will also be collected as metadata for post-hoc analysis, but these are not scored directly.

The 1991 eruption of Mount Pinatubo produced a well-observed transient perturbation to the climate system that uniquely tests a model's forced response within the historical period. We evaluate whether models reproduce a realistic volcanic signature during the post-Pinatubo period by examining the combination of reduced surface downwelling shortwave radiation and the associated surface cooling. Rather than prescribing a separate volcanic forcing experiment, we assess this directly from the historical simulation, comparing ensemble-mean anomalies of surface SW↓ and near-surface temperature against BSRN-derived and observational estimates for the 1991--1993 period. We note that data-driven models trained on historical data may reproduce the Pinatubo signal through overfitting rather than physical representation of the aerosol--radiation interaction; requiring the correct co-variation of radiative and thermal anomalies, rather than in isolation, provides a partial guard against this.

For the last quarter of the 20th century anthropogenic aerosols produced a pronounced hemispheric asymmetry in historical warming: sulfate emissions concentrated in the Northern Hemisphere mid-latitudes partially offset greenhouse gas warming there, resulting in a weaker NH warming trend and a transient southward shift of the ITCZ relative to a GHG-only world \cite<see>{stevens_rethinking_2015}. This asymmetry provides an observable fingerprint of aerosol forcing that can be evaluated directly from the historical simulation without requiring additional experiments. We assess the interhemispheric temperature contrast (NH minus SH mean surface temperature) trend over the 1950--1985 period, during which aerosol forcing increased most rapidly, and compare it against observational estimates from HadCRUT and Berkeley Earth. Models should reproduce the observed NH cooling tendency (or reduced warming relative to the SH) during this period, and the correct sign of the associated tropical precipitation shift. As a scalar trend diagnostic, it is evaluated, like the global warming level, through the ensemble-consistency test rather than ranked. This test complements the Tier I hist-aer experiment by evaluating the observable imprint of aerosol forcing as it manifests in the full historical simulation, providing a check on whether models achieve realistic historical warming through a plausible balance of greenhouse gas and aerosol forcing rather than through compensating errors.

\subsection{Baselines and reference scores}
\label{sec:baselines}

\begin{figure}[htbp]
    \centering
    \includegraphics[width=\linewidth]{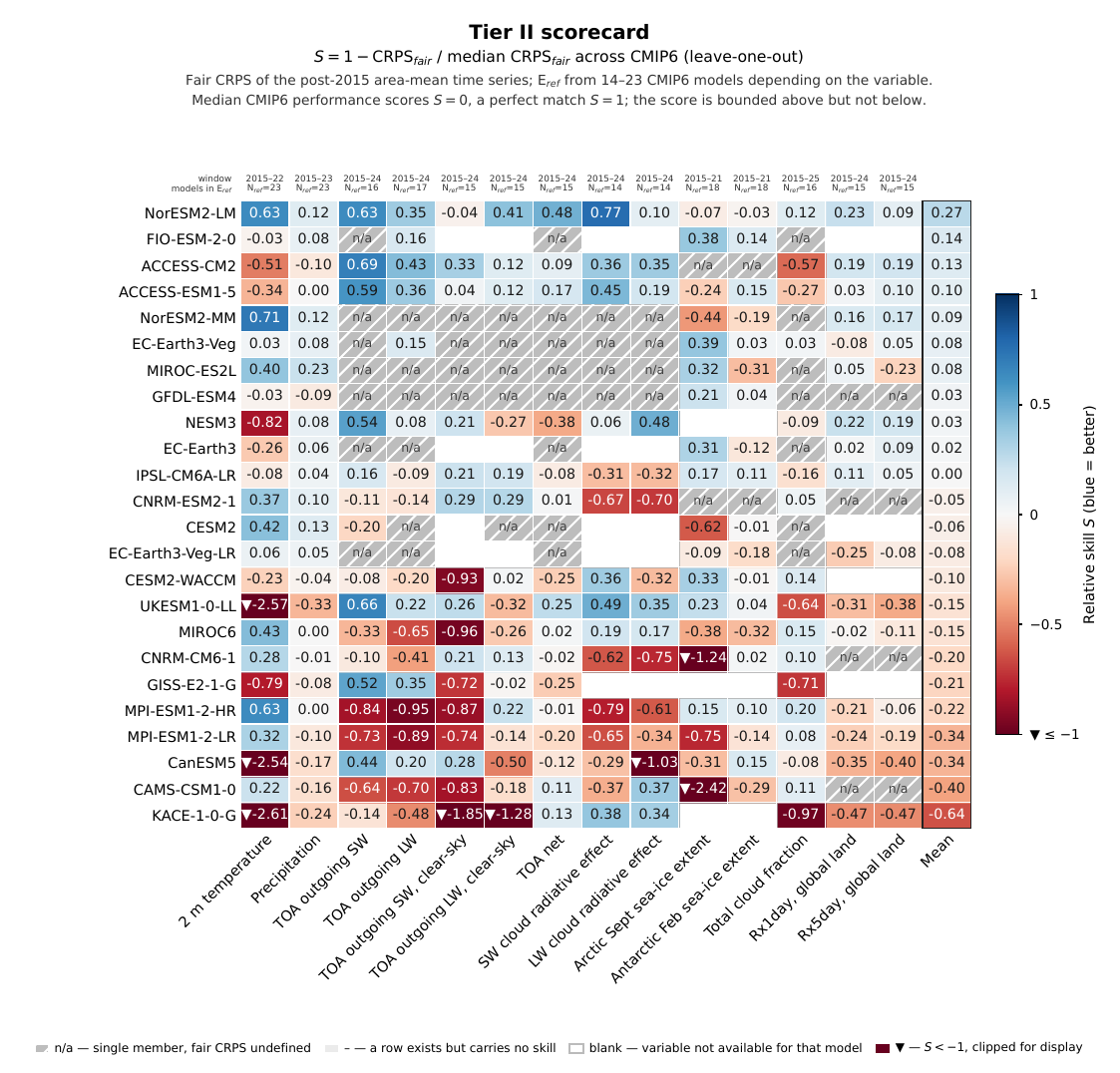}
    \caption{Example Tier~II scorecard for the CMIP6 ensemble over the post-2015 test window, computed from their historical and SSP2-4.5 simulations. Each cell gives the relative skill score $S = 1 - \mathcal{E}_\text{model}/\mathcal{E}_\text{ref}$ (Equation~\ref{eq:skill}) for one model and one variable, where $\mathcal{E}$ is the fair CRPS (Equation~\ref{eq:fair_crps}) of the annual global-mean anomaly series over the test window (September Arctic and February Antarctic extent for the sea-ice columns; the land-area mean of the per-gridpoint annual maximum 1-day and 5-day precipitation, Rx1day and Rx5day, for the two extremes columns), and $\mathcal{E}_\text{ref}$ is the median $\mathcal{E}$ over the other CMIP6 models (leave-one-out; 14--23 models depending on the variable, as indicated above each column). Every ensemble member, and the reference, is expressed as an anomaly about its own 1985--2014 climatology (2001--2014 for the CERES- and IMERG-referenced columns and 2003--2014 for cloud fraction). Each model's ensemble comprises its \texttt{r*i1p1f*} members, capped at the first 10; the 24 CMIP6 models with at least two such historical+SSP2-4.5 members are shown. Cells marked n/a are variables for which that model has only one member, so that the fair CRPS is undefined; blank cells indicate variables not available for that model. Median performance therefore scores $S = 0$, a perfect match to observations $S = 1$, and performance worse than the median $S < 0$; the score is bounded above but not below, and values below $-1$ are clipped to the darkest shade. Models are ordered by the mean relative score across variables (rightmost column). The references are HadCRUT5 (2~m temperature), GPCP (precipitation), CERES-EBAF (TOA fluxes, and the cloud radiative effects $\text{swcre} = \text{rsutcs} - \text{rsut}$ and $\text{lwcre} = \text{rlutcs} - \text{rlut}$), HadISST (sea-ice extent), MODIS Aqua MYD08\_M3 Collection~6.1 (total cloud fraction) and IMERG V07 Final daily precipitation on a 1$^\circ$ grid (Rx1day, Rx5day); see Table~\ref{tab:tier2_variables}. Where the reference record ends before 2025 the window ends earlier: 2022 for 2~m temperature, 2023 for precipitation, 2024 for the CERES fluxes and the precipitation extremes, and 2021 for sea ice.}
    \label{fig:benchmark}
\end{figure}

To make scores interpretable across model classes we anchor each headline skill score to a single reference derived from the CMIP6 multi-model ensemble, and provide two further baselines as points of comparison on the same scale. The headline reference score $\mathcal{E}_\text{ref}$ is the median of the per-model fair CRPS over the CMIP6 models that submitted historical and SSP2-4.5 simulations, each model being scored on its own ensemble: its realizations with physics and initialization index 1 and any forcing index (\texttt{r*i1p1f*}), capped at the first 10. When a CMIP6 model is itself being scored it is excluded from the median (leave-one-out), and models with a single qualifying member, for which the fair CRPS is undefined, are excluded altogether. We use the median model rather than a pooled mixture of all CMIP6 members because the spread of such a mixture reflects structural inter-model disagreement rather than internal variability; it is therefore over-dispersed, and would let almost any submission centred on the observations clear $S = 0$. The median CMIP6 model represents current community capability and is the reference against which new submissions are naturally judged. Against it we report two additional diagnostic baselines: a climatology baseline that predicts the 1985--2014 monthly mean for every test year, establishing the floor below which a submission has demonstrated no skill; and a multi-linear pattern-scaling baseline in which regional temperature and precipitation anomalies are reconstructed as the product of a global-mean temperature trajectory (from a simple two-layer energy balance model calibrated to observations through 2014) and a fixed CMIP6 multi-model-mean response pattern, representing the simplest defensible emulator for forced large-scale change. Each submission therefore has one headline skill score per scorecard entry relative to $\mathcal{E}_\text{ref}$ (Equation~\ref{eq:skill}); the climatology and pattern-scaling comparisons are reported alongside it so that performance can also be read against a no-skill floor and a simple emulator, and the individual CMIP6 models are shown for context. An example set of scores for the CMIP6 ensemble against the observational datasets described in Table~\ref{tab:tier2_variables}, relative to the median CMIP6 model, is shown in Fig.~\ref{fig:benchmark}.

Several features of this example scorecard are worth noting. Global-mean 2~m temperature discriminates strongly between models ($S$ from $-2.6$ to $+0.7$), with CanESM5, UKESM1-0-LL and KACE-1-0-G warming too fast over the test window; on the surface temperature series (2~m temperature, and the skin and sea surface temperatures, which behave similarly and are omitted from the figure) the climatology baseline, which predicts no warming, scores $S \approx -1.5$ to $-1.8$ and the pattern-scaling baseline $S \approx +0.3$ to $+0.4$. Global-mean precipitation and net TOA flux, by contrast, discriminate only weakly (median $|S|$ of 0.09 and 0.13 respectively): GPCP's interannual variability exceeds the models' ensemble spread, and every model under-predicts the increase in Earth's energy imbalance observed by CERES over 2015--2024 \cite<cf.>{loebSatelliteOceanData2021} by about half. Both are genuine results, but for precipitation they argue for adding an explicit observational-uncertainty term to the score. Global-mean cloud condensed and ice water paths, evaluated against MODIS, barely discriminate ($\mathcal{E}_\text{ref}$ of 0.002--0.004~kg~m$^{-2}$, since the post-2015 global-mean anomalies are small against MODIS's own variability) and are omitted from the figure, whereas total cloud fraction and the SW and LW cloud radiative effects do discriminate between models.

Figure~\ref{fig:benchmark} is the time-resolved score of Section~\ref{sec:scoring}: each member is reduced to anomalies about its \emph{own} 1985--2014 climatology, so the mean-state bias is removed and the score measures the post-2015 trajectory and the calibration of the ensemble spread. This is the headline score by design: the benchmark target is a \emph{change} relative to a recent baseline (Section~\ref{sec:task}), so it is the forced response and the spread about it, not the absolute base state, that we most want a submission to get right; a model with a biased climatology but a faithful response is more useful for projection than the converse, and base-state fidelity is in any case already tested by the Tier~I gates and by the aggregated score below. The complementary aggregated score, in which each member's post-2015 climatology is expressed as an anomaly about the \emph{observed} 1985--2014 climatology and projected onto the reference's fixed pre-2015 EOFs, retains the mean-state bias and therefore scores the spatial pattern of the recent climatology, bias included; its scorecard for the same models is shown in Appendix~\ref{app:eofscorecard}. The two rank models differently (NorESM2-MM and CESM2 lead the pattern score, NorESM2-LM the trajectory score), and the pattern score is more coherent across variables for a given model because bias patterns are physically correlated across fields, whereas each trajectory score rests on only 8--11 annual values.

\subsection{Aggregation and reporting}
\label{sec:aggregation}
We deliberately do not collapse a model's performance to a single number. As in established benchmarks in adjacent fields (notably WeatherBench~2, which reports per-variable, per-lead-time scorecards rather than an overall ranking on the grounds that no single metric captures forecast performance; \citeNP{raspWeatherBench2Benchmark2024}), results are reported as a multi-dimensional scorecard across tiers, variables, and regions. Within the scorecard, each entry is expressed as a skill score relative to a reference baseline,
\begin{equation}
    S = 1 - \frac{\mathcal{E}_\text{model}}{\mathcal{E}_\text{baseline}},
    \label{eq:skill}
\end{equation}
where $\mathcal{E}$ is the fair CRPS of the relevant quantity, time-resolved or aggregated (Section~\ref{sec:scoring}), and the baseline for the headline score is the CMIP6 reference score $\mathcal{E}_\text{ref}$ (Section~\ref{sec:baselines}). When a CMIP6 model is itself being scored, it is excluded from the median that defines $\mathcal{E}_\text{ref}$, so that no model is evaluated against a reference that contains itself. Thus $S = 0$ denotes performance equal to the median CMIP6 model, $S = 1$ a perfect score, and $S < 0$ performance worse than the median CMIP6 model; the same expression evaluated against the climatology and pattern-scaling references provides the two additional diagnostic comparisons, making explicit whether a submission has at least beaten a no-skill baseline and the simplest emulator. The tiers are combined by logical structure as opposed to weighted averaging: Tier~I functions as a gate, in that all applicable required tests must pass for a model to be scored at all, while Tiers~II and III are reported as a vector of skill scores with per-tier and per-variable summaries; there is thus one skill score per scorecard entry, but no aggregation across variables or tiers into an overall rank. Pooling normalized scores across variables would raise the effective sample per model, one remedy when observations of a single quantity are few, but each entry already averages the fair CRPS over many spatial or temporal instances, so we retain the per-variable resolution rather than pool. Entries are additionally labelled as held-out (computed only from post-2015 or otherwise reserved data) or in-sample (computed over the historical period), so that real out-of-sample skill is never conflated with adequacy on data a submission may have trained on. This preserves information that a single rank would obscure (for example, a model that is skillful on temperature but poor on precipitation, or one that fits present-day observations well but fails the out-of-distribution tests), which is precisely the trade-off structure the benchmark is designed to expose. The software implementation of the benchmark, a wrapper around the ClimateEval community diagnostic framework build on ESMValTool \cite{eyringEarthSystemModel2020}, is described in \ref{app:implementation}.

\section{Tier III: Out-of-Distribution and Predictive Skill}
\label{sec:tier3}

The tests in Tiers I and II evaluate models against present-day physical constraints and the recent observational record. While necessary, they are not sufficient: a model that reproduces the recent past may do so for the wrong reasons, or through correlations that break down under novel forcing. Tier III addresses this by testing models against climates that are substantially different from the training period, providing the most direct assessment of extrapolation skill available.

We include two complementary categories of out-of-distribution tests. First, paleoclimate simulations test whether models can reproduce the climate of periods with very different boundary conditions, for which proxy observations exist. Second, perfect-model experiments test whether data-driven models can predict the future climate of existing ESMs when trained only on their present-day output. Table~\ref{tab:tier3} summarizes these tests, their experiments, reference targets, and scoring.

\begin{table}[ht]
\caption{Summary of Tier~III out-of-distribution tests. Paleoclimate simulations follow the PMIP4 protocols; perfect-model and large-ensemble tests are most relevant for data-driven models. Scoring uses the seasonal aggregated-diagnostic (fair CRPS) framework of Section~\ref{sec:scoring} except where a specific threshold is stated.}
\label{tab:tier3}
\small
\begin{tabular}{p{3.0cm}p{2.2cm}p{4.4cm}p{2.4cm}l}
\hline
Test & Experiment & Reference / target & Metric & Status \\
\hline
Last Interglacial & lig127k & SST \cite{osmanUpdatedCompilationLast2026,hoffmanRegionalGlobalSeasurface2017}; land $T$ \cite{otto-bliesnerLargescaleFeaturesLast2021}; precip. \cite{scussoliniAgreementReconstructedModeled2019,kaushalSISALv3GlobalSpeleothem2024} & Fair CRPS vs proxies & Extended \\
Last Glacial Maximum & lgm & SST \cite{tierneyClimatesInformOur2020,osmanGloballyResolvedSurface2021}; land $T$ \cite{bartleinPollenbasedContinentalClimate2011,cleatorNewMultivariableBenchmark2020} & Fair CRPS vs proxies & Required \\
Mid-Holocene & midHolocene & N. Africa JJAS precip. \cite{bartleinPollenbasedContinentalClimate2011,harrisonMidHoloceneSaharan2015}; SST \cite{osmanGloballyResolvedSurface2021}; land $T$ \cite{kaufmanGlobalDatabaseHolocene2020} & Fair CRPS + monsoon pass/fail ($\geq0.5$~mm~day$^{-1}$) & Extended \\
Perfect model (CESM2) & historical $\to$ SSP2-4.5 & Held-out ESM daily tas, pr & Tier~II metrics & Required (ML only) \\
Perfect model (MPI-ESM, GISS ModelE2) & historical $\to$ SSP2-4.5 & Held-out ESM daily tas, pr & Tier~II metrics & Extended (ML only) \\
Large-ensemble spread & hist + SSP2-4.5 ($>$20 members) & CESM-LE inter-member variability & Variance ratio; spatial correlation & Extended \\
\hline
\end{tabular}
\end{table}

\subsection{Paleoclimate Evaluation}
\begin{figure}[htbp]
    \centering
    \includegraphics[width=0.85\linewidth]{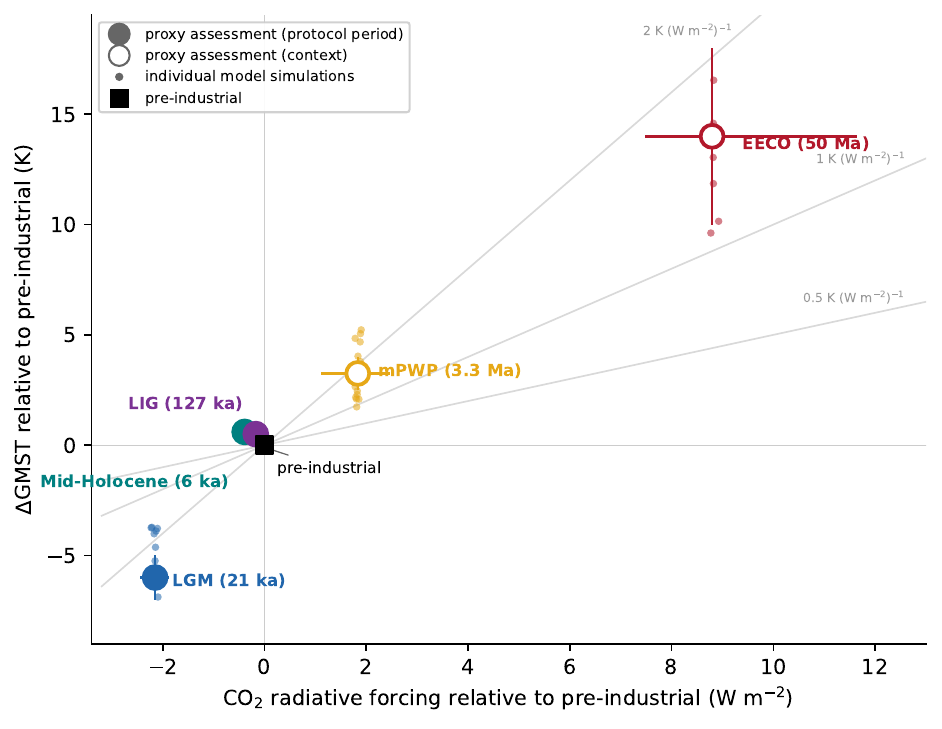}
    \caption{Paleoclimate periods as a lever on the Earth-system response to forcing. Global-mean surface temperature anomalies (relative to pre-industrial) against CO$_2$ radiative forcing, for proxy-based assessments (large circles, with assessed ranges) and individual model simulations (small circles: PMIP4 LGM, PlioMIP mPWP, and DeepMIP EECO ensembles), from the data underlying Figure 7.19 of \citeA{forsterEarthsEnergyBudget2021}, as archived by \citeA{luntIPCCAR6Fig719Data2023}. The LIG assessment is from \citeA{fischerPalaeoclimateConstraintsImpact2018} and the Mid-Holocene from \citeA{kaufmanGlobalDatabaseHolocene2020}. Grey lines indicate constant Earth-system response ratios. Filled circles denote the periods evaluated in Tier III (LGM, LIG, Mid-Holocene); open circles show deeper-time context periods excluded from the protocol because their boundary conditions are less well known. Model simulations are plotted at the central assessed CO$_2$ forcing of each period, computed as $5.35\ln(C/C_0)$ with $C_0 = 284$~ppm. Note, each period also has significant, and distinct non-CO2 forcing contributions as well.}
    \label{fig:paleolever}
\end{figure}

Paleoclimate records provide the only direct observations of the Earth system under CO$_2$ concentrations, orbital forcing, and other boundary conditions substantially different from today, and have been shown to provide valuable constraints on climate sensitivity and feedback processes \cite{schmidtUsingPalaeoclimateComparisons2014,tierneyClimatesInformOur2020,zhuLGMPaleoclimateConstraints2022,sherwoodAssessmentEarthsClimate2020}. We include three paleo periods that span a range of climatic states and for which the proxy data coverage and quality is sufficient for meaningful model evaluation, and for which the boundary conditions are well known (ruling out time periods before 800,000 years ago, for which CO$_2$ and other greenhouse gases, and orbital forcing, are not known to sufficient accuracy): the Last Interglacial (LIG, ~127 ka), the Last Glacial Maximum (LGM, ~21 ka), and the Mid-Holocene (Mid-H, ~6 ka). These periods are also targets of the PMIP4 protocol, meaning that boundary conditions and experimental design are well established. Recent work has highlighted the importance of accounting for the `pattern effect' when using paleo data to constrain sensitivity \cite{cooperLastGlacialMaximum2024,luntPaleoclimateDataProvide2024}, and our evaluation accounts for this by comparing spatial patterns of temperature change rather than relying solely on global-mean sensitivity estimates. Figure~\ref{fig:paleolever} illustrates the leverage these periods collectively provide: the LGM anchors the low-CO$_2$ end of the relationship between CO$_2$ forcing and global temperature, while the orbitally-forced LIG and Mid-Holocene probe the response to a rearrangement of insolation at near-pre-industrial CO$_2$. Deeper-time warm periods (the mid-Pliocene Warm Period and Early Eocene Climatic Optimum) would extend this lever much further, but fall outside the protocol for the boundary-condition reasons above; they are shown for context.

For each period, we define the boundary conditions, the proxy datasets for comparison, and the metrics used for scoring. We apply the same probabilistic scoring framework as in Tier II; because the paleoclimate comparisons are seasonal climatological anomalies rather than time series, they are scored as aggregated diagnostics (Section~\ref{sec:scoring}), with the substantially larger observational uncertainties of the proxy reconstructions included and the model spread estimated from non-overlapping blocks of the equilibrated portion of the corresponding paleoclimate simulation, since these experiments require only a single continuous member. The proxy reference datasets for each period are listed in Table~\ref{tab:tier3} and described in ~\ref{app:paleodata}.

Last Interglacial (${\sim}127$ ka): The LIG was the most recent period with global temperatures significantly warmer than pre-industrial (estimated $+0.5$~K global mean; \citeNP{fischerPalaeoclimateConstraintsImpact2018}), driven primarily by orbital forcing that produced strong Northern Hemisphere high-latitude warming, reduced Arctic sea ice, and a smaller Greenland ice sheet. Models are evaluated against compilations of SST, terrestrial temperature, and terrestrial precipitation (\ref{app:paleodata}).

Last Glacial Maximum (${\sim}21$ ka): The LGM represents the most recent cold extreme, with global temperatures approximately 5--7 K below present, greatly expanded ice sheets, lower sea level, and atmospheric CO$_2$ of ${\sim}190$ ppm. It provides a strong constraint on the upper bound of climate sensitivity \cite{zhuLGMPaleoclimateConstraints2022,cooperLastGlacialMaximum2024}: because the LGM's large negative forcing scales the global cooling by the model's sensitivity, an excessively sensitive model produces an unrealistically cold LGM that is inconsistent with the proxy record, so this low-CO$_2$ state constrains the high-sensitivity tail more tightly than the near-present-day record can. Given this demonstrated leverage, including its role in understanding the recent high-sensitivity configuration of CESM2 as an unintended consequence of exposed microphysical formulation imperfections  \cite{zhuLGMPaleoclimateConstraints2022}, the LGM simulation is required, whereas the orbitally-forced Last Interglacial and Mid-Holocene experiments are extended. The LGM is evaluated against large SST proxy compilations together with pollen-based land temperatures (\ref{app:paleodata}).

Mid-Holocene (${\sim}6$~ka): The Mid-Holocene represents a modest perturbation from present-day climate driven primarily by orbital changes, with a strengthened boreal summer monsoon system, ``green Sahara'' conditions, and relatively small global-mean temperature changes but large regional, and in particular seasonal, shifts. The most robust and diagnostically valuable signal is the intensification and northward expansion of the West African Monsoon, which produced substantially wetter conditions across the Sahara and Sahel. This is one of the largest signals of the Mid-Holocene in the proxy record and one that models have consistently underestimated across successive PMIP generations \cite<e.g.,>[Figure 3.11]{ipccClimateChange20212021}, making it a stringent test of a model's ability to represent monsoon dynamics under altered orbital forcing. We therefore evaluate the simulated JJAS precipitation anomaly averaged over North Africa (10--30$^\circ$N, 20$^\circ$W--30$^\circ$E) against pollen- and lake-status-based reconstructions \cite{bartleinPollenbasedContinentalClimate2011,harrisonMidHoloceneSaharan2015}, with the specific requirement that models produce a positive JJAS precipitation anomaly of at least 0.5~mm~day$^{-1}$ area-averaged over this region relative to the pre-industrial simulation.

\subsection{Perfect-Model Experiments}
A complementary set of tests, most relevant for data-driven models, are ``perfect model'' experiments. These involve training a model on the current-climate (historical) output of an existing ESM and then evaluating against future, held-out simulations of that same ESM. This provides a concrete test of the task at hand: given only present-day climate data, can the model predict the climate response to forcing it has not seen? While conceptually similar to the original ClimateBench emulation protocol \cite{watson-parrisClimateBenchV10Benchmark2022}, the critical difference is that training is restricted to the historical period, making the future prediction out-of-distribution.

A model might succeed at this task for a single ESM by chance, particularly if that ESM's future climate is only modestly different from its historical climate. The likelihood of success by chance drops rapidly when evaluated against multiple ESMs with different sensitivities, since a higher-sensitivity model will produce future warming that is further from any model's historical training data. We therefore provide training and evaluation data from three ESMs that span the CMIP6 range of climate sensitivity: CESM2 [ECS $\approx$ 5.2 K], MPI-ESM [ECS $\approx$ 3.63 K], and GISS ModelE2 [ECS $\approx$ 2.31 K]. Evaluation against CESM2 is required and evaluation against the remaining two is optional but encouraged, with the additional scores reported alongside the required one. For each ESM, we provide a subset of variables from the CMIP6 historical experiment for training, and daily temperature and precipitation fields from the SSP2-4.5 scenario as the held-out test data (see Data Availability). Scoring uses the same metrics defined in Tier II.

We note that the perfect-model framework also accommodates fully data-driven emulators that have been trained on an ESM's future projections, provided they have not been trained on any output from the three test ESMs during the ClimateBench2 test period. Such emulators may use future simulations from other ESMs as training data, but any ESM used for training must have been initialized before the ClimateBench2 test period to avoid indirect data leakage through shared observational constraints. This restriction ensures that the perfect-model evaluation remains a fair test of out-of-distribution generalization regardless of whether the submitted model is a climate simulator or an emulator of one. The submission requirements, including the model card through which training-data provenance is declared and attested, are described in Section~\ref{sec:discussion}.

Finally, we test models' ability to represent variability under forcing by asking submissions to produce a large ensemble comparable to the CESM Large Ensemble (LE). While Tier I tests unforced internal variability through a long piControl simulation, the large-ensemble test probes a distinct and complementary question: can the model produce realistic spread as compared to the ESM spread across initial-condition perturbations in the presence of a common forced signal? This is an extension of the perfect-model experiments above, which test the forced response itself and adds increased robustness as well as a tougher generalization test. A model that has learned only the ensemble-mean forced signal will generate near-identical members, while one that correctly captures the interaction between forcing and internal variability should produce ensemble spread consistent with that of the target ESM in both amplitude and spatial structure (i.e. well calibrated). We evaluate the ensemble spread using the variance ratio and spatial correlation of inter-member variability for temperature and precipitation, compared against the equivalent statistics from the CESM-LE.

\subsection{Preventing data leakage and gaming}

A benchmark is only as informative as its train/test separation. We define three categories of data that must not influence model development: (i) all observational products covering 2015 onward, including reanalysis fields derived in part from post-2015 observations; (ii) the daily SSP2-4.5 fields from CESM2, MPI-ESM, and GISS ModelE2 used in the perfect-model experiment; and (iii) the held-out paleoclimate proxy compilations listed in Tier III. Submitters are required to provide a model card declaring the data sources, time spans, and reanalysis products used during training and tuning, and to attest that none of the held-out categories above were used. For models that use ERA5 or other reanalyses that ingest post-2015 observations, training must terminate at 2014 and this must be documented. We further encourage, but do not require, that submissions include results from an ablation in which a randomly selected pre-2015 decade is additionally held out, providing an internal check that scores degrade gracefully on data that was never seen.

\section{Discussion}
\label{sec:discussion}

\subsection{Scope and design choices}

A deliberate choice in this protocol that bears specific justification is the omission of large-scale circulation metrics from the scored evaluation. In traditional climate modelling, the dynamical response (jet shifts, Hadley cell weakening, storm track changes) is a primary pathway through which forcing produces regional temperature and precipitation changes. It is therefore natural to argue that a model which gets the dynamics wrong cannot reliably predict regional outcomes. However, the scope of ClimateBench v2 explicitly encompasses model architectures that do not resolve atmospheric dynamics in any traditional sense, learning instead a direct mapping from forcing to regional climate response. For such models, evaluating circulation is neither possible nor meaningful. Rather than maintain parallel evaluation tracks for models with and without dynamics we adopt the principle of evaluating all models on their predictive outcomes. For the same reason, the Tier~I tests that diagnose a resolved circulation, such as midlatitude geostrophic balance, gate entry only for models that produce the corresponding fields and are recorded as not applicable by architecture otherwise (Section~\ref{sec:task}), so that a model predicting only surface fields is not excluded at the entry ticket. This is an intentionally agnostic stance: we make no claim that dynamics are unimportant for physical understanding, only that they are not the object of prediction here. We acknowledge the risk that a model could produce skillful hindcasts through spurious correlations that break down under novel forcing, particularly for regional precipitation where circulation changes are the proximate cause. The Tier I physical consistency tests and Tier III out-of-distribution evaluations are designed in part to guard against this failure mode. We note that evaluation of dynamical fidelity is a focus of the AIMIP protocol \cite{AIMIP}, which provides a complementary protocol for evaluating atmospheric circulation models.

A central design goal of this protocol is that the same metrics apply to all model classes (physics-based ESMs, neural GCMs, hybrid models, and purely data-driven emulators) without privileging any particular architecture. In practice, the model classes differ in maturity, failure modes, and the degree to which physical constraints are built in rather than learned. Physics-based models satisfy conservation laws by construction but rely on empirical parameterizations which risk overfitting to the present climate through extensive tuning against the historical record. Data-driven models face the complementary risk of learning spurious correlations that generalize poorly under novel forcing, but offer the potential to learn representations directly from data without the assumptions embedded in traditional parameterizations. The Tier I physical consistency tests are designed to catch the most egregious failures of data-driven models (e.g. violation of energy conservation), while the Tier III out-of-distribution tests probe whether any model, physical or otherwise, has actually learned the forced response rather than memorizing the training distribution. We stress that the protocol does not assume one class is superior; it provides a common evidential basis on which the relative merits of different approaches can be assessed transparently.

\subsection{Submission, scoring, and benchmark integrity}

All submissions must be provided on a common grid and in CF-compliant NetCDF format, following the conventions established by CMIP and adopted by AIMIP. We specify a standard output variable list and grid (see Data Availability) to minimize the burden on submitters and ensure reproducibility of scores. Benchmark scores will be computed by a central evaluation pipeline, publicly available on GitHub, that ingests standardized submissions and produces a multi-metric scorecard for each model. We intend to host a public-facing web portal that displays scorecard results alongside interactive visualizations of model performance across Tiers, variables, and regions. The portal will present results as a multi-dimensional scorecard rather than a single-number leaderboard, reflecting the view that model skill is inherently multi-faceted and that collapsing it to a single rank obscures important trade-offs. We anticipate updating the observational reference datasets annually as new data become available, and versioning the protocol to ensure that scores remain comparable across submissions.

Each submission must include: a model card describing architecture, training data, training and tuning procedures, and computational cost (training and per-projection inference); at least three ensemble members (ten preferred) for the updated historical and perfect-model experiments, with the one-member equilibrium and Tier~I exceptions specified in Table~\ref{tab:simulations}; output in CF-compliant NetCDF on the common evaluation grid; and a free-text statement of which Tier I physical constraints are imposed by construction versus learned from data, and which, if any, are not applicable by architecture. Submissions that cannot provide ensembles for experiments requiring them, that have used post-2015 observational data in training, or that do not produce the required output variables (other than those not applicable by architecture) will be reported separately as "non-conforming" rather than ranked against compliant submissions.

It has been noted that some modelling centers may also want to calibrate against the test data to produce the very best model. Following standard practice, further tuning/calibration can still be performed using the test data, but the scores for comparison must all be calculated before this. For the latest CMIP7 models that have been developed prior to this protocol scores will be computed and included, but models which used observations from the test period in their tuning and/or calibration will carry an asterisk denoting this caveat. 

A risk inherent to any successful benchmark is that, once it becomes a development target, models begin to optimize for the benchmark rather than for the underlying scientific goal. We mitigate this in three ways. First, the multi-tier structure means that excelling on the Tier II observational metrics alone does not produce a strong result: a model must also pass the Tier I physical-consistency checks to be scored at all, and because results are reported as a vector rather than collapsed to one number (Section~\ref{sec:aggregation}), its out-of-distribution skill in Tier III, which cannot be achieved by overfitting present-day statistics, is shown alongside its Tier II scores rather than traded off against them. Second, we will rotate a fraction of the Tier II evaluation each year as new observations accumulate, so that the exact test set is not static. Indeed, the information content of the held-out record itself grows as observations accumulate (Section~\ref{sec:task}): unlike static benchmarks, which decay as they are progressively overfit, this one strengthens with age. Third, the benchmark will be versioned (ClimateBench v2.0, v2.1, ...), with new versions adding tests that have been shown to discriminate between submissions that scored similarly on the previous version. We commit to publishing the rationale for each protocol change so that the evolution of the benchmark itself is transparent and auditable.

\subsection{Outlook}

In many other fields such benchmarks have led to phenomenal improvements in performance. For example, ImageNet led to the transformation of deep learning, the Critical Assessment of protein Structure Prediction (CASP) project directly led to the recent AlphaFold breakthroughs \cite{jumperHighlyAccurateProtein2021}, and WeatherBench has had a similar effect for weather forecasting \cite{raspWeatherBenchBenchmarkData2020}. We hope to stimulate similar, and urgently needed, improvements in climate models, fully recognizing the fundamentally different character of benchmarks for climate models: climate projections are out-of-distribution predictions, and clear and transparent benchmarks for their success are not available on practically useful timescales, unlike in CASP or weather forecasting \cite{shawOtherClimateCrisis2025}.

Motivated by the rapid development of data-driven models of the climate system, future work should develop new approaches to ensure the physicality of such models. One potentially fruitful approach would be to explicitly invoke existing theories of the climate system during their training and testing. For example, using radiative-convective equilibrium \cite{manabeThermalEquilibriumAtmosphere1967} and diffusive moist energy balance models \cite{northTheoryEnergyBalance1975} to provide prior constraints on the global- and zonal-mean climate states of such models during their training. Providing data-driven models with this minimal theoretical scaffolding will allow them to better leverage existing observations, and help build trust in their predictions of a changing climate.

\section{Conclusions}
\label{sec:conclusion}

Several limitations should be acknowledged. The protocol's narrow focus on 2050 temperature and precipitation excludes important aspects of climate projections including ice-sheet dynamics, sea-level rise, and ecosystem impacts. The post-2015 test period, while containing the best available observations and substantial warming, is short relative to the timescales of interest. Paleoclimate proxy uncertainties remain large, and the perfect-model framework tests generalization across model physics rather than against the real climate system. These are deliberate scope choices that keep the first version of the benchmark tractable; we intend to expand the protocol as observational constraints improve and community experience accumulates.

Looking ahead, a natural extension of this protocol is incorporating carbon cycle feedbacks, a critical and poorly constrained component of the Earth system response to forcing. Future versions of ClimateBench should evaluate land and ocean carbon fluxes against observational products such as the Global Carbon Project and SOCAT, with paleoclimate constraints on carbon cycle sensitivity providing natural extensions to the existing Tier III framework. Beyond expanding physical scope, standardized and machine-readable scoring pipelines will increasingly support agentic model development workflows, in which systems iteratively propose and refine models (or components) against benchmark targets. Providing a stable, versioned API for benchmark evaluation is therefore a priority, enabling ClimateBench to serve not only as a scorecard for completed models but as an active, multi-objective target in the development of next-generation climate models; any scalarization of the scorecard vector for such optimization is then an explicit choice of the developer, not of the benchmark.

In other scientific fields, standardized benchmarks have catalyzed rapid progress by providing clear targets, enabling reproducible comparison, and lowering barriers to entry. Climate science has lacked an equivalent. We hope that ClimateBench v2 begins to fill this gap, offering a transparent, extensible framework in which any climate model can be evaluated on its merits. All evaluation code, observational datasets, and perfect-model data are publicly available, and we invite the community to submit models, propose protocol improvements, and help build the evidential basis needed for accurate and trustworthy climate projection.

\appendix
\section{Supplementary tables}

\begin{table}[ht]
\caption{Required and extended simulations across all three tiers, with ensemble requirements and estimated duration.}
\label{tab:simulations}
\small
\begin{tabular}{p{4.5cm}lllp{2cm}}
\hline
Simulation & Tier(s) & Ensemble members & Duration (years) & Status \\
\hline
\multicolumn{5}{l}{\textbf{Tier I -- Entry Ticket}} \\
piControl & I, II & 1 & $\geq$500 & Required \\
amip & I & 1 & $\sim$36 & Required \\
amip-4xCO2 & I & 1 & $\sim$36 & Required \\
SST patch (East Pacific) & I & 1 & $\sim$30 & Extended \\
SST patch (West Pacific) & I & 1 & $\sim$30 & Extended \\
abrupt-4xCO$_2$ & I & 1 & $\geq$150 & Required \\
hist-aer & I & 1 & $\sim$165 & Required \\
\hline
\multicolumn{5}{l}{\textbf{Tier II -- Evaluation \& Benchmarking}} \\
historical (updated) & I, II & $\geq$3 (ideally 10) & $\sim$175 $\times$ N & Required \\
\hline
\multicolumn{5}{l}{\textbf{Tier III -- Out of Distribution}} \\
lig127k & III & 1 & $\geq$500 & Extended \\
lgm & III & 1 & $\geq$500 & Required \\
midHolocene & III & 1 & $\geq$500 & Extended \\
Perfect model (CESM2) & III & $\geq$3 (ideally 10) & $\sim$85 $\times$ N & Required (ML only) \\
Perfect model (MPI-ESM) & III & $\geq$3 (ideally 10) & $\sim$85 $\times$ N & Extended (ML only) \\
Perfect model (GISS ModelE2) & III & $\geq$3 (ideally 10) & $\sim$85 $\times$ N & Extended (ML only) \\
Large ensemble (hist + SSP2-4.5) & III & $>$20 & $\sim$175 $\times$ N & Extended \\
\hline
\multicolumn{5}{l}{Estimated minimum (required only): $\sim$2,200 simulation years} \\
\multicolumn{5}{l}{Estimated total including extended: $\sim$7,000+ simulation years} \\
\hline
\end{tabular}
\end{table}


\begin{longtable}{p{4.5cm}lp{2.5cm}ll}
\caption{Required output variables by tier, with CMIP short names and temporal resolution. }
\label{tab:variables} \\
\hline
Variable & Short name(s) & Temporal averaging & Tier(s) & Status \\
\hline
\endfirsthead
\hline
Variable & Short name(s) & Temporal averaging & Tier(s) & Status \\
\hline
\endhead
\hline
\endfoot
\multicolumn{5}{l}{\textbf{Surface \& Atmospheric Temperature}} \\
2m air temperature & tas & Daily & I, II, III & Core \\
Surface temperature (skin) & ts & Monthly & II & Core \\
Sea surface temperature & tos & Monthly & I, III & Core \\
Near-surface temp range & dtr & Daily & II & Extended \\
\hline
\multicolumn{5}{l}{\textbf{Precipitation \& Water Cycle}} \\
Precipitation & pr & Daily & I, II, III & Core \\
Precipitation (sub-daily) & pr & 3-hourly & II & Extended \\
Evaporation & evspsbl & Monthly & I & Core \\
Integrated water vapor & prw & Monthly & II & Extended \\
Integrated water vapor (sub-daily) & prw & 3-hourly & II & Extended \\
Air temperature (pressure levels) & ta & Monthly & I & Core \\
Specific humidity (pressure levels) & hus & Monthly & I & Core \\
Geopotential height (pressure levels) & zg & Monthly & I & Core \\
Surface air pressure & ps & Monthly & I & Core \\
\hline
\multicolumn{5}{l}{\textbf{Radiative Fluxes}} \\
TOA outgoing LW & rlut & Monthly & I, II & Core \\
TOA outgoing LW (clear-sky) & rlutcs & Monthly & I, II & Core \\
TOA outgoing SW & rsut & Monthly & I, II & Core \\
TOA outgoing SW (clear-sky) & rsutcs & Monthly & I, II & Core \\
TOA incoming SW & rsdt & Monthly & I & Core \\
Surface downwelling SW & rsds & Monthly & I, II & Core \\
Surface downwelling LW & rlds & Monthly & I, II & Core \\
Surface upwelling SW & rsus & Monthly & I, II & Core \\
Surface upwelling LW & rlus & Monthly & I, II & Core \\
TOA all- and clear-sky LW/SW fluxes (sub-daily) & rlut, rlutcs, rsut, rsutcs & 3-hourly & II & Extended \\
\hline
\multicolumn{5}{l}{\textbf{Surface Energy Fluxes}} \\
Sensible heat flux & hfss & Monthly & I, II & Core \\
Latent heat flux & hfls & Monthly & I, II & Core \\
\hline
\multicolumn{5}{l}{\textbf{Cryosphere}} \\
Sea ice concentration & siconc & Monthly & II & Core \\
Sea ice extent & siextentn / siextents & Monthly & II & Core \\
\hline
\multicolumn{5}{l}{\textbf{Ocean}} \\
Ocean temperature (3D) & thetao & Annual & II & Extended \\
Ocean salinity (3D) & so & Annual & II & Extended \\
Sea surface height & zos & Monthly & II & Extended \\
\hline
\multicolumn{5}{l}{\textbf{Cloud Properties}} \\
Total cloud fraction & clt & Monthly & II & Extended \\
Liquid water path & clwvi & Monthly & II & Extended \\
Ice water path & clivi & Monthly & II & Extended \\
Cloud-top temperature & (non-standard) & Monthly & II & Extended \\
Cloud-top pressure & (non-standard) & Monthly & II & Extended \\
\hline
\multicolumn{5}{l}{\textbf{Circulation Diagnostics}} \\
Sea-level pressure & psl & Monthly & I & Core \\
Zonal wind (850\,hPa) & ua & Daily & I & Core \\
Geopotential height (850\,hPa) & zg & Daily & I & Core \\
Geopotential height (500\,hPa) & zg500 & Monthly & I & Extended \\
\hline
\multicolumn{5}{p{14cm}}{Core variables are required of every submission that produces the corresponding field; tests that depend on a field a model does not produce are recorded as not applicable by architecture (Section~\ref{sec:task}).} \\
\end{longtable}

\section{Tier I test specifications}
\label{app:tier1}

This appendix gives the precise definitions, formulae, and thresholds for the Tier~I tests summarized in Table~\ref{tab:tier1} and motivated in Section~\ref{sec:tier1}. 

\subsection{Energy balance}
Both criteria are computed over the piControl simulation from the net top-of-atmosphere (TOA) radiative imbalance, $N(t) = \text{rsdt} - \text{rsut} - \text{rlut}$. We define the long-term mean imbalance,
\begin{equation}
    \mu = \frac{1}{T} \int_0^T N(t)\, dt,
    \label{eq:mean_imbalance}
\end{equation}
and the drift rate,
\begin{equation}
    \delta = \frac{d \langle N \rangle_{10}}{dt},
    \label{eq:drift_rate}
\end{equation}
where $\langle \cdot \rangle_{10}$ denotes a 10-year running mean. We require $|\mu| < 0.5$~W~m$^{-2}$ and $|\delta| < 0.02$~W~m$^{-2}$~decade$^{-1}$. The bound on $|\mu|$ reflects the residual control-run imbalances of several tenths of a W~m$^{-2}$ documented across the CMIP5 and CMIP6 ensembles \cite{hobbsEnergyConservationAnalysis2016,irvingMassEnergyConservation2021}, while keeping the control imbalance below the observed present-day Earth energy imbalance of 0.5--1.0~W~m$^{-2}$ \cite{loebSatelliteOceanData2021}. Both quantities are diagnosed from standard CMIP TOA radiative output and calculated over the last 100 years of the piControl simulation.

\subsection{Budget closures}
In steady state (piControl), the global-mean water budget must close,
\begin{equation}
    \mathcal{R}_W = |\langle P \rangle - \langle E \rangle| < 0.05~\text{mm~day}^{-1},
    \label{eq:water_budget}
\end{equation}
where $P$ is precipitation, $E$ is evaporation. The residual of the global-mean atmospheric energy budget must satisfy:
\begin{equation}
    \mathcal{R}_A = |\langle Q_\text{rad} \rangle - (\langle LP \rangle + \langle \text{SHF} \rangle)| < 2~\text{W~m}^{-2},
    \label{eq:atm_energy}
\end{equation}
where $L$ is the latent heat of vaporization, SHF is positive upward from the surface, and $Q_\text{rad}$ is the positive atmospheric radiative cooling rate, diagnosed from CMIP fluxes as $Q_\text{rad}=(\text{rsds}-\text{rsus})+(\text{rlds}-\text{rlus})-(\text{rsdt}-\text{rsut}-\text{rlut})$. The brackets denote a global, multidecadal mean over the piControl. With these sign conventions, latent heating plus sensible heating balances atmospheric radiative cooling in steady state. For models that do not include a traditional hydrological cycle we will evaluate this closure using (semi-empirical) global constraints \cite{allenConstraintsFutureChanges2002}.

\subsection{Diagnostic covariances}
Three relationships must be reproduced quantitatively: (i) the forced clear-sky longwave feedback, diagnosed from the abrupt-4xCO2 experiment as the ordinary-least-squares slope $\beta_\text{LW}$ of annual global-mean $\Delta\text{rlutcs}$ on annual global-mean $\Delta T_s$ over the first 150 years (anomalies relative to piControl, as for the Gregory regression used for ECS), required to fall within $\pm 25\%$ of $1.9$~W~m$^{-2}$~K$^{-1}$, the robust global-mean value of this feedback in GCMs \cite{zhangLinearityOutgoingLongwave2020}; locally the feedback is negative over the tropical oceans, and the GCM value falls somewhat below the ${\approx}2.2$~W~m$^{-2}$~K$^{-1}$ of fixed-relative-humidity column calculations \cite{kollEarthsOutgoingLongwave2018}. We deliberately use the forced response: a regression of interannual monthly anomalies over the historical period measures a different quantity, dominated by the covariance of tropical SST and column water vapour; (ii) midlatitude geostrophic balance in the historical simulation, evaluated as the spatial correlation $\rho(u, u_g)$ between the daily 850~hPa zonal wind and its geostrophic estimate $u_g = -f^{-1}\,\partial \Phi/\partial y$, with $\Phi$ the geopotential on the 850~hPa surface, averaged over $30$--$60^\circ$ latitude and required to exceed $0.9$; and (iii) the tropical precipitation--buoyancy relationship of \citeA{chouEvaluatingRichGetRicherMechanism2009}, evaluated in the historical simulation as the regression of monthly precipitation anomalies on column-integrated moist static energy anomalies over the deep tropics ($20^\circ$S--$20^\circ$N) over 1983--2014, the period of the GPCP/ERA5 reference, and required to match the reference slope to within $\pm 30\%$. Because the slope depends on the pressure levels used to integrate the moist static energy, the reference slope is not a fixed constant but is computed from ERA5 on the submission's own pressure-level set.

\subsection{Causal response tests}
Paired SST patch experiments warm a localized region of the Eastern (\textsc{patch-EP}) and Western (\textsc{patch-WP}) Tropical Pacific by 1~K relative to amip, following the Greens Function Model Intercomparison Project (GFMIP) protocol. The metric is the contrast in the global-mean TOA radiative response per K of local warming,
\begin{equation}
    \Delta \lambda = \frac{\Delta R_\text{EP}}{\Delta \overline{T_s}_\text{,EP}} - \frac{\Delta R_\text{WP}}{\Delta \overline{T_s}_\text{,WP}},
    \label{eq:patch_kernel}
\end{equation}
where $\Delta R$ is the change in global-mean net downward TOA radiative flux relative to amip and $\Delta \overline{T}_s$ the change in area-mean surface temperature over the perturbed patch, which is robustly positive ($\sim 1$--$3$~W~m$^{-2}$~K$^{-1}$) in CMIP6 models and observational reconstructions; we require $\Delta \lambda > 0.5$~W~m$^{-2}$~K$^{-1}$. The amip-4xCO2 experiment tests the fast atmospheric response to CO$_2$ at fixed SST, with the diagnosed effective radiative forcing required to fall within the assessed likely range of $6.5$--$9.0$~W~m$^{-2}$ \cite{forsterEarthsEnergyBudget2021}.

\subsection{ENSO criteria}
\label{app:enso}
The Niño-3.4 SST index (monthly anomalies over $5^\circ$S--$5^\circ$N, $170^\circ$W--$120^\circ$W, with the seasonal cycle removed), computed over $\geq 100$ years of piControl, must satisfy:
\begin{enumerate}
    \item \textbf{Amplitude:} the standard deviation of the deseasonalized monthly Ni\~no-3.4 anomalies satisfies $\sigma_\text{N34} \in [0.5, 1.4]$~K, a generous envelope (roughly $-40\%$ to $+65\%$) around the observed HadISST value of $\sim 0.85$~K.
    \item \textbf{Spectral character:} the integrated power in the 2--7~yr band exceeds that in the 1--2~yr band,
    \begin{equation}
        \frac{\int_{1/7}^{1/2} S(f)\, df}{\int_{1/2}^{1} S(f)\, df} > 1.5,
        \label{eq:enso_spectral}
    \end{equation}
    where $S(f)$ is the power spectral density of the Niño-3.4 index and $f$ is frequency in yr$^{-1}$.
    \item \textbf{Tropical teleconnections:} local monthly anomalies of surface temperature and precipitation are regressed onto the standardized Ni\~no-3.4 index at each grid point over 30$^\circ$S--30$^\circ$N, and the centred spatial correlation between the modelled and observed regression patterns is required to exceed 0.7 (equivalently, a spatial $R^2 > 0.5$), evaluated against HadISST/ERA5 (temperature) and GPCP (precipitation). This pattern criterion captures both the sign and the approximate magnitude of the teleconnection in a single statistic.
\end{enumerate}

\subsection{MJO criterion}
From the Wheeler--Kiladis space--time spectrum of equatorial ($15^\circ$S--$15^\circ$N) daily precipitation, the ratio of eastward- to westward-propagating power for zonal wavenumbers 1--3 and periods 30--90 days must exceed 1.5,
\begin{equation}
    R_\text{MJO} = \frac{\sum_{k=1}^{3} \int_{1/90}^{1/30} S^+(k, \omega)\, d\omega}{\sum_{k=1}^{3} \int_{1/90}^{1/30} S^-(k, \omega)\, d\omega} > 1.5,
    \label{eq:mjo}
\end{equation}
where $S^+$ and $S^-$ denote eastward- and westward-propagating power, respectively.

\subsection{Meridional heat transport partitioning}
The implied oceanic and atmospheric heat transports, diagnosed from TOA and surface energy fluxes in the piControl simulation, must show peak oceanic transport in the tropics (1.5--2.0~PW near 15--20$^\circ$) and peak atmospheric transport in the mid-latitudes (4--5~PW, located at $41\pm2.5^\circ$N, i.e.\ within 38.5--43.5$^\circ$N, centred on the observed peak near $41^\circ$N), consistent with estimates from Estimating the Circulation and Climate of the Ocean (ECCO) and ERA5. The transports are obtained by cumulative meridional integration of the zonally integrated energy-flux divergence without first removing the global-mean imbalance, and peak latitudes are estimated at sub-grid-scale resolution from the vertex of a parabola fitted through the gridded maximum and its two neighbours, so that the reported latitudes are not quantized to the evaluation grid.

\subsection{ITCZ--EFE relationship}
Zonal-mean precipitation and the vertically integrated meridional atmospheric energy transport are averaged into a climatological seasonal cycle (12 monthly means over the historical period). For each climatological month the latitude of the zonal-mean precipitation maximum (the ITCZ latitude) and the cross-equatorial atmospheric energy transport, which sets the latitude of the energy flux equator (EFE) where the meridional energy transport changes sign, are diagnosed. Because the ITCZ responds to the seasonally varying transport with a lag, the ITCZ latitude is lagged two months behind the cross-equatorial transport before the twelve monthly values are regressed \cite{donohoeRelationshipITCZLocation2013}. The magnitude of the regression slope must lie within $\pm 30\%$ of the observed value (approximately $3^\circ$ latitude per petawatt of cross-equatorial energy transport, i.e.\ 2.1--3.9$^\circ$~PW$^{-1}$; \citeNP{donohoeSeasonalCycleAtmospheric2013}) and the magnitude of the correlation coefficient must exceed 0.85 across the twelve months; both are negative for northward-positive transport, since the ITCZ lies in the hemisphere that exports energy across the equator.

\subsection{Aerosol forcing diagnosis}
\label{app:aerosolerf}
Because hist-aer is a coupled, transient experiment rather than a fixed-SST time slice, the aerosol effective radiative forcing (ERF) cannot be read directly as a top-of-atmosphere flux difference \cite{smithEffectiveRadiativeForcing2020}. We instead invert a simple energy-balance model, $\Delta N = F + \lambda\,\Delta T$: with the climate feedback parameter $\lambda$ taken as the slope of the model's own abrupt-4xCO2 Gregory regression \cite{gregoryNewMethodDiagnosing2004} (negative for a stable climate), the aerosol ERF is recovered as $F_\text{aer}(t) = \Delta N(t) - \lambda\,\Delta T(t)$, where $\Delta N$ and $\Delta T$ are the net top-of-atmosphere imbalance and global-mean surface temperature anomalies of hist-aer relative to piControl, evaluated as a decadal mean centred on 2015. This requires only standard CMIP output, but rests on two assumptions that should be borne in mind: the aerosol feedback parameter is taken equal to the CO$_2$-derived $\lambda$ (a forcing efficacy of unity), with pattern-effect differences entering as a second-order correction, and any control-run drift must be removed using the parallel piControl segment before forming the anomalies. The breadth of the acceptance range in Table~\ref{tab:tier1} is intended to absorb these methodological uncertainties. Where a model additionally provides online double-radiation-call diagnostics \cite{ghanTechnicalNoteEstimating2013}, or fixed-SST ERF estimates following RFMIP practice \cite{smithEffectiveRadiativeForcing2020}, these are preferred over the energy-balance inversion and are requested as extended output.

\section{Idealized test-window demonstration}
\label{app:testwindow}

The demonstration in Figure~\ref{fig:testwindow} (Section~\ref{sec:task}) follows \citeA{watson-parrisIntegratingTopDown2025}. A probabilistic version of the FaIR impulse-response model \cite{leachFaIRv200GeneralizedImpulse2021,bouabidFaIRGPBayesianEnergy2024} is conditioned on the global-mean surface temperature simulated by NorESM2-LM \cite{selandOverviewNorESM22020} under SSP2-4.5, with a single ensemble member playing the role of observations. The equilibrium climate sensitivity and the amplitude of the aerosol forcing are treated as uncertain and updated by Bayesian inference, while the response timescales are held fixed as the sensitivity is varied. The constraint statistics are treated as independent Gaussians with internal-variability uncertainties estimated from the NorESM2-LM ensemble, and the aerosol forcing amplitude is normalized to $-1.0$~W~m$^{-2}$ (2005--2014 mean) at unit scale. The posterior variance reduction quoted in the main text is evaluated at the fixed 2050 target, $T(2041$--$2050) - T(1990$--$2020)$; the range $42$--$54\%$ spans the three available realizations of internal variability in the NorESM2-LM ensemble. Because the constraint statistics share a common observational baseline period, they are in practice correlated, and treating them as independent Gaussians likely modestly overstates the strength of the constraint; together with the small number of realizations and the two-parameter uncertainty treatment, the quoted variance reductions should be read as illustrative of the mechanism rather than as precise estimates.

\section{Aggregated (EOF-projection) Tier~II scorecard}
\label{app:eofscorecard}

\begin{figure}[htbp]
    \centering
    \includegraphics[width=\linewidth]{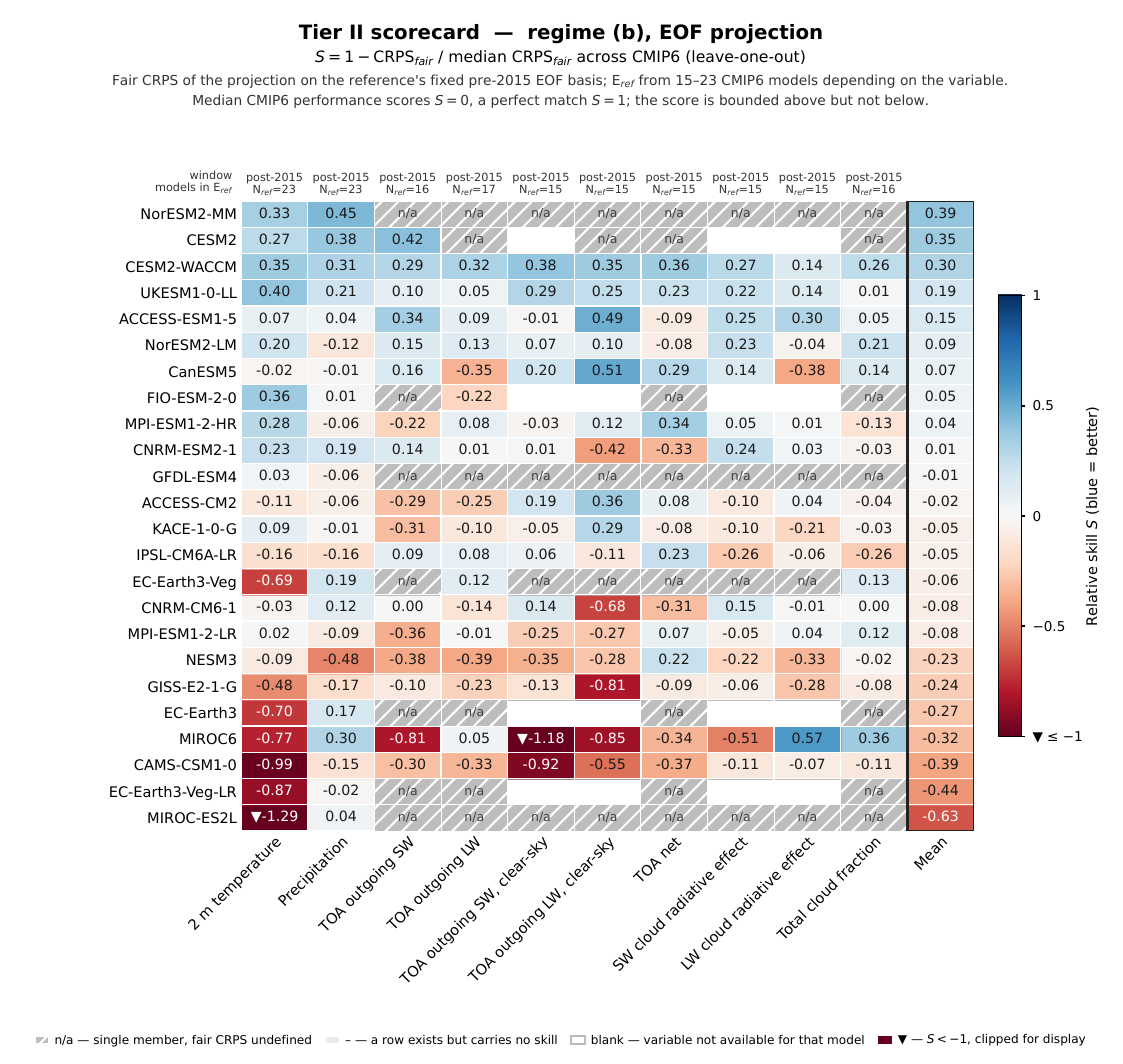}
    \caption{Tier~II scorecard for the aggregated (spatial-pattern) score of Section~\ref{sec:scoring}, for the same 24 CMIP6 models and ensemble definition as Figure~\ref{fig:benchmark}. For each variable the post-2015 time-mean field of every member is expressed as an anomaly about the \emph{reference's} 1985--2014 climatology (2001--2014 for the CERES-referenced columns, 2003--2014 for cloud fraction), projected onto the reference's EOFs computed over its pre-2015 record (the leading modes reaching 90\% of the variance, at most 20), each coefficient standardized by its pre-2015 observed standard deviation, and $\mathcal{E}$ is the equal-weight mean over coefficients of the fair CRPS; $S$ and $\mathcal{E}_\text{ref}$ are defined as in Figure~\ref{fig:benchmark}. Because the anomaly is taken about the observed rather than the model's own climatology, the mean-state bias pattern remains in the scored field. Sea-ice extent, a scalar series, has no projection and is omitted.}
    \label{fig:eofscorecard}
\end{figure}

\section{Paleoclimate proxy reference datasets}
\label{app:paleodata}

Note, for all time periods we use the raw proxy data presented in these papers instead of the assimilated global products also presented in some of them, as the global products are influenced by the spatial patterns and co-variances in the climate models used in the assimilation.

\textit{Last Interglacial:} \citeA{osmanUpdatedCompilationLast2026} provide an updated compilation of LIG SST (building on the original compilation of \citeA{hoffmanRegionalGlobalSeasurface2017}) data. For terrestrial temperatures, comparisons can be made to the compilation of \citeA{otto-bliesnerLargescaleFeaturesLast2021}, used to validate the PMIP4 simulations. For terrestrial precipitation responses, the compilation of \citeA{scussoliniAgreementReconstructedModeled2019}, provide constraints on the water cycle.

\textit{Last Glacial Maximum:} For SST, \citeA{tierneyClimatesInformOur2020} and \citeA{osmanGloballyResolvedSurface2021} both provide large proxy datasets for the LGM time slice. Land temperatures can be assessed with the pollen compilation of \citeA{bartleinPollenbasedContinentalClimate2011} and/or \citeA{cleatorNewMultivariableBenchmark2020} (noting that the latter is a data assimilation product and so depends in part on covariance structures from PMIP3 LGM experiments). 

\textit{Mid-Holocene:} For SST, the compilation of \citeA{osmanGloballyResolvedSurface2021} provides time-slice averages for the 5--7~ka period. Terrestrial temperature constraints are available from the Temp12K compilation \cite{kaufmanGlobalDatabaseHolocene2020}, and pollen-based reconstructions of both temperature and water balance from \citeA{bartleinPollenbasedContinentalClimate2011}.

\section{Benchmark implementation}
\label{app:implementation}

The benchmarking framework for ClimateBench2 is implemented as a wrapper around ClimateEval, drawing on diagnostics available in that framework to assemble the benchmark suites (defined as YAML files; see below). The generic diagnostics are contributed to ClimateEval; the protocol layer (thresholds, probabilistic scoring, baselines, and the leaderboard) is in the ClimateBench2 repository.

ClimateEval is a community diagnostic repository for the evaluation of Earth System Models during model development which encourages broad participation. Developed based on ICONEval \cite{ICONEval}, it can be run on every new simulation, giving modelers immediate feedback and guiding iterative model development. It complements existing tools such as ESMValTool \cite{eyringEarthSystemModel2020} and the CMIP Rapid Evaluation Framework \cite{hoffmannRapidEvaluationFramework2025}, which focus on the evaluation of CMIP multi-model ensembles, by providing a flexible environment for individual modeling groups to systematically test and diagnose their simulations during model development or before participation in a Model Intercomparison Project.

ClimateEval employs a suite-based architecture where diagnostics (e.g., spatial maps, time series, vertical profiles), variables, and reference datasets (e.g., ERA5 reanalysis, GPCP precipitation, CERES-EBAF radiative fluxes) are specified through YAML files, allowing flexible and reproducible evaluation workflows. Diagnostics include ``simple'' diagnostics, which create plots for a single dataset and calculate metrics relative to a reference dataset, and ``complex'' diagnostics, which are more flexible and can process arbitrary combinations of multiple datasets. The modular approach also facilitates the use of custom diagnostics, variables, and reference datasets; ClimateBench selects from this growing pool to define the benchmark. When executed, the suites process input data through standardized preprocessing steps such as unit conversion, spatial regridding, regional masking, and calculation of statistics. This preprocessing is based on ESMValCore \cite{righiEarthSystemModel2020}, which provides a set of robust and well-established preprocessing functions. Any missing reference data is automatically downloaded if required.

\section*{Open Research Section}

ClimateBench2 and ClimateEval are hosted under the climate-federation GitHub organization (\url{https://github.com/climate-federation}). Both are released under the Apache License, version 2.0. The latest release of ClimateBench2 and ClimateEval will be made publicly available on Zenodo on publication. 

The corresponding observational datasets used for calculation of the metrics are available here, and the perfect model training and test data will also be made available on publication.

Following AGU policy on the use of AI tools, we disclose that Anthropic's Claude was used at several stages of this work. It was used in drafting and revising the manuscript, including rewording and restructuring sections in response to co-author comments; in proposing candidate formulations, equations, and tolerances for several of the diagnostics specified in Tier~I and Tier~II; in identifying and verifying literature and constructing bibliography entries; and in writing analysis and figure-generation code, including the data pipeline underlying Figure~\ref{fig:testsplit} and parts of the benchmark evaluation code released with this protocol. No AI tool was used to generate or substitute for observational or model data; all data shown derive from the archived sources given in the Open Research section. All AI-assisted text, code, metric definitions, and citations were reviewed, tested, and where necessary corrected by the authors, who take full responsibility for the content of the manuscript and the released software.

\acknowledgments
DWP acknowledges generous support through a Google Academic Research Award and from U.S. National Science Foundation award 2441832. AP, MS, and VE acknowledge funding by the European Research Council (ERC) Synergy Grant ``Understanding and Modeling the Earth System with Machine Learning'' (USMILE) under the Horizon 2020 Research and Innovation program (Grant Agreement No. 855187). VE was additionally supported by the Deutsche Forschungsgemeinschaft (DFG, German Research Foundation) through the Gottfried Wilhelm Leibniz Prize awarded to Veronika Eyring (Reference No. EY 22/2-1).  IRS acknowledges funding from the National Center for Atmospheric Research, which is a major facility sponsored by the NSF under Cooperative Agreement No. 1852977. HH was supported by the Met Office Hadley Centre Climate Programme funded by DSIT. WDC acknowledges support from DOE's BER/RGMA-funded CASCADE scientific focus area.

\bibliography{references}

@article{valdiviesoAssessmentAirSea2017,
  author = {Valdivieso, M. and Haines, K. and Balmaseda, M. and Chang, Y.-S. and Drevillon, M. and Ferry, N. and Fujii, Y. and K{\"o}hl, A. and Storto, A. and Toyoda, T. and Wang, X. and Waters, J. and Xue, Y. and Yin, Y. and Barnier, B. and Hernandez, F. and Kumar, A. and Lee, T. and Masina, S. and Peterson, K. A.},
  title = {An assessment of air--sea heat fluxes from ocean and coupled reanalyses},
  journal = {Climate Dynamics},
  volume = {49},
  number = {3},
  pages = {983--1008},
  year = {2017},
  doi = {10.1007/s00382-015-2843-3}
}

@article{willsForcedComponentEstimation2026,
  title = {Forced {{Component Estimation Statistical Method Intercomparison Project}} ({{ForceSMIP}})},
  author = {Wills, Robert C. J. and Deser, Clara and McKinnon, Karen A. and Phillips, Adam and {Po-Chedley}, Stephen and Sippel, Sebastian and Merrifield, Anna L. and B{\^o}ne, Constantin and Bonfils, C{\'e}line and {Camps-Valls}, Gustau and Cropper, Stephen and Connolly, Charlotte and Duan, Shiheng and Durand, Homer and Feigin, Alexander and Fernandez, M. A. and Gastineau, Guillaume and Gavrilov, Andrei and Gordon, Emily and G{\"u}nther, Moritz and H{\"o}ver, Maren and Kravtsov, Sergey and Kuo, Yan-Ning and Lien, Justin and Madakumbura, Gavin D. and Mankovich, Nathan and Newman, Matthew and Rader, Jamin and Shi, Jia-Rui and Shin, Sang-Ik and Varando, Gherardo},
  year = 2026,
  month = mar,
  doi = {10.1175/JCLI-D-25-0326.1},
  urldate = {2026-08-21},
  chapter = {Journal of Climate},
  langid = {english},
}

@article{tokarskaPastWarmingTrend2020,
  author  = {Tokarska, Katarzyna B. and Stolpe, Martin B. and Sippel, Sebastian and Fischer, Erich M. and Smith, Christopher J. and Lehner, Flavio and Knutti, Reto},
  title   = {Past warming trend constrains future warming in {CMIP6} models},
  journal = {Science Advances},
  volume  = {6}, number = {12}, pages = {eaaz9549}, year = {2020},
  doi     = {10.1126/sciadv.aaz9549}
}

@article{cowtanRobustComparisonClimate2015,
  author  = {Cowtan, Kevin and Hausfather, Zeke and Hawkins, Ed and Jacobs, Peter and Mann, Michael E. and Miller, Sonya K. and Steinman, Byron A. and Stolpe, Martin B. and Way, Robert G.},
  title   = {Robust comparison of climate models with observations using blended land air and ocean sea surface temperatures},
  journal = {Geophysical Research Letters},
  volume  = {42}, number = {15}, pages = {6526--6534}, year = {2015},
  doi     = {10.1002/2015GL064888}
}

@article{richardsonReconciledClimateResponse2016,
  author  = {Richardson, Mark and Cowtan, Kevin and Hawkins, Ed and Stolpe, Martin B.},
  title   = {Reconciled climate response estimates from climate models and the energy budget of {Earth}},
  journal = {Nature Climate Change},
  volume  = {6}, number = {10}, pages = {931--935}, year = {2016},
  doi     = {10.1038/nclimate3066}
}

@article{allenConstraintsFutureChanges2002,
  title = {Constraints on Future Changes in Climate and the Hydrologic Cycle},
  author = {Allen, Myles R and Ingram, William J},
  year = 2002,
  month = sep,
  journal = {Nature Publishing Group},
  volume = {419},
  number = {6903},
  pages = {224},
  issn = {1476-4687},
  doi = {10.1038/nature01092},
}

@article{Hall19z,
	author = {Hall, Alex and Cox, Peter and Huntingford, Chris and Klein, Stephen},
	journal = {Nature Climate Change},
	pages = {269--278},
	title = {Progressing emergent constraints on future climate change},
	volume = {4},
	year = {2019}}

@article{Williamson21a,
	author = {Williamson, Mark S and Thackeray, Chad W and Cox, Peter M and Hall, Alex and Huntingford, Chris and Nijsse, Femke JMM},
	journal = {Rev. Mod. Phys.},
	pages = {025004},
	title = {Emergent constraints on climate sensitivities},
	volume = {93},
	year = {2021}}

@article{Schneider21b,
	author = {Tapio Schneider and Nadir Jeevanjee and Robert Socolow},
	journal = {Physics Today},
	pages = {44--51},
	title = {Accelerating progress in climate science},
	volume = {74},
	year = {2021}}

@article{allenQuantifyingUncertaintyForecasts2000,
  title = {Quantifying the Uncertainty in Forecasts of Anthropogenic Climate Change},
  author = {Allen, Myles R and Stott, Peter A and Mitchell, John F B and Schnur, Reiner and Delworth, Thomas L},
  year = 2000,
  journal = {Nature},
  volume = {407},
  number = {6804},
  pages = {617--620},
  issn = {0028-0836},
  doi = {10.1038/35036559}
}

@article{arakawa2004cumulus,
  title={The cumulus parameterization problem: Past, present, and future},
  author={Arakawa, Akio},
  journal={Journal of climate},
  volume={17},
  number={13},
  pages={2493--2525},
  year={2004}
}

@article{arcomanoHybridAtmosphericModel2023,
  title = {A {{Hybrid Atmospheric Model Incorporating Machine Learning Can Capture Dynamical Processes Not Captured}} by {{Its Physics-Based Component}}},
  author = {Arcomano, Troy and Szunyogh, Istvan and Wikner, Alexander and Hunt, Brian R. and Ott, Edward},
  year = 2023,
  journal = {Geophysical Research Letters},
  volume = {50},
  number = {8},
  pages = {e2022GL102649},
  issn = {1944-8007},
  doi = {10.1029/2022GL102649},
  urldate = {2026-03-03},
  copyright = {\copyright{} 2023 The Authors.},
  langid = {english}
}

@article{bartleinPollenbasedContinentalClimate2011,
  title = {Pollen-Based Continental Climate Reconstructions at 6 and 21 Ka: A Global Synthesis},
  author = {Bartlein, Patrick J. and Harrison, Sandy P. and Brewer, Simon and Connor, Simon and Davis, Basil A. S. and Gajewski, Konrad and Guiot, Jo{\"e}l and {Harrison-Prentice}, Thomas I. and Henderson, Andrew and Peyron, Odile and Prentice, I. Colin and Scholze, Marko and Sepp{\"a}, Heikki and Shuman, Bryan and Sugita, Shinya and Thompson, Robert S. and Viau, Andr{\'e} E. and Williams, John and Wu, Haibin},
  year = 2011,
  journal = {Climate Dynamics},
  volume = {37},
  pages = {775--802},
  doi = {10.1007/s00382-010-0904-1}
}

@misc{bauerLearningClimateSensitivity2025,
  title = {Learning {{Climate Sensitivity}} from {{Future Observations}}, {{Fast}} and {{Slow}}},
  author = {Bauer, Adam Michael and Proistosescu, Cristian and Droegemeier, Kelvin K.},
  year = 2025,
  month = jul,
  journal = {arXiv.org},
  urldate = {2026-02-13},
  howpublished = {https://arxiv.org/abs/2507.15767v1},
  langid = {english}
}

@article{bouabidFaIRGPBayesianEnergy2024,
  title = {{{FaIRGP}}: {{A Bayesian Energy Balance Model}} for {{Surface Temperatures Emulation}}},
  author = {Bouabid, Shahine and Sejdinovic, Dino and Watson-Parris, Duncan},
  year = 2024,
  journal = {Journal of Advances in Modeling Earth Systems},
  volume = {16},
  number = {6},
  issn = {1942-2466},
  doi = {10.1029/2023ms003926}
}

@article{chouEvaluatingRichGetRicherMechanism2009,
  title = {Evaluating the ``{{Rich-Get-Richer}}'' {{Mechanism}} in {{Tropical Precipitation Change}} under {{Global Warming}}},
  author = {Chou, Chia and Neelin, J. David and Chen, Chao-An and Tu, Jien-Yi},
  year = 2009,
  journal = {Journal of Climate},
  volume = {22},
  number = {8},
  pages = {1982--2005},
  doi = {10.1175/2008JCLI2471.1}
}

@article{cleatorNewMultivariableBenchmark2020,
  title = {A New Multivariable Benchmark for {{Last Glacial Maximum}} Climate Simulations},
  author = {Cleator, Sean F. and Harrison, Sandy P. and Nichols, Nancy K. and Prentice, I. Colin and Roulstone, Ian},
  year = 2020,
  journal = {Climate of the Past},
  volume = {16},
  pages = {699--712},
  doi = {10.5194/cp-16-699-2020}
}

@article{cooperLastGlacialMaximum2024,
  title = {Last {{Glacial Maximum}} Pattern Effects Reduce Climate Sensitivity Estimates},
  author = {Cooper, Vincent T. and Armour, Kyle C. and Hakim, Gregory J. and Tierney, Jessica E. and Osman, Matthew B. and Proistosescu, Cristian and Dong, Yue and Burls, Natalie J. and Andrews, Timothy and Amrhein, Daniel E. and Zhu, Jiang and Dong, Wenhao and Ming, Yi and Chmielowiec, Philip},
  year = 2024,
  month = apr,
  journal = {Science Advances},
  volume = {10},
  number = {16},
  pages = {eadk9461},
  publisher = {American Association for the Advancement of Science},
  doi = {10.1126/sciadv.adk9461},
  urldate = {2026-02-13}
}

@article{cooperPaleoclimatePatternEffects2026,
  title = {Paleoclimate Pattern Effects Help Constrain Climate Sensitivity and 21st-Century Warming},
  author = {Cooper, Vincent T. and Armour, Kyle C. and Hakim, Gregory J. and Tierney, Jessica E. and Burls, Natalie J. and Proistosescu, Cristian and Andrews, Timothy and Dong, Wenhao and Dvorak, Michelle T. and Feng, Ran and Osman, Matthew B. and Dong, Yue},
  year = 2026,
  month = jan,
  journal = {Proceedings of the National Academy of Sciences},
  volume = {123},
  number = {4},
  pages = {e2511370123},
  publisher = {Proceedings of the National Academy of Sciences},
  doi = {10.1073/pnas.2511370123},
  urldate = {2026-02-13}
}

@article{cresswell-clayDeepLearningEarth2025,
  title = {A {{Deep Learning Earth System Model}} for {{Efficient Simulation}} of the {{Observed Climate}}},
  author = {{Cresswell-Clay}, Nathaniel and Liu, Bowen and Durran, Dale R. and Liu, Zihui and Espinosa, Zachary I. and Moreno, Raul A. and Karlbauer, Matthias},
  year = 2025,
  journal = {AGU Advances},
  volume = {6},
  number = {4},
  pages = {e2025AV001706},
  issn = {2576-604X},
  doi = {10.1029/2025AV001706},
  urldate = {2026-01-29},
  copyright = {\copyright{} 2025. The Author(s).},
  langid = {english}
}

@article{donohoeSeasonalCycleAtmospheric2013,
  title = {The {{Seasonal Cycle}} of {{Atmospheric Heating}} and {{Temperature}}},
  author = {Donohoe, Aaron and Battisti, David S.},
  year = 2013,
  journal = {Journal of Climate},
  volume = {26},
  number = {14},
  pages = {4962--4980},
  doi = {10.1175/JCLI-D-12-00713.1}
}

@article{eyringEarthSystemModel2020,
  title = {Earth {{System Model Evaluation Tool}} ({{ESMValTool}}) v2.0 -- an Extended Set of Large-Scale Diagnostics for Quasi-Operational and Comprehensive Evaluation of {{Earth}} System Models in {{CMIP}}},
  author = {Eyring, Veronika and Bock, Lisa and Lauer, Axel and Righi, Mattia and Schlund, Manuel and Andela, Bouwe and Arnone, Enrico and Bellprat, Omar and Br{\"o}tz, Bj{\"o}rn and Caron, Louis-Philippe and Carvalhais, Nuno and Cionni, Irene and Cortesi, Nicola and Crezee, Bas and Davin, Edouard L. and Davini, Paolo and Debeire, Kevin and {de Mora}, Lee and Deser, Clara and Docber, David and Haber, Laura and Hagemann, Stefan and Hassler, Birgit and Hunter, Alistair and Kadow, Christopher and Kindermann, Stephan and Koirala, Sujan and Koldunov, Nikolay and Lejeune, Quentin and Magari{\~n}o, Valerio and Maher, Nicola and Manzini, Elisa and M{\"u}ller, Rolf and Nabel, Julia and Nam, Christine and Niess, Simone and Predoi, Valeriu and Rember, Erich and Righi, Mattia and S{\'e}n{\'e}si, St{\'e}phane and S{\"o}rensson, Anna and Stacke, Tobias and Steinert, Nicola J. and Swaminathan, Ranjini and Torber, Tobias and {Vegas-Regidor}, Javier and {von Hardenberg}, Jost and Weigel, Katja and Zimmermann, Klaus},
  year = 2020,
  journal = {Geoscientific Model Development},
  volume = {13},
  number = {7},
  pages = {3383--3438},
  doi = {10.5194/gmd-13-3383-2020}
}

@incollection{eyringHumanInfluenceClimate2021,
  type = {Book {{Section}}},
  title = {Human {{Influence}} on the {{Climate System}}},
  booktitle = {Climate {{Change}} 2021: {{The Physical Science Basis}}. {{Contribution}} of {{Working Group I}} to the {{Sixth Assessment Report}} of the {{Intergovernmental Panel}} on {{Climate Change}}},
  author = {Eyring, V. and Gillett, N.P. and Achuta Rao, K.M. and Barimalala, R. and Barreiro Parrillo, M. and Bellouin, N. and Cassou, C. and Durack, P.J. and Kosaka, Y. and McGregor, S. and Min, S. and Morgenstern, O. and Sun, Y.},
  editor = {{Masson-Delmotte}, V. and Zhai, P. and Pirani, A. and Connors, S.L. and P{\'e}an, C. and Berger, S. and Caud, N. and Chen, Y. and Goldfarb, L. and Gomis, M.I. and Huang, M. and Leitzell, K. and Lonnoy, E. and Matthews, J.B.R. and Maycock, T.K. and Waterfield, T. and Yelek{\c c}i, O. and Yu, R. and Zhou, B.},
  year = 2021,
  pages = {423--552},
  publisher = {Cambridge University Press},
  address = {Cambridge, United Kingdom and New York, NY, USA},
  doi = {10.1017/9781009157896.005}
}

@article{eyringOverviewCoupledModel2016,
  title = {Overview of the {{Coupled Model Intercomparison Project Phase}} 6 ({{CMIP6}}) Experimental Design and Organization},
  author = {Eyring, Veronika and Bony, Sandrine and Meehl, Gerald A and Senior, Catherine A and Stevens, Bjorn and Stouffer, Ronald J and Taylor, Karl E},
  year = 2016,
  journal = {Geoscientific Model Development},
  volume = {9},
  number = {5},
  pages = {1937--1958},
  issn = {1991-959X},
  doi = {10.5194/gmd-9-1937-2016},
  langid = {english}
}

@article{eyringTakingClimateModel2019,
  title = {Taking Climate Model Evaluation to the next Level},
  author = {Eyring, Veronika and Cox, Peter M. and Flato, Gregory M. and Gleckler, Peter J. and Abramowitz, Gab and Caldwell, Peter and Collins, William D. and Gier, Bettina K. and Hall, Alex D. and Hoffman, Forrest M. and Hurtt, George C. and Jahn, Alexandra and Jones, Chris D. and Klein, Stephen A. and Krasting, John P. and Kwiatkowski, Lester and Lorenz, Ruth and Maloney, Eric and Meehl, Gerald A. and Pendergrass, Angeline G. and Pincus, Robert and Ruane, Alex C. and Russell, Joellen L. and Sanderson, Benjamin M. and Santer, Benjamin D. and Sherwood, Steven C. and Simpson, Isla R. and Stouffer, Ronald J. and Williamson, Mark S.},
  year = 2019,
  month = feb,
  journal = {Nature Climate Change},
  volume = {9},
  number = {2},
  pages = {102--110},
  publisher = {Nature Publishing Group},
  issn = {1758-6798},
  doi = {10.1038/s41558-018-0355-y},
  urldate = {2026-02-13},
  copyright = {2019 Springer Nature Limited},
  langid = {english}
}

@article{brientReducingUncertaintiesClimate2020,
	title = {Reducing {Uncertainties} in {Climate} {Projections} with {Emergent} {Constraints}: {Concepts}, {Examples} and {Prospects}},
	volume = {37},
	issn = {1861-9533},
	shorttitle = {Reducing {Uncertainties} in {Climate} {Projections} with {Emergent} {Constraints}},
	url = {https://doi.org/10.1007/s00376-019-9140-8},
	doi = {10.1007/s00376-019-9140-8},
	language = {en},
	number = {1},
	urldate = {2026-05-19},
	journal = {Advances in Atmospheric Sciences},
	author = {Brient, Florent},
	month = jan,
	year = {2020},
	pages = {1--15},
}

@incollection{forsterEarthsEnergyBudget2021,
  title = {The {{Earth}}'s {{Energy Budget}}, {{Climate Feedbacks}}, and {{Climate Sensitivity}}},
  author = {Forster, P. and Storelvmo, T. and Armour, K. and Collins, W. and Dufresne, J. L. and Frame, D. and Lunt, D. J. and Mauritsen, T. and Palmer, M. D. and Watanabe, M. and Wild, M. and Zhang, H.},
  editor = {{V. Masson-Delmotte} and A. Pirani and S. L. Connors and C. P{\'e}an and S. Berger and N. Caud,Y. Chen and L. Goldfarb and M. I. Gomis and M. Huang and K. Leitzell and E. Lonnoy and J.B.R. Matthews and T. K. Maycock and T. Waterfield and O. Yelek{\c c}i and R. Yu and B. Zhou and P. Zhai},
  year = 2021,
  series = {Climate {{Change}} 2021: {{The Physical Science Basis}}. {{Contribution}} of {{Working Group I}} to the {{Sixth Assessment Report}} of the {{Intergovernmental Panel}} on {{Climate Change}}},
  publisher = {Cambridge University Press}
}

@article{ghanTechnicalNoteEstimating2013,
  title = {Technical {{Note}}: {{Estimating}} Aerosol Effects on Cloud Radiative Forcing},
  author = {Ghan, S. J.},
  year = 2013,
  journal = {Atmospheric Chemistry and Physics},
  volume = {13},
  number = {19},
  pages = {9971--9974},
  issn = {1680-7316},
  doi = {10.5194/acp-13-9971-2013},
  langid = {english}
}

@article{gillettDetectionAttributionModel2016,
  title = {The {{Detection}} and {{Attribution Model Intercomparison Project}} ({{DAMIP}} v1.0) Contribution to {{CMIP6}}},
  author = {Gillett, Nathan P. and Shiogama, Hideo and Funke, Bernd and Hegerl, Gabriele and Knutti, Reto and Matthes, Katja and Santer, Benjamin D. and Stone, Daithi and Tebaldi, Claudia},
  year = 2016,
  journal = {Geoscientific Model Development},
  volume = {9},
  number = {10},
  pages = {3685--3697},
  doi = {10.5194/gmd-9-3685-2016}
}

@article{glecklerPerformanceMetricsClimate2008,
  title = {Performance Metrics for Climate Models},
  author = {Gleckler, P. J. and Taylor, K. E. and Doutriaux, C.},
  year = 2008,
  journal = {Journal of Geophysical Research: Atmospheres},
  volume = {113},
  number = {D6},
  issn = {2156-2202},
  doi = {10.1029/2007JD008972},
  urldate = {2026-02-13},
  copyright = {Copyright 2008 by the American Geophysical Union.},
  langid = {english}
}

@article{gregoryNewMethodDiagnosing2004,
  title = {A New Method for Diagnosing Radiative Forcing and Climate Sensitivity},
  author = {Gregory, J. M. and Ingram, W. J. and Palmer, M. A. and Jones, G. S. and Stott, P. A. and Thorpe, R. B. and Lowe, J. A. and Johns, T. C. and Williams, K. D.},
  year = 2004,
  journal = {Geophysical Research Letters},
  volume = {31},
  number = {3},
  issn = {0094-8276},
  doi = {10.1029/2003gl018747}
}

@article{hawkinsPotentialNarrowUncertainty2009,
  title = {The {{Potential}} to {{Narrow Uncertainty}} in {{Regional Climate Predictions}}},
  author = {Hawkins, Ed and Sutton, Rowan},
  year = 2009,
  journal = {Bulletin of the American Meteorological Society},
  volume = {90},
  number = {8},
  pages = {1095--1108},
  issn = {0003-0007},
  doi = {10.1175/2009bams2607.1}
}

@article{hoffmanRegionalGlobalSeasurface2017,
  title = {Regional and Global Sea-Surface Temperatures during the Last Interglaciation},
  author = {Hoffman, Jeremy S. and Clark, Peter U. and Parnell, Andrew C. and He, Feng},
  year = 2017,
  journal = {Science},
  volume = {355},
  number = {6322},
  pages = {276--279},
  doi = {10.1126/science.aai8464}
}

@book{ipccClimateChange20212021,
  type = {Book},
  title = {Climate {{Change}} 2021: {{The Physical Science Basis}}. {{Contribution}} of {{Working Group I}} to the {{Sixth Assessment Report}} of the {{Intergovernmental Panel}} on {{Climate Change}}},
  author = {{IPCC}},
  year = 2021,
  volume = {In Press},
  publisher = {Cambridge University Press},
  address = {Cambridge, United Kingdom and New York, NY, USA},
  doi = {10.1017/9781009157896}
}

@article{joshiMechanismsLandSea2008,
  title = {Mechanisms for the Land/Sea Warming Contrast Exhibited by Simulations of Climate Change},
  author = {Joshi, Manoj M. and Gregory, Jonathan M. and Webb, Mark J. and Sexton, David M. H. and Johns, Tim C.},
  year = 2008,
  journal = {Climate Dynamics},
  volume = {30},
  pages = {455--465},
  doi = {10.1007/s00382-007-0306-1}
}

@article{jumperHighlyAccurateProtein2021,
  title = {Highly Accurate Protein Structure Prediction with {{AlphaFold}}},
  author = {Jumper, John and Evans, Richard and Pritzel, Alexander and Green, Tim and Figurnov, Michael and Ronneberger, Olaf and Tunyasuvunakool, Kathryn and Bates, Russ and {\v Z}{\'i}dek, Augustin and Potapenko, Anna and Bridgland, Alex and Meyer, Clemens and Kohl, Simon A. A. and Ballard, Andrew J. and Cowie, Andrew and {Romera-Paredes}, Bernardino and Nikolov, Stanislav and Jain, Rishub and Adler, Jonas and Back, Trevor and Petersen, Stig and Reiman, David and Clancy, Ellen and Zielinski, Michal and Steinegger, Martin and Pacholska, Michalina and Berghammer, Tamas and Bodenstein, Sebastian and Silver, David and Vinyals, Oriol and Senior, Andrew W. and Kavukcuoglu, Koray and Kohli, Pushmeet and Hassabis, Demis},
  year = 2021,
  month = aug,
  journal = {Nature},
  volume = {596},
  number = {7873},
  pages = {583--589},
  publisher = {Nature Publishing Group},
  issn = {1476-4687},
  doi = {10.1038/s41586-021-03819-2},
  urldate = {2026-03-03},
  copyright = {2021 The Author(s)},
  langid = {english}
}

@article{kangResponseITCZExtratropical2008,
  title = {The {{Response}} of the {{ITCZ}} to {{Extratropical Thermal Forcing}}: {{Idealized Slab-Ocean Experiments}} with a {{GCM}}},
  author = {Kang, Sarah M. and Held, Isaac M. and Frierson, Dargan M. W. and Zhao, Ming},
  year = 2008,
  journal = {Journal of Climate},
  volume = {21},
  number = {14},
  pages = {3521--3532},
  doi = {10.1175/2007JCLI2146.1}
}

@article{kangTropicalResponseExtratropical2009,
  title = {The {{Tropical Response}} to {{Extratropical Thermal Forcing}} in an {{Idealized GCM}}: {{The Importance}} of {{Radiative Feedbacks}} and {{Convective Parameterization}}},
  author = {Kang, Sarah M. and Frierson, Dargan M. W. and Held, Isaac M.},
  year = 2009,
  journal = {Journal of the Atmospheric Sciences},
  volume = {66},
  number = {9},
  pages = {2812--2827},
  doi = {10.1175/2009JAS2924.1}
}

@article{kaufmanGlobalDatabaseHolocene2020,
  title = {A Global Database of {{Holocene}} Paleotemperature Records},
  author = {Kaufman, Darrell and McKay, Nicholas and Routson, Cody and Erb, Michael and Davis, Basil and Heiri, Oliver and Jaccard, Samuel and Tierney, Jessica and D{\"a}twyler, Christoph and Axford, Yarrow and Brussel, Thomas and Cartapanis, Olivier and Chase, Brian and Dawson, Andria and {de Vernal}, Anne and Engels, Stefan and Jonkers, Lukas and Marsicek, Jeremiah and {Moffa-S{\'a}nchez}, Paola and Morrill, Carrie and Orsi, Anais and Rehfeld, Kira and Saunders, Krystyna and Sommer, Philipp S. and Thomas, Elizabeth and Tonello, Marcela and T{\'o}th, M{\'o}nika and Vachula, Richard and Andreev, Andrei and Bertrand, Sebastien and Biskaborn, Boris and Bringu{\'e}, Manuel and Brooks, Stephen and Caniup{\'a}n, Magaly and Chevalier, Manuel and Cwynar, Les and {Emile-Geay}, Julien and Fegyveresi, John and Feurdean, Angelica and Finsinger, Walter and Fortin, Marie-Claude and Foster, Louise and Fox, Mathew and Gajewski, Konrad and Grosjean, Martin and Hausmann, Sonja and Heinrichs, Markus and Holmes, Naomi and Ilyashuk, Boris and Ilyashuk, Elena and Juggins, Steve and Khider, Deborah and Koinig, Karin and Langdon, Peter and {Larocque-Tobler}, Isabelle and Li, Jianyong and Lotter, Andr{\'e} and Luoto, Tomi and Mackay, Anson and Magyari, Eniko and Malevich, Steven and Mark, Bryan and Massaferro, Julieta and Montade, Vincent and Nazarova, Larisa and Novenko, Elena and Pa{\v r}il, Petr and Pearson, Emma and Peros, Matthew and Pienitz, Reinhard and P{\l}{\'o}ciennik, Mateusz and Porinchu, David and Potito, Aaron and Rees, Andrew and Reinemann, Scott and Roberts, Stephen and Rolland, Nicolas and Salonen, Sakari and Self, Angela and Sepp{\"a}, Heikki and Shala, Shyhrete and {St-Jacques}, Jeannine-Marie and Stenni, Barbara and Syrykh, Liudmila and Tarrats, Pol and Taylor, Karen and {van den Bos}, Valerie and Velle, Gaute and Wahl, Eugene and Walker, Ian and Wilmshurst, Janet and Zhang, Enlou and Zhilich, Snezhana},
  year = 2020,
  month = apr,
  journal = {Scientific Data},
  volume = {7},
  number = {1},
  pages = {115},
  publisher = {Nature Publishing Group},
  issn = {2052-4463},
  doi = {10.1038/s41597-020-0445-3},
  urldate = {2026-03-03},
  copyright = {2020 The Author(s)},
  langid = {english}
}

@article{kaushalSISALv3GlobalSpeleothem2024,
  title = {{{SISALv3}}: A Global Speleothem Stable Isotope and Trace Element Database},
  shorttitle = {{{SISALv3}}},
  author = {Kaushal, Nikita and Lechleitner, Franziska A. and Wilhelm, Micah and Azennoud, Khalil and B{\"u}hler, Janica C. and Braun, Kerstin and Ait Brahim, Yassine and Baker, Andy and Burstyn, Yuval and {Comas-Bru}, Laia and Fohlmeister, Jens and Goldsmith, Yonaton and Harrison, Sandy P. and Hatvani, Istv{\'a}n G. and Rehfeld, Kira and Ritzau, Magdalena and Skiba, Vanessa and Stoll, Heather M. and Sz{\H u}cs, J{\'o}zsef G. and Tanos, P{\'e}ter and Treble, Pauline C. and Azevedo, Vitor and Baker, Jonathan L. and Borsato, Andrea and Chawchai, Sakonvan and Columbu, Andrea and Endres, Laura and Hu, Jun and Kern, Zolt{\'a}n and Kimbrough, Alena and Ko{\c c}, Koray and Markowska, Monika and Martrat, Belen and Masood Ahmad, Syed and Nehme, Carole and Novello, Valdir Felipe and {P{\'e}rez-Mej{\'i}as}, Carlos and Ruan, Jiaoyang and Sekhon, Natasha and Sinha, Nitesh and Tadros, Carol V. and Tiger, Benjamin H. and Warken, Sophie and Wolf, Annabel and Zhang, Haiwei and {SISAL Working Group members}},
  year = 2024,
  month = apr,
  journal = {Earth System Science Data},
  volume = {16},
  number = {4},
  pages = {1933--1963},
  publisher = {Copernicus GmbH},
  issn = {1866-3508},
  doi = {10.5194/essd-16-1933-2024},
  urldate = {2026-03-03},
  langid = {english}
}

@article{kiehlTwentiethCenturyClimate2007,
  title = {Twentieth Century Climate Model Response and Climate Sensitivity},
  author = {Kiehl, Jeffrey T.},
  year = 2007,
  journal = {Geophysical Research Letters},
  volume = {34},
  number = {22},
  issn = {1944-8007},
  doi = {10.1029/2007gl031383}
}

@article{knuttiChallengesCombiningProjections2010,
  title = {Challenges in {{Combining Projections}} from {{Multiple Climate Models}}},
  author = {Knutti, Reto and Furrer, Reinhard and Tebaldi, Claudia and Cermak, Jan and Meehl, Gerald A.},
  year = 2010,
  journal = {Journal of Climate},
  volume = {23},
  number = {10},
  pages = {2739--2758},
  issn = {0894-8755},
  doi = {10.1175/2009jcli3361.1},
  langid = {english}
}

@article{knuttiClimateModelProjection2017,
  title = {A Climate Model Projection Weighting Scheme Accounting for Performance and Interdependence},
  author = {Knutti, Reto and Sedl{\'a}{\v c}ek, Jan and Sanderson, Benjamin M. and Lorenz, Ruth and Fischer, Erich M. and Eyring, Veronika},
  year = 2017,
  journal = {Geophysical Research Letters},
  volume = {44},
  number = {4},
  pages = {1909--1918},
  issn = {1944-8007},
  doi = {10.1002/2016GL072012},
  urldate = {2026-02-13},
  copyright = {\copyright 2017. American Geophysical Union. All Rights Reserved.},
  langid = {english}
}

@article{kochkovNeuralGeneralCirculation2024,
  title = {Neural General Circulation Models for Weather and Climate},
  author = {Kochkov, Dmitrii and Yuval, Janni and Langmore, Ian and Norgaard, Peter and Smith, Jamie and Mooers, Griffin and Kl{\"o}wer, Milan and Lottes, James and Rasp, Stephan and D{\"u}ben, Peter and Hatfield, Sam and Battaglia, Peter and {Sanchez-Gonzalez}, Alvaro and Willson, Matthew and Brenner, Michael P. and Hoyer, Stephan},
  year = 2024,
  journal = {Nature},
  volume = {632},
  number = {8027},
  pages = {1060--1066},
  issn = {0028-0836},
  doi = {10.1038/s41586-024-07744-y}
}

@article{lauerEarthSystemModel2020,
  title = {Earth {{System Model Evaluation Tool}} ({{ESMValTool}}) v2.0 -- Diagnostics for Emergent Constraints and Future Projections from {{Earth}} System Models in {{CMIP}}},
  author = {Lauer, Axel and Eyring, Veronika and Bellprat, Omar and Bock, Lisa and Gier, Bettina K. and Hunter, Alasdair and Lorenz, Ruth and {P{\'e}rez-Zan{\'o}n}, N{\'u}ria and Righi, Mattia and Schlund, Manuel and Senftleben, Daniel and Weigel, Katja and Zechlau, Sabrina},
  year = 2020,
  month = sep,
  journal = {Geoscientific Model Development},
  volume = {13},
  number = {9},
  pages = {4205--4228},
  publisher = {Copernicus GmbH},
  issn = {1991-959X},
  doi = {10.5194/gmd-13-4205-2020},
  urldate = {2026-02-13},
  langid = {english}
}

@article{leachFaIRv200GeneralizedImpulse2021,
  title = {{{FaIRv2}}.0.0: A Generalized Impulse Response Model for Climate Uncertainty and Future Scenario Exploration},
  author = {Leach, Nicholas J. and Jenkins, Stuart and Nicholls, Zebedee and Smith, Christopher J. and Lynch, John and Cain, Michelle and Walsh, Tristram and Wu, Bill and Tsutsui, Junichi and Allen, Myles R.},
  year = 2021,
  journal = {Geoscientific Model Development},
  volume = {14},
  number = {5},
  pages = {3007--3036},
  doi = {10.5194/gmd-14-3007-2021}
}

@incollection{leeFutureGlobalClimate2021,
  type = {Book {{Section}}},
  title = {Future {{Global Climate}}: {{Scenario-Based Projections}} and {{Near-Term Information}}},
  booktitle = {Climate {{Change}} 2021: {{The Physical Science Basis}}. {{Contribution}} of {{Working Group I}} to the {{Sixth Assessment Report}} of the {{Intergovernmental Panel}} on {{Climate Change}}},
  author = {Lee, J.-Y. and Marotzke, J. and Bala, G. and Cao, L. and Corti, S. and Dunne, J.P. and Engelbrecht, F. and Fischer, E. and Fyfe, J.C. and Jones, C. and Maycock, A. and Mutemi, J. and Ndiaye, O. and Panickal, S. and Zhou, T.},
  editor = {{Masson-Delmotte}, V. and Zhai, P. and Pirani, A. and Connors, S.L. and P{\'e}an, C. and Berger, S. and Caud, N. and Chen, Y. and Goldfarb, L. and Gomis, M.I. and Huang, M. and Leitzell, K. and Lonnoy, E. and Matthews, J.B.R. and Maycock, T.K. and Waterfield, T. and Yelek{\c c}i, O. and Yu, R. and Zhou, B.},
  year = 2021,
  pages = {553--672},
  publisher = {Cambridge University Press},
  address = {Cambridge, United Kingdom and New York, NY, USA},
  doi = {10.1017/9781009157896.006}
}

@article{luntPaleoclimateDataProvide2024,
  title = {Paleoclimate Data Provide Constraints on Climate Models' Large-Scale Response to Past {{CO2}} Changes},
  author = {Lunt, Daniel J. and {Otto-Bliesner}, Bette L. and Brierley, Chris and Haywood, Alan and Inglis, Gordon N. and Izumi, Kenji and Kageyama, Masa and Kaufman, Darrell and Mauritsen, Thorsten and McClymont, Erin L. and Salzmann, Ulrich and Steinig, Sebastian and Tierney, Jessica E. and Zhao, Anni and Zhu, Jiang},
  year = 2024,
  month = aug,
  journal = {Communications Earth \& Environment},
  volume = {5},
  number = {1},
  pages = {419},
  publisher = {Nature Publishing Group},
  issn = {2662-4435},
  doi = {10.1038/s43247-024-01531-3},
  urldate = {2026-02-13},
  copyright = {2024 The Author(s)},
  langid = {english}
}

@article{manabeThermalEquilibriumAtmosphere1967,
  title = {Thermal {{Equilibrium}} of the {{Atmosphere}} with a {{Given Distribution}} of {{Relative Humidity}}},
  author = {Manabe, Syukuro and Wetherald, Richard T.},
  year = 1967,
  journal = {Journal of the Atmospheric Sciences},
  volume = {24},
  number = {3},
  pages = {241--259},
  issn = {0022-4928},
  doi = {10.1175/1520-0469(1967)024<0241:teotaw>2.0.co;2}
}

@article{marshallMeanClimateVariability2007,
  title = {Mean {{Climate}} and {{Variability}} of the {{Atmosphere}} and {{Ocean}} on an {{Aquaplanet}}},
  author = {Marshall, John and Ferreira, David and Campin, Jean-Michel and Enderton, Daniel},
  year = 2007,
  journal = {Journal of the Atmospheric Sciences},
  volume = {64},
  number = {12},
  pages = {4270--4286},
  doi = {10.1175/2007JAS2226.1}
}

@article{mathisonRapidapplicationEmissionstoimpactsTool2025,
  title = {A Rapid-Application Emissions-to-Impacts Tool for Scenario Assessment: {{Probabilistic Regional Impacts}} from {{Model}} Patterns and {{Emissions}} ({{PRIME}})},
  shorttitle = {A Rapid-Application Emissions-to-Impacts Tool for Scenario Assessment},
  author = {Mathison, Camilla and Burke, Eleanor J. and Munday, Gregory and Jones, Chris D. and Smith, Chris J. and Steinert, Norman J. and Wiltshire, Andy J. and Huntingford, Chris and Kovacs, Eszter and Gohar, Laila K. and Varney, Rebecca M. and McNeall, Douglas},
  year = 2025,
  month = mar,
  journal = {Geoscientific Model Development},
  volume = {18},
  number = {5},
  pages = {1785--1808},
  publisher = {Copernicus GmbH},
  issn = {1991-959X},
  doi = {10.5194/gmd-18-1785-2025},
  urldate = {2026-03-03},
  langid = {english}
}

@article{merrifieldClimateModelSelection2023,
  title = {Climate Model {{Selection}} by {{Independence}}, {{Performance}}, and {{Spread}} ({{ClimSIPS}} v1.0.1) for Regional Applications},
  author = {Merrifield, Anna L. and Brunner, Lukas and Lorenz, Ruth and Humphrey, Vincent and Knutti, Reto},
  year = 2023,
  month = aug,
  journal = {Geoscientific Model Development},
  volume = {16},
  number = {16},
  pages = {4715--4747},
  publisher = {Copernicus GmbH},
  issn = {1991-959X},
  doi = {10.5194/gmd-16-4715-2023},
  urldate = {2026-02-13},
  langid = {english}
}

@article{nowackOpinionWhyAll2025,
  title = {Opinion: {{Why}} All Emergent Constraints Are Wrong but Some Are Useful -- a Machine Learning Perspective},
  shorttitle = {Opinion},
  author = {Nowack, Peer and {Watson-Parris}, Duncan},
  year = 2025,
  month = feb,
  journal = {Atmospheric Chemistry and Physics},
  volume = {25},
  number = {4},
  pages = {2365--2384},
  publisher = {Copernicus GmbH},
  issn = {1680-7316},
  doi = {10.5194/acp-25-2365-2025},
  urldate = {2025-07-23},
  langid = {english}
}

@article{osmanGloballyResolvedSurface2021,
  title = {Globally Resolved Surface Temperatures since the {{Last Glacial Maximum}}},
  author = {Osman, Matthew B. and Tierney, Jessica E. and Zhu, Jiang and Tardif, Robert and Hakim, Gregory J. and King, Jonathan and Poulsen, Christopher J.},
  year = 2021,
  journal = {Nature},
  volume = {599},
  number = {7884},
  pages = {239--244},
  doi = {10.1038/s41586-021-03984-4}
}

@article{osmanUpdatedCompilationLast2026,
  title = {Updated Compilation of {{Last Interglacial}} Sea Surface Temperatures},
  author = {Osman, Matthew B.},
  year = 2026
}

@article{otto-bliesnerLargescaleFeaturesLast2021,
  title = {Large-Scale Features of {{Last Interglacial}} Climate: Results from Evaluating the Lig127k Simulations for the {{Coupled Model Intercomparison Project}} ({{CMIP6}})--{{Paleoclimate Modeling Intercomparison Project}} ({{PMIP4}})},
  author = {{Otto-Bliesner}, Bette L. and Brady, Esther C. and Zhao, Aixue and Brierley, Chris M. and Axford, Yarrow and Capron, Emilie and Goelzer, Heiko and Kageyama, Masa and Kelley, Maxwell and Lunt, Daniel J. and Phipps, Steven J. and Pop, Sorin and Rosenbloom, Nan and Singh, Hansi A. and Stone, Emma J.},
  year = 2021,
  journal = {Climate of the Past},
  volume = {17},
  pages = {63--94},
  doi = {10.5194/cp-17-63-2021}
}

@article{pithanArcticAmplificationDominated2014,
  title = {Arctic Amplification Dominated by Temperature Feedbacks in Contemporary Climate Models},
  author = {Pithan, Felix and Mauritsen, Thorsten},
  year = 2014,
  journal = {Nature Geoscience},
  volume = {7},
  number = {3},
  pages = {181--184},
  doi = {10.1038/ngeo2071}
}

@incollection{plantonAnnexIVModes2021,
  title = {Annex {{IV}}: {{Modes}} of {{Variability}}},
  booktitle = {Climate {{Change}} 2021: {{The Physical Science Basis}}. {{Contribution}} of {{Working Group I}} to the {{Sixth Assessment Report}} of the {{Intergovernmental Panel}} on {{Climate Change}}},
  author = {Planton, Yohan Y. and Guilyardi, Eric and Wittenberg, Andrew T. and Lee, Jiwoo and Gleckler, Peter J. and Baehr, Johanna and Bellouin, Nicolas and Danabasoglu, Gokhan and Durack, Paul J. and Gerber, Edwin P. and Gillett, Nathan P. and Gregory, Jonathan M. and Kosaka, Yu and Maher, Nicola and Panickal, Swapna and Phillips, Adam S. and Robock, Alan and Scaife, Adam A.},
  year = 2021,
  publisher = {Cambridge University Press},
  address = {Cambridge, United Kingdom and New York, NY, USA}
}

@article{randall100YearsEarth2019,
  title = {100 {{Years}} of {{Earth System Model Development}} 100 {{Years}} of {{Earth System Model Development}}},
  author = {Randall, David A. and Bitz, Cecilia M. and Danabasoglu, Gokhan and Denning, A. Scott and Gent, Peter R. and Gettelman, Andrew and Griffies, Stephen M. and Lynch, Peter and Morrison, Hugh and Pincus, Robert and Thuburn, John},
  year = 2019,
  journal = {Meteorological Monographs},
  volume = {59},
  pages = {12.1-12.66},
  issn = {0065-9401},
  doi = {10.1175/amsmonographs-d-18-0018.1}
}

@article{raspDeepLearningRepresent2018,
  title = {Deep Learning to Represent Subgrid Processes in Climate Models.},
  author = {Rasp, Stephan and Pritchard, Michael S and Gentine, Pierre},
  year = 2018,
  journal = {Proceedings of the National Academy of Sciences of the United States of America},
  volume = {115},
  number = {39},
  pages = {9684--9689},
  issn = {0027-8424},
  doi = {10.1073/pnas.1810286115}
}

@article{raspWeatherBenchBenchmarkData2020,
  title = {{{WeatherBench}}: {{A Benchmark Data Set}} for {{Data}}-{{Driven Weather Forecasting}}},
  author = {Rasp, Stephan and Dueben, Peter D. and Scher, Sebastian and Weyn, Jonathan A. and Mouatadid, Soukayna and Thuerey, Nils},
  year = 2020,
  journal = {Journal of Advances in Modeling Earth Systems},
  volume = {12},
  number = {11},
  issn = {1942-2466},
  doi = {10.1029/2020ms002203}
}

@article{schlundConstrainingUncertaintyProjected2020,
  title = {Constraining {{Uncertainty}} in {{Projected Gross Primary Production With Machine Learning}}},
  author = {Schlund, Manuel and Eyring, Veronika and {Camps-Valls}, Gustau and Friedlingstein, Pierre and Gentine, Pierre and Reichstein, Markus},
  year = 2020,
  journal = {Journal of Geophysical Research: Biogeosciences},
  volume = {125},
  number = {11},
  pages = {e2019JG005619},
  issn = {2169-8961},
  doi = {10.1029/2019JG005619},
  urldate = {2026-02-13},
  copyright = {\copyright 2020. The Authors.},
  langid = {english}
}

@article{schmidtUsingPalaeoclimateComparisons2014,
  title = {Using Palaeo-Climate Comparisons to Constrain Future Projections in {{CMIP5}}},
  author = {Schmidt, G. A. and Annan, J. D. and Bartlein, P. J. and Cook, B. I. and Guilyardi, E. and Hargreaves, J. C. and Harrison, S. P. and Kageyama, M. and LeGrande, A. N. and Konecky, B. and Lovejoy, S. and Mann, M. E. and {Masson-Delmotte}, V. and Risi, C. and Thompson, D. and Timmermann, A. and Tremblay, L.-B. and Yiou, P.},
  year = 2014,
  journal = {Climate of the Past},
  volume = {10},
  number = {1},
  pages = {221--250},
  doi = {10.5194/cp-10-221-2014}
}

@article{schneiderMigrationsDynamicsIntertropical2014,
  title = {Migrations and Dynamics of the Intertropical Convergence Zone},
  author = {Schneider, Tapio and Bischoff, Tobias and Haug, Gerald H.},
  year = 2014,
  journal = {Nature},
  volume = {513},
  number = {7516},
  issn = {0028-0836},
  doi = {10.1038/nature13636}
}

@article{schneiderOpinionOptimizingClimate2024,
  title = {Opinion: {{Optimizing}} Climate Models with Process Knowledge, Resolution, and Artificial Intelligence},
  author = {Schneider, Tapio and Leung, L. Ruby and Wills, Robert C. J.},
  year = 2024,
  journal = {Atmospheric Chemistry and Physics},
  volume = {24},
  number = {12},
  pages = {7041--7062},
  doi = {10.5194/acp-24-7041-2024}
}

@article{scussoliniAgreementReconstructedModeled2019,
  title = {Agreement between Reconstructed and Modeled Boreal Precipitation of the {{Last Interglacial}}},
  author = {Scussolini, Paolo and Bakker, Pepijn and Guo, Chuanyu and Stepanek, Christian and Zhang, Qiong and Braconnot, Pascale and Cao, Jian and Guarino, Maria-Vittoria and Coumou, Dim and Prange, Matthias and Ward, Philip J. and Renssen, Hans and Kageyama, Masa and {Otto-Bliesner}, Bette and Aerts, Jeroen C. J. H.},
  year = 2019,
  journal = {Science Advances},
  volume = {5},
  number = {11},
  pages = {eaax7047},
  doi = {10.1126/sciadv.aax7047}
}

@article{shawOtherClimateCrisis2025,
  title = {The Other Climate Crisis},
  author = {Shaw, Tiffany A. and Stevens, Bjorn},
  year = 2025,
  journal = {Nature},
  volume = {639},
  number = {8056},
  pages = {877--887},
  issn = {0028-0836},
  doi = {10.1038/s41586-025-08680-1}
}

@article{sherwoodAssessmentEarthsClimate2020,
  title = {An {{Assessment}} of {{Earth}}'s {{Climate Sensitivity Using Multiple Lines}} of {{Evidence}}},
  author = {Sherwood, S. C. and Webb, M. J. and Annan, J. D. and Armour, K. C. and Forster, P. M. and Hargreaves, J. C. and Hegerl, G. and Klein, S. A. and Marvel, K. D. and Rohling, E. J. and Watanabe, M. and Andrews, T. and Braconnot, P. and Bretherton, C. S. and Foster, G. L. and Hausfather, Z. and Heydt, A. S. and Knutti, R. and Mauritsen, T. and Norris, J. R. and Proistosescu, C. and Rugenstein, M. and Schmidt, G. A. and Tokarska, K. B. and Zelinka, M. D.},
  year = 2020,
  journal = {Reviews of Geophysics},
  volume = {58},
  number = {4},
  pages = {e2019RG000678},
  issn = {8755-1209},
  doi = {10.1029/2019rg000678}
}

@article{simpsonConfrontingEarthSystem2025,
  title = {Confronting {{Earth System Model}} Trends with Observations},
  author = {Simpson, Isla R. and Shaw, Tiffany A. and Ceppi, Paulo and Clement, Amy C. and Fischer, Erich and Grise, Kevin M. and Pendergrass, Angeline G. and Screen, James A. and Wills, Robert C. J. and Woollings, Tim and Blackport, Russell and Kang, Joonsuk M. and {Po-Chedley}, Stephen},
  year = 2025,
  month = mar,
  journal = {Science Advances},
  volume = {11},
  number = {11},
  pages = {eadt8035},
  publisher = {American Association for the Advancement of Science},
  doi = {10.1126/sciadv.adt8035},
  urldate = {2026-03-03}
}

@article{smithEffectiveRadiativeForcing2020,
  title = {Effective Radiative Forcing and Adjustments in {{CMIP6}} Models},
  author = {Smith, Christopher J. and Kramer, Ryan J. and Myhre, Gunnar and Alterskj{\ae}r, Kari and Collins, William and Sima, Adriana and Boucher, Olivier and Dufresne, Jean-Louis and Nabat, Pierre and Michou, Martine and Yukimoto, Seiji and Cole, Jason and Paynter, David and Shiogama, Hideo and O'Connor, Fiona M. and Robertson, Eddy and Wiltshire, Andy and Andrews, Timothy and Hannay, C{\'e}cile and Miller, Ron and Nazarenko, Larissa and Kirkev{\aa}g, Alf and Olivi{\'e}, Dirk and Fiedler, Stephanie and Lewinschal, Anna and Mackallah, Chloe and Dix, Martin and Pincus, Robert and Forster, Piers M.},
  year = 2020,
  journal = {Atmospheric Chemistry and Physics},
  volume = {20},
  number = {16},
  pages = {9591--9618},
  doi = {10.5194/acp-20-9591-2020}
}

@article{stevens_rethinking_2015,
  title = {Rethinking the {{Lower Bound}} on {{Aerosol Radiative Forcing}}},
  author = {Stevens, Bjorn},
  year = 2015,
  month = mar,
  journal = {Journal of Climate},
  volume = {28},
  number = {12},
  pages = {4794--4819},
  issn = {0894-8755},
  doi = {10.1175/JCLI-D-14-00656.1},
  urldate = {2015-09-22},
  langid = {english}
}

@article{suttonLandSeaWarming2007,
  title = {Land/Sea Warming Ratio in Response to Climate Change: {{IPCC AR4}} Model Results and Comparison with Observations},
  author = {Sutton, Rowan T. and Dong, Buwen and Gregory, Jonathan M.},
  year = 2007,
  journal = {Geophysical Research Letters},
  volume = {34},
  number = {2},
  doi = {10.1029/2006GL028164}
}

@article{tebaldiEmulatorsClimateModel2025,
  title = {Emulators of {{Climate Model Output}}},
  author = {Tebaldi, C. and Selin, N. E. and Ferrari, R. and Flierl, G.},
  year = 2025,
  month = oct,
  journal = {Annual Review of Environment and Resources},
  volume = {50},
  number = {Volume 50, 2025},
  pages = {709--737},
  publisher = {Annual Reviews},
  issn = {1543-5938, 1545-2050},
  doi = {10.1146/annurev-environ-012125-085838},
  urldate = {2026-03-03},
  langid = {english}
}

@article{tierneyClimatesInformOur2020,
  title = {Past Climates Inform Our Future},
  author = {Tierney, Jessica E. and Poulsen, Christopher J. and Monta{\~n}ez, Isabel P. and Bhattacharya, Tripti and Feng, Ran and Ford, Heather L. and H{\"o}nisch, B{\"a}rbel and Inglis, Gordon N. and Petersen, Sierra V. and Sagoo, Navjit and Tabor, Clay R. and Thirumalai, Kaustubh and Zhu, Jiang and Burls, Natalie J. and Foster, Gavin L. and Godd{\'e}ris, Yves and Huber, Brian T. and Ivany, Linda C. and Kirtland Turner, Sandra and Lunt, Daniel J. and McElwain, Jennifer C. and Mills, Benjamin J. W. and {Otto-Bliesner}, Bette L. and Ridgwell, Andy and Zhang, Yi Ge},
  year = 2020,
  month = nov,
  journal = {Science},
  volume = {370},
  number = {6517},
  pages = {eaay3701},
  publisher = {American Association for the Advancement of Science},
  doi = {10.1126/science.aay3701},
  urldate = {2026-02-13}
}

@article{watson-parrisClimateBenchV10Benchmark2022,
  title = {{{ClimateBench}} v1.0: {{A Benchmark}} for {{Data}}-{{Driven Climate Projections}}},
  author = {Watson-Parris, D. and Rao, Y. and Olivi{\'e}, D. and Seland, {\O}. and Nowack, P. and Camps-Valls, G. and Stier, P. and Bouabid, S. and Dewey, M. and Fons, E. and Gonzalez, J. and Harder, P. and Jeggle, K. and Lenhardt, J. and Manshausen, P. and Novitasari, M. and Ricard, L. and Roesch, C.},
  year = 2022,
  journal = {Journal of Advances in Modeling Earth Systems},
  volume = {14},
  number = {10},
  issn = {1942-2466},
  doi = {10.1029/2021ms002954}
}

@article{watson-parrisMachineLearningWeather2021,
  title = {Machine Learning for Weather and Climate Are Worlds Apart},
  author = {{Watson-Parris}, D.},
  year = 2021,
  journal = {Philosophical Transactions of the Royal Society A},
  volume = {379},
  number = {2194},
  pages = {20200098},
  issn = {1364-503X},
  doi = {10.1098/rsta.2020.0098}
}

@article{watt-meyerACE2AccuratelyLearning2025,
  title = {{{ACE2}}: Accurately Learning Subseasonal to Decadal Atmospheric Variability and Forced Responses},
  shorttitle = {{{ACE2}}},
  author = {{Watt-Meyer}, Oliver and Henn, Brian and McGibbon, Jeremy and Clark, Spencer K. and Kwa, Anna and Perkins, W. Andre and Wu, Elynn and Harris, Lucas and Bretherton, Christopher S.},
  year = 2025,
  month = may,
  journal = {npj Climate and Atmospheric Science},
  volume = {8},
  number = {1},
  pages = {205},
  publisher = {Nature Publishing Group},
  issn = {2397-3722},
  doi = {10.1038/s41612-025-01090-0},
  urldate = {2026-03-03},
  copyright = {2025 The Author(s)},
  langid = {english}
}

@article{williamsDependenceFastChanges2023,
  title = {Dependence of Fast Changes in Global and Local Precipitation on the Geographical Location of Absorbing Aerosol},
  author = {Williams, Andrew I L and {Watson-Parris}, Duncan and Dagan, Guy and Stier, Philip},
  year = 2023,
  journal = {Journal of Climate},
  pages = {1--38},
  issn = {0894-8755},
  doi = {10.1175/jcli-d-23-0022.1}
}

@article{zhuLGMPaleoclimateConstraints2022,
  title = {{{LGM Paleoclimate Constraints Inform Cloud Parameterizations}} and {{Equilibrium Climate Sensitivity}} in {{CESM2}}},
  author = {Zhu, Jiang and Otto-Bliesner, Bette L. and Brady, Esther C. and Gettelman, Andrew and Bacmeister, Julio T. and Neale, Richard B. and Poulsen, Christopher J. and Shaw, Jonah K. and McGraw, Zachary S. and Kay, Jennifer E.},
  year = 2022,
  journal = {Journal of Advances in Modeling Earth Systems},
  volume = {14},
  number = {4},
  issn = {1942-2466},
  doi = {10.1029/2021ms002776}
}

@article{eyringStrategyProcessOriented2005,
  author  = {Eyring, V. and Harris, N. R. P. and Rex, M. and Shepherd, T. G. and Fahey, D. W. and Amanatidis, G. T. and Austin, J. and Chipperfield, M. P. and Dameris, M. and Forster, P. M. De F. and Gettelman, A. and Graf, H. F. and Nagashima, T. and Newman, P. A. and Pawson, S. and Prather, M. J. and Pyle, J. A. and Salawitch, R. J. and Santer, B. D. and Waugh, D. W.},
  title   = {A Strategy for Process-Oriented Validation of Coupled Chemistry--Climate Models},
  journal = {Bulletin of the American Meteorological Society},
  volume  = {86},
  number  = {8},
  pages   = {1117--1134},
  year    = {2005},
  doi     = {10.1175/BAMS-86-8-1117}
}

@article{dunneEvolvingCMIP2025,
  author  = {Dunne, John P. and Hewitt, Helene T. and Arblaster, Julie M. and Bonou, Fr{\'e}d{\'e}ric and Boucher, Olivier and Cavazos, Tereza and Dingley, Beth and Durack, Paul J. and Hassler, Birgit and Juckes, Martin and Miyakawa, Tomoki and Mizielinski, Matt and Naik, Vaishali and Nicholls, Zebedee and O'Rourke, Eleanor and Pincus, Robert and Sanderson, Benjamin M. and Simpson, Isla R. and Taylor, Karl E.},
  title   = {An evolving Coupled Model Intercomparison Project phase 7 (CMIP7) and Fast Track in support of future climate assessment},
  journal = {Geoscientific Model Development},
  volume  = {18},
  number  = {19},
  pages   = {6671--6700},
  year    = {2025},
  doi     = {10.5194/gmd-18-6671-2025}
}

@article{collierInternationalLandBenchmarking2018,
  author  = {Collier, Nathan and Hoffman, Forrest M. and Lawrence, David M. and Keppel-Aleks, Gretchen and Koven, Charles D. and Riley, William J. and Mu, Mingquan and Randerson, James T.},
  title   = {The International Land Model Benchmarking (ILAMB) System: Design, Theory, and Implementation},
  journal = {Journal of Advances in Modeling Earth Systems},
  volume  = {10},
  number  = {11},
  pages   = {2731--2754},
  year    = {2018},
  doi     = {10.1029/2018MS001354}
}

@article{bockQuantifyingProgressCMIP2020,
  author  = {Bock, L. and Lauer, A. and Schlund, M. and Barreiro, M. and Bellouin, N. and Jones, C. and Meehl, G. A. and Predoi, V. and Roberts, M. J. and Eyring, V.},
  title   = {Quantifying Progress Across Different CMIP Phases With the ESMValTool},
  journal = {Journal of Geophysical Research: Atmospheres},
  volume  = {125},
  number  = {21},
  pages   = {e2019JD032321},
  year    = {2020},
  doi     = {10.1029/2019JD032321}
}

@article{eyringImprovedRoutineEvaluation2016,
  author  = {Eyring, Veronika and Gleckler, Peter J. and Heinze, Christoph and Stouffer, Ronald J. and Taylor, Karl E. and Balaji, V. and Guilyardi, Eric and Joussaume, Sylvie and Kindermann, Stephan and Lawrence, Bryan N. and Meehl, Gerald A. and Righi, Mattia and Williams, Dean N.},
  title   = {Towards improved and more routine Earth system model evaluation in CMIP},
  journal = {Earth System Dynamics},
  volume  = {7},
  number  = {4},
  pages   = {813--830},
  year    = {2016},
  doi     = {10.5194/esd-7-813-2016}
}

@article{hoffmannRapidEvaluationFramework2025,
  author  = {Hoffman, Forrest M. and Hassler, Birgit and Swaminathan, Ranjini and Lewis, Jared and Andela, Bouwe and Collier, Nathaniel and Heged{\H{u}}s, D{\'o}ra and Lee, Jiwoo and Pascoe, Charlotte and Pfl{\"u}ger, Mika and Stockhause, Martina and Ullrich, Paul and Xu, Min and Bock, Lisa and Chun, Felicity and Gier, Bettina K. and Kelley, Douglas I. and Lauer, Axel and Lenhardt, Julien and Schlund, Manuel and Sreeush, Mohanan G. and Weigel, Katja and Blockley, Ed and Beadling, Rebecca and Beucher, Romain and Dugassa, Demiso D. and Lembo, Valerio and Lu, Jianhua and Brands, Swen and Tjiputra, Jerry and Malinina, Elizaveta and Medeiros, Brian and Scoccimarro, Enrico and Walton, Jeremy and Kershaw, Philip and Marquez, Andr{\'e} L. and Roberts, Malcolm J. and O'Rourke, Eleanor and Dingley, Elisabeth and Turner, Briony and Hewitt, Helene and Dunne, John P.},
  title   = {Rapid Evaluation Framework for the CMIP7 Assessment Fast Track},
  journal = {EGUsphere [preprint]},
  year    = {2025},
  note    = {Preprint, under review for Geoscientific Model Development},
  doi     = {10.5194/egusphere-2025-2685}
}

@article{reichlerHowWellDo2008,
  author  = {Reichler, Thomas and Kim, Junsu},
  title   = {How Well Do Coupled Models Simulate Today's Climate?},
  journal = {Bulletin of the American Meteorological Society},
  volume  = {89},
  number  = {3},
  pages   = {303--312},
  year    = {2008},
  doi     = {10.1175/BAMS-89-3-303}
}

@article{wenzelProjectedLandPhotosynthesis2016,
  author  = {Wenzel, Sabrina and Cox, Peter M. and Eyring, Veronika and Friedlingstein, Pierre},
  title   = {Projected land photosynthesis constrained by changes in the seasonal cycle of atmospheric CO$_2$},
  journal = {Nature},
  volume  = {538},
  number  = {7626},
  pages   = {499--501},
  year    = {2016},
  doi     = {10.1038/nature19772}
}

@article{wenzelEmergentConstraintsClimate2014,
  author  = {Wenzel, Sabrina and Cox, Peter M. and Eyring, Veronika and Friedlingstein, Pierre},
  title   = {Emergent constraints on climate-carbon cycle feedbacks in the CMIP5 Earth system models},
  journal = {Journal of Geophysical Research: Biogeosciences},
  volume  = {119},
  number  = {5},
  pages   = {794--807},
  year    = {2014},
  doi     = {10.1002/2013JG002591}
}

@article{righiEarthSystemModel2020,
  author  = {Righi, Mattia and Andela, Bouwe and Eyring, Veronika and Lauer, Axel and Predoi, Valeriu and Schlund, Manuel and Vegas-Regidor, Javier and Bock, Lisa and Br{\"o}tz, Bj{\"o}rn and de Mora, Lee and Diblen, Faruk and Dreyer, Laura and Drost, Niels and Earnshaw, Paul and Hassler, Birgit and Koldunov, Nikolay and Little, Bill and Loosveldt Tomas, Saskia and Zimmermann, Klaus},
  title   = {Earth System Model Evaluation Tool (ESMValTool) v2.0 -- technical overview},
  journal = {Geoscientific Model Development},
  volume  = {13},
  number  = {3},
  pages   = {1179--1199},
  year    = {2020},
  doi     = {10.5194/gmd-13-1179-2020}
}

@article{kollEarthsOutgoingLongwave2018,
  author  = {Koll, Daniel D. B. and Cronin, Timothy W.},
  title   = {Earth's outgoing longwave radiation linear due to H$_2$O greenhouse effect},
  journal = {Proceedings of the National Academy of Sciences},
  volume  = {115},
  number  = {41},
  pages   = {10293--10298},
  year    = {2018},
  doi     = {10.1073/pnas.1809868115}
}

@article{stephensUpdateEarthsEnergy2012,
  author  = {Stephens, Graeme L. and Li, Juilin and Wild, Martin and Clayson, Carol Anne and Loeb, Norman and Kato, Seiji and L'Ecuyer, Tristan and Stackhouse, Paul W. and Lebsock, Matthew and Andrews, Timothy},
  title   = {An update on Earth's energy balance in light of the latest global observations},
  journal = {Nature Geoscience},
  volume  = {5},
  number  = {10},
  pages   = {691--696},
  year    = {2012},
  doi     = {10.1038/ngeo1580}
}

@article{waliserMJOSimulationDiagnostics2009,
  author  = {Waliser, D. and Sperber, K. and Hendon, H. and Kim, D. and Maloney, E. and Wheeler, M. and Weickmann, K. and Zhang, C. and Donner, L. and Gottschalck, J. and Higgins, W. and Kang, I.-S. and Legler, D. and Moncrieff, M. and Vitart, F. and Wang, B. and Wang, W. and Woolnough, S. and Maloney, E. and Schubert, S. and Stern, W.},
  title   = {MJO Simulation Diagnostics},
  journal = {Journal of Climate},
  volume  = {22},
  number  = {11},
  pages   = {3006--3030},
  year    = {2009},
  doi     = {10.1175/2008JCLI2731.1}
}

@article{andrewsDependenceGlobalCloud2018,
  author  = {Andrews, Timothy and Webb, Mark J.},
  title   = {The Dependence of Global Cloud and Lapse Rate Feedbacks on the Spatial Structure of Tropical Pacific Warming},
  journal = {Journal of Climate},
  volume  = {31},
  number  = {2},
  pages   = {641--654},
  year    = {2018},
  doi     = {10.1175/JCLI-D-17-0087.1}
}

@article{byrneLandOceanWarming2013,
  author  = {Byrne, Michael P. and O'Gorman, Paul A.},
  title   = {Land--Ocean Warming Contrast over a Wide Range of Climates: Convective Quasi-Equilibrium Theory and Idealized Simulations},
  journal = {Journal of Climate},
  volume  = {26},
  number  = {12},
  pages   = {4000--4016},
  year    = {2013},
  doi     = {10.1175/JCLI-D-12-00262.1}
}

@article{croninAirSeaFluxes2019,
  author  = {Cronin, Meghan F. and Gentemann, Chelle L. and Edson, James and Ueki, Iwao and Bourassa, Mark and Brown, Shannon and Clayson, Carol Anne and Fairall, Chris W. and Farrar, J. Thomas and Gille, Sarah T. and Gulev, Sergey and Josey, Simon A. and Kato, Seiji and Katsumata, Masaki and Kent, Elizabeth and Krug, Marjolaine and Minnett, Peter J. and Parfitt, Rhys and Pinker, Rachel T. and Stackhouse, Paul W. and Swart, Sebastiaan and Tomita, Hiroyuki and Vandemark, Douglas and Weller, A. Robert and Yoneyama, Kunio and Yu, Lisan and Zhang, Dongxiao},
  title   = {Air-Sea Fluxes With a Focus on Heat and Momentum},
  journal = {Frontiers in Marine Science},
  volume  = {6},
  pages   = {430},
  year    = {2019},
  doi     = {10.3389/fmars.2019.00430}
}

@article{deserHowWellWe2017,
  author  = {Deser, Clara and Simpson, Isla R. and McKinnon, Karen A. and Phillips, Adam S.},
  title   = {The Northern Hemisphere Extratropical Atmospheric Circulation Response to ENSO: How Well Do We Know It and How Do We Evaluate Models Accordingly?},
  journal = {Journal of Climate},
  volume  = {30},
  number  = {13},
  pages   = {5059--5082},
  year    = {2017},
  doi     = {10.1175/JCLI-D-16-0844.1}
}

@article{deserInsightsEarthSystem2020,
  author  = {Deser, C. and Lehner, F. and Rodgers, K. B. and Ault, T. and Delworth, T. L. and DiNezio, P. N. and Fiore, A. and Frankignoul, C. and Fyfe, J. C. and Horton, D. E. and Kay, J. E. and Knutti, R. and Lovenduski, N. S. and Marotzke, J. and McKinnon, K. A. and Minobe, S. and Randerson, J. and Screen, J. A. and Simpson, I. R. and Ting, M.},
  title   = {Insights from Earth system model initial-condition large ensembles and future prospects},
  journal = {Nature Climate Change},
  volume  = {10},
  number  = {4},
  pages   = {277--286},
  year    = {2020},
  doi     = {10.1038/s41558-020-0731-2}
}

@article{harrisonMidHoloceneSaharan2015,
  author  = {Harrison, S. P. and Bartlein, P. J. and Izumi, K. and Li, G. and Annan, J. and Hargreaves, J. and Braconnot, P. and Kageyama, M.},
  title   = {Evaluation of CMIP5 palaeo-simulations to improve climate projections},
  journal = {Nature Climate Change},
  volume  = {5},
  number  = {8},
  pages   = {735--743},
  year    = {2015},
  doi     = {10.1038/nclimate2649}
}

@article{knuttiBeyondEquilibriumClimate2017,
  author  = {Knutti, Reto and Rugenstein, Maria A. A. and Hegerl, Gabriele C.},
  title   = {Beyond equilibrium climate sensitivity},
  journal = {Nature Geoscience},
  volume  = {10},
  number  = {10},
  pages   = {727--736},
  year    = {2017},
  doi     = {10.1038/ngeo3017}
}

@article{maherLargeEnsembleClimate2021,
  author  = {Maher, Nicola and Milinski, Sebastian and Ludwig, Ralf},
  title   = {Large ensemble climate model simulations: introduction, overview, and future prospects for utilising multiple types of large ensemble},
  journal = {Earth System Dynamics},
  volume  = {12},
  number  = {2},
  pages   = {401--418},
  year    = {2021},
  doi     = {10.5194/esd-12-401-2021}
}

@article{northTheoryEnergyBalance1975,
  author  = {North, Gerald R.},
  title   = {Theory of Energy-Balance Climate Models},
  journal = {Journal of the Atmospheric Sciences},
  volume  = {32},
  number  = {11},
  pages   = {2033--2043},
  year    = {1975},
  doi     = {10.1175/1520-0469(1975)032<2033:TOEBCM>2.0.CO;2}
}

@article{yuGlobalAirSea2019,
  author  = {Yu, Lisan},
  title   = {Global Air--Sea Fluxes of Heat, Fresh Water, and Momentum: Energy Budget Closure and Unanswered Questions},
  journal = {Annual Review of Marine Science},
  volume  = {11},
  number  = {1},
  pages   = {227--248},
  year    = {2019},
  doi     = {10.1146/annurev-marine-010816-060704}
}

@article{fischerPalaeoclimateConstraintsImpact2018,
  author  = {Fischer, Hubertus and Meissner, Katrin J. and Mix, Alan C. and Abram, Nerilie J. and Austermann, Jacqueline and Brovkin, Victor and Capron, Emilie and Colombaroli, Daniele and Daniau, Anne-Laure and Dyez, Kelsey A. and Felis, Thomas and Finkelstein, Sarah A. and Jaccard, Samuel L. and McClymont, Erin L. and Rovere, Alessio and Sutter, Johannes and Wolff, Eric W. and Affolter, St{\'e}phane and Bakker, Pepijn and Ballesteros-C{\'a}novas, Juan Antonio and Barbante, Carlo and Caley, Thibaut and Carlson, Anders E. and Churakova, Olga and Cortese, Giuseppe and Cumming, Brian F. and Davis, Basil A. S. and de Vernal, Anne and Emile-Geay, Julien and Fritz, Sherilyn C. and Gierz, Paul and Gottschalk, Julia and Holloway, Max D. and Joos, Fortunat and Kucera, Michal and Loutre, Marie-France and Lunt, Daniel J. and Marcisz, Katarzyna and Marlon, Jennifer R. and Martinez, Philippe and Masson-Delmotte, Valerie and Nehrbass-Ahles, Christoph and Otto-Bliesner, Bette L. and Raible, Christoph C. and Risebrobakken, Bj{\o}rg and S{\'a}nchez Go{\~n}i, Mar{\'i}a F. and Saleem Arrigo, Jennifer and Sarnthein, Michael and Sjolte, Jesper and Stocker, Thomas F. and Velasquez Alvarez, Patricio A. and Tinner, Willy and Valdes, Paul J. and Vogel, Hendrik and Wanner, Heinz and Yan, Qing and Yu, Zicheng and Ziegler, Martin and Zhou, Liping},
  title   = {Palaeoclimate constraints on the impact of 2\,{\textdegree}C anthropogenic warming and beyond},
  journal = {Nature Geoscience},
  volume  = {11},
  number  = {7},
  pages   = {474--485},
  year    = {2018},
  doi     = {10.1038/s41561-018-0146-0}
}

@article{AIMIP,
  author  = {Henn, Brian and Bretherton, Christopher S. and Koldunov, Nikolay and Lessig, Christian and Molina, Maria J. and Arcomano, Troy and Watt-Meyer, Oliver and Couairon, Guillaume and Singh, Renu and Brunstein, Robert and Hasson, Yana and Jost, Antonia and Brenowitz, Noah and Manshausen, Peter and Cresswell-Clay, Nathaniel and Durran, Dale and Hall, Kyle Joseph Chen and Yuval, Janni and Kochkov, Dmitrii and Hoyer, Stephan and Lopez-Gomez, Ignacio},
  title   = {AIMIP Phase 1: systematic evaluations of AI weather and climate models},
  journal = {EGUsphere [preprint]},
  year    = {2026},
  note    = {Preprint, under review for Geoscientific Model Development},
  doi     = {10.5194/egusphere-2026-2709}
}

@article{hakimDynamicalTestsDeep2024,
  author  = {Hakim, Gregory J. and Masanam, Sanjit},
  title   = {Dynamical Tests of a Deep Learning Weather Prediction Model},
  journal = {Artificial Intelligence for the Earth Systems},
  volume  = {3},
  number  = {3},
  year    = {2024},
  doi     = {10.1175/AIES-D-23-0090.1}
}

@article{duffyHighECSCESM22026,
  author  = {Duffy, Margaret L. and Simpson, Isla R. and Medeiros, Brian and Zhu, Jiang and McCluskey, Christina S. and Herrington, Adam R. and Gettelman, Andrew and Otto-Bliesner, Bette L. and Fasullo, John T. and Lauritzen, Peter H. and Neale, Richard B. and Lawrence, David M.},
  title   = {Is the High ECS in CESM2 Degrading Transient Climate Change Projections Over the 21st Century?},
  journal = {Journal of Advances in Modeling Earth Systems},
  volume  = {18},
  number  = {1},
  pages   = {e2025MS004967},
  year    = {2026},
  doi     = {10.1029/2025MS004967}
}

@article{ICONEval,
  author  = {Lauer, Axel and Schlund, Manuel and Bock, Lisa and Hassler, Birgit and Behrens, Gunnar and Gier, Bettina and Lindenlaub, Lukas and Lorenz, Stephan and Malles, Jan-Hendrik and M{\"u}ller, Wolfgang A. and van Pham, Trang and Weigel, Katja and Zeng, Guang and Eyring, Veronika},
  title   = {An ESMValTool-based framework for sanity checks, physical consistency and climate fidelity during model development -- ICONEval v1.0},
  journal = {EGUsphere [preprint]},
  year    = {2026},
  note    = {Preprint, under review for Geoscientific Model Development},
  doi     = {10.5194/egusphere-2026-2288}
}

@article{raspWeatherBench2Benchmark2024,
  title   = {{WeatherBench 2}: A Benchmark for the Next Generation of Data-Driven Global Weather Models},
  author  = {Rasp, Stephan and Hoyer, Stephan and Merose, Alexander and Langmore, Ian and Battaglia, Peter and Russell, Tyler and Sanchez-Gonzalez, Alvaro and Yang, Vivian and Carver, Rob and Agrawal, Shreya and Chantry, Matthew and Bouallegue, Zied Ben and Dueben, Peter and Bromberg, Carla and Sisk, Jared and Barrington, Luke and Bell, Aaron and Sha, Fei},
  journal = {Journal of Advances in Modeling Earth Systems},
  volume  = {16},
  number  = {6},
  pages   = {e2023MS004019},
  year    = {2024},
  doi     = {10.1029/2023MS004019}
}

@article{watson-parrisIntegratingTopDown2025,
  title   = {Integrating Top-Down Energetic Constraints With Bottom-Up Process-Based Constraints for More Accurate Projections of Future Warming},
  author  = {{Watson-Parris}, Duncan},
  year    = {2025},
  journal = {Geophysical Research Letters},
  volume  = {52},
  pages   = {e2024GL114269},
  doi     = {10.1029/2024GL114269}
}

@article{selandOverviewNorESM22020,
  title   = {Overview of the Norwegian Earth System Model ({NorESM2}) and key climate response of {CMIP6} {DECK}, historical, and scenario simulations},
  author  = {Seland, {\O}yvind and Bentsen, Mats and Olivi{\'e}, Dirk and Toniazzo, Thomas and Gjermundsen, Ada and Graff, Lise Seland and Debernard, Jens Boldingh and Gupta, Alok Kumar and He, Yan-Chun and Kirkev{\aa}g, Alf and Schwinger, J{\"o}rg and Tjiputra, Jerry and Aas, Kjetil Schanke and Bethke, Ingo and Fan, Yuanchao and Griesfeller, Jan and Grini, Alf and Guo, Chuncheng and Ilicak, Mehmet and Karset, Inger Helene Hafsahl and Landgren, Oskar and Liakka, Johan and Moseid, Kine Onsum and Nummelin, Aleksi and Spensberger, Clemens and Tang, Hui and Zhang, Zhongshi and Heinze, Christoph and Iversen, Trond and Schulz, Michael},
  year    = {2020},
  journal = {Geoscientific Model Development},
  volume  = {13},
  number  = {12},
  pages   = {6165--6200},
  doi     = {10.5194/gmd-13-6165-2020}
}

@article{hausfatherEvaluatingPerformancePast2020,
  title   = {Evaluating the Performance of Past Climate Model Projections},
  author  = {Hausfather, Zeke and Drake, Henri F. and Abbott, Tristan and Schmidt, Gavin A.},
  year    = {2020},
  journal = {Geophysical Research Letters},
  volume  = {47},
  number  = {1},
  pages   = {e2019GL085378},
  doi     = {10.1029/2019GL085378}
}

@misc{brenowitzClimateBottleGenerative2025,
  title         = {Climate in a Bottle: Towards a Generative Foundation Model for the Kilometer-Scale Global Atmosphere},
  author        = {Brenowitz, Noah D. and Ge, Tao and Subramaniam, Akshay and Manshausen, Peter and Gupta, Aayush and Hall, David M. and Mardani, Morteza and Vahdat, Arash and Kashinath, Karthik and Pritchard, Michael S.},
  year          = {2025},
  eprint        = {2505.06474},
  archivePrefix = {arXiv},
  howpublished  = {arXiv:2505.06474}
}

@article{zhangIndicesMonitoringChanges2011,
  title   = {Indices for Monitoring Changes in Extremes Based on Daily Temperature and Precipitation Data},
  author  = {Zhang, Xuebin and Alexander, Lisa and Hegerl, Gabriele C. and Jones, Philip and Tank, Albert Klein and Peterson, Thomas C. and Trewin, Blair and Zwiers, Francis W.},
  year    = {2011},
  journal = {WIREs Climate Change},
  volume  = {2},
  number  = {6},
  pages   = {851--870},
  doi     = {10.1002/wcc.147}
}

@article{perkinsEvaluationAR4Climate2007,
  title   = {Evaluation of the {AR4} Climate Models' Simulated Daily Maximum Temperature, Minimum Temperature, and Precipitation over {Australia} Using Probability Density Functions},
  author  = {Perkins, Sarah E. and Pitman, Andy J. and Holbrook, Neil J. and McAneney, John},
  year    = {2007},
  journal = {Journal of Climate},
  volume  = {20},
  number  = {17},
  pages   = {4356--4376},
  doi     = {10.1175/JCLI4253.1}
}

@misc{luntIPCCAR6Fig719Data2023,
  title        = {Global Mean Temperature Anomaly in Models and Observations from Five Time Periods: Data Underlying {IPCC AR6 WGI} Figure 7.19},
  author       = {Lunt, Daniel J.},
  year         = {2023},
  publisher    = {NERC EDS Centre for Environmental Data Analysis},
  doi          = {10.5285/9ce84c3a242e4b999c24dc1647c89794}
}

@article{ferroFairScoresEnsemble2014,
  title = {Fair Scores for Ensemble Forecasts},
  author = {Ferro, C. A. T.},
  year = {2014},
  journal = {Quarterly Journal of the Royal Meteorological Society},
  volume = {140},
  number = {683},
  pages = {1917--1923},
  doi = {10.1002/qj.2270}
}

@article{lenssenImprovementsGISTEMP2019,
  author  = {Lenssen, Nathan J. L. and Schmidt, Gavin A. and Hansen, James E. and Menne, Matthew J. and Persin, Avraham and Ruedy, Reto and Zyss, Daniel},
  title   = {Improvements in the {GISTEMP} Uncertainty Model},
  journal = {Journal of Geophysical Research: Atmospheres},
  volume  = {124},
  number  = {12},
  pages   = {6307--6326},
  year    = {2019},
  doi     = {10.1029/2018JD029522}
}

@article{rohdeBerkeleyEarthTemperature2020,
  author  = {Rohde, Robert A. and Hausfather, Zeke},
  title   = {The {Berkeley Earth} Land/Ocean Temperature Record},
  journal = {Earth System Science Data},
  volume  = {12},
  number  = {4},
  pages   = {3469--3479},
  year    = {2020},
  doi     = {10.5194/essd-12-3469-2020}
}

@article{moriceUpdatedAssessmentNear2021,
  author  = {Morice, Colin P. and Kennedy, John J. and Rayner, Nick A. and Winn, J. P. and Hogan, Emma and Killick, R. E. and Dunn, R. J. H. and Osborn, Tim J. and Jones, Phil D. and Simpson, I. R.},
  title   = {An Updated Assessment of Near-Surface Temperature Change From 1850: The {HadCRUT5} Data Set},
  journal = {Journal of Geophysical Research: Atmospheres},
  volume  = {126},
  number  = {3},
  pages   = {e2019JD032361},
  year    = {2021},
  doi     = {10.1029/2019JD032361}
}

@article{huangImprovementsNOAAGlobalTemp2020,
  author  = {Huang, Boyin and Menne, Matthew J. and Boyer, Tim and Freeman, Eric and Gleason, Byron E. and Lawrimore, Jay H. and Liu, Chunying and Rennie, J. Jared and Schreck, Carl J. and Sun, Fengying and Vose, Russell and Williams, Claude N. and Yin, Xungang and Zhang, Huai-Min},
  title   = {Uncertainty Estimates for Sea Surface Temperature and Land Surface Air Temperature in {NOAAGlobalTemp} Version 5},
  journal = {Journal of Climate},
  volume  = {33},
  number  = {4},
  pages   = {1351--1379},
  year    = {2020},
  doi     = {10.1175/JCLI-D-19-0395.1}
}

@article{caesarLargescaleChangesObserved2006,
  author  = {Caesar, J. and Alexander, L. and Vose, R.},
  title   = {Large-scale changes in observed daily maximum and minimum temperatures: Creation and analysis of a new gridded data set},
  journal = {Journal of Geophysical Research: Atmospheres},
  volume  = {111},
  number  = {D5},
  pages   = {D05101},
  year    = {2006},
  doi     = {10.1029/2005JD006280}
}

@article{raynerGlobalAnalysesSea2003,
  author  = {Rayner, N. A. and Parker, D. E. and Horton, E. B. and Folland, C. K. and Alexander, L. V. and Rowell, D. P. and Kent, E. C. and Kaplan, A.},
  title   = {Global analyses of sea surface temperature, sea ice, and night marine air temperature since the late nineteenth century},
  journal = {Journal of Geophysical Research: Atmospheres},
  volume  = {108},
  number  = {D14},
  pages   = {4407},
  year    = {2003},
  doi     = {10.1029/2002JD002670}
}

@article{harrisVersion4CRU2020,
  author  = {Harris, Ian and Osborn, Timothy J. and Jones, Phil and Lister, David},
  title   = {Version 4 of the {CRU TS} monthly high-resolution gridded multivariate climate dataset},
  journal = {Scientific Data},
  volume  = {7},
  number  = {1},
  pages   = {109},
  year    = {2020},
  doi     = {10.1038/s41597-020-0453-3}
}

@article{adlerGlobalPrecipitationClimatology2018,
  author  = {Adler, Robert F. and Sapiano, Mathew R. P. and Huffman, George J. and Wang, Jian-Jian and Gu, Guojun and Bolvin, David and Chiu, Long and Schneider, Udo and Becker, Andreas and Nelkin, Eric and Xie, Pingping and Ferraro, Ralph and Shin, Dong-Bin},
  title   = {The {Global Precipitation Climatology Project} ({GPCP}) Monthly Analysis (New Version 2.3) and a Review of 2017 Global Precipitation},
  journal = {Atmosphere},
  volume  = {9},
  number  = {4},
  pages   = {138},
  year    = {2018},
  doi     = {10.3390/atmos9040138}
}

@incollection{huffmanIntegratedMultisatelliteRetrievals2020,
  author    = {Huffman, George J. and Bolvin, David T. and Braithwaite, Dan and Hsu, Kuo-Lin and Joyce, Robert J. and Kidd, Christopher and Nelkin, Eric J. and Sorooshian, Soroosh and Stocker, Erich F. and Tan, Jackson and Wolff, David B. and Xie, Pingping},
  title     = {Integrated Multi-satellite Retrievals for the {Global Precipitation Measurement} ({GPM}) Mission ({IMERG})},
  booktitle = {Satellite Precipitation Measurement},
  series    = {Advances in Global Change Research},
  volume    = {67},
  editor    = {Levizzani, V. and Kidd, C. and Kirschbaum, D. B. and Kummerow, C. D. and Nakamura, K. and Turk, F. J.},
  publisher = {Springer},
  address   = {Cham},
  pages     = {343--353},
  year      = {2020},
  doi       = {10.1007/978-3-030-24568-9_19}
}

@article{beckMSWEPV2Global2019,
  author  = {Beck, Hylke E. and Wood, Eric F. and Pan, Ming and Fisher, Colby K. and Miralles, Diego G. and van Dijk, Albert I. J. M. and McVicar, Tim R. and Adler, Robert F.},
  title   = {{MSWEP} V2 Global 3-Hourly 0.1$^{\circ}$ Precipitation: Methodology and Quantitative Assessment},
  journal = {Bulletin of the American Meteorological Society},
  volume  = {100},
  number  = {3},
  pages   = {473--500},
  year    = {2019},
  doi     = {10.1175/BAMS-D-17-0138.1}
}

@article{loebCloudsEarthRadiant2018,
  author  = {Loeb, Norman G. and Doelling, David R. and Wang, Hailan and Su, Wenying and Nguyen, Cathy and Corbett, Joseph G. and Liang, Lusheng and Mitrescu, Cristian and Rose, Fred G. and Kato, Seiji},
  title   = {Clouds and the {Earth's Radiant Energy System} ({CERES}) {Energy Balanced and Filled} ({EBAF}) Top-of-Atmosphere ({TOA}) Edition-4.0 Data Product},
  journal = {Journal of Climate},
  volume  = {31},
  number  = {2},
  pages   = {895--918},
  year    = {2018},
  doi     = {10.1175/JCLI-D-17-0208.1}
}

@article{rutanCERESSynopticProduct2015,
  author  = {Rutan, David A. and Kato, Seiji and Doelling, David R. and Rose, Fred G. and Nguyen, Le Trang and Caldwell, Thomas E. and Loeb, Norman G.},
  title   = {{CERES} Synoptic Product: Methodology and Validation of Surface Radiant Flux},
  journal = {Journal of Atmospheric and Oceanic Technology},
  volume  = {32},
  number  = {6},
  pages   = {1121--1143},
  year    = {2015},
  doi     = {10.1175/JTECH-D-14-00165.1}
}

@article{pastorelloFLUXNET2015DatasetONEFlux2020,
  author  = {Pastorello, Gilberto and Trotta, Carlo and Canfora, Eleonora and Chu, Housen and Christianson, Danielle and Cheah, You-Wei and others},
  title   = {The {FLUXNET2015} dataset and the {ONEFlux} processing pipeline for eddy covariance data},
  journal = {Scientific Data},
  volume  = {7},
  number  = {1},
  pages   = {225},
  year    = {2020},
  doi     = {10.1038/s41597-020-0534-3}
}

@inproceedings{sendOceanSITES2010,
  author    = {Send, Uwe and Weller, Robert A. and Wallace, Douglas and Chavez, Francisco and Lampitt, Richard and Dickey, Tommy and Honda, Makio and Nittis, Kostas and Lukas, Roger and McPhaden, Michael and Feely, Richard},
  title     = {{OceanSITES}},
  booktitle = {Proceedings of OceanObs'09: Sustained Ocean Observations and Information for Society},
  editor    = {Hall, J. and Harrison, D. E. and Stammer, D.},
  volume    = {2},
  publisher = {European Space Agency},
  year      = {2010},
  note      = {ESA Publication WPP-306; conference held in Venice, Italy, 21--25 September 2009},
  doi       = {10.5270/OceanObs09.cwp.79}
}

@misc{fettererSeaIceIndex2017,
  author       = {Fetterer, F. and Knowles, K. and Meier, W. N. and Savoie, M. and Windnagel, A. K.},
  title        = {Sea Ice Index, Version 3},
  howpublished = {NSIDC: National Snow and Ice Data Center, Boulder, Colorado, USA},
  year         = {2017},
  note         = {Data set G02135},
  doi          = {10.7265/N5K072F8}
}

@article{lavergneVersion2EUMETSAT2019,
  author  = {Lavergne, Thomas and S{\o}rensen, Atle Macdonald and Kern, Stefan and Tonboe, Rasmus and Notz, Dirk and Aaboe, Signe and Bell, Louisa and Dybkj{\ae}r, Gorm and Eastwood, Steinar and Gabarro, Carolina and Heygster, Georg and Killie, Mari Anne and Brandt Kreiner, Matilde and Lavelle, John and Saldo, Roberto and Sandven, Stein and Pedersen, Leif Toudal},
  title   = {Version 2 of the {EUMETSAT OSI SAF} and {ESA CCI} sea-ice concentration climate data records},
  journal = {The Cryosphere},
  volume  = {13},
  number  = {1},
  pages   = {49--78},
  year    = {2019},
  doi     = {10.5194/tc-13-49-2019}
}

@article{chengImprovedEstimatesOcean2017,
  author  = {Cheng, Lijing and Trenberth, Kevin E. and Fasullo, John and Boyer, Tim and Abraham, John and Zhu, Jiang},
  title   = {Improved estimates of ocean heat content from 1960 to 2015},
  journal = {Science Advances},
  volume  = {3},
  number  = {3},
  pages   = {e1601545},
  year    = {2017},
  doi     = {10.1126/sciadv.1601545}
}

@article{levitusWorldOceanHeat2012,
  author  = {Levitus, S. and Antonov, J. I. and Boyer, T. P. and Baranova, O. K. and Garcia, H. E. and Locarnini, R. A. and Mishonov, A. V. and Reagan, J. R. and Seidov, D. and Yarosh, E. S. and Zweng, M. M.},
  title   = {World ocean heat content and thermosteric sea level change (0--2000 m), 1955--2010},
  journal = {Geophysical Research Letters},
  volume  = {39},
  number  = {10},
  pages   = {L10603},
  year    = {2012},
  doi     = {10.1029/2012GL051106}
}

@article{platnickMODISCloudOptical2017,
  author  = {Platnick, Steven and Meyer, Kerry G. and King, Michael D. and Wind, Galina and Amarasinghe, Nandana and Marchant, Benjamin and Arnold, G. Thomas and Zhang, Zhibo and Hubanks, Paul A. and Holz, Robert E. and Yang, Ping and Ridgway, William L. and Riedi, Jerome},
  title   = {The {MODIS} Cloud Optical and Microphysical Products: Collection 6 Updates and Examples From {Terra} and {Aqua}},
  journal = {IEEE Transactions on Geoscience and Remote Sensing},
  volume  = {55},
  number  = {1},
  pages   = {502--525},
  year    = {2017},
  doi     = {10.1109/TGRS.2016.2610522}
}

@article{youngInternationalSatelliteCloud2018,
  author  = {Young, Alisa H. and Knapp, Kenneth R. and Inamdar, Anand and Hankins, William and Rossow, William B.},
  title   = {The {International Satellite Cloud Climatology Project} {H}-Series climate data record product},
  journal = {Earth System Science Data},
  volume  = {10},
  number  = {1},
  pages   = {583--593},
  year    = {2018},
  doi     = {10.5194/essd-10-583-2018}
}

@article{stephensCloudSatMissionATrain2002,
  author  = {Stephens, Graeme L. and Vane, Deborah G. and Boain, Ronald J. and Mace, Gerald G. and Sassen, Kenneth and Wang, Zhien and Illingworth, Anthony J. and O'Connor, Ewan J. and Rossow, William B. and Durden, Stephen L. and Miller, Steven D. and Austin, Richard T. and Benedetti, Angela and Mitrescu, Cristian and {{The CloudSat Science Team}}},
  title   = {The {CloudSat} Mission and the {A-Train}: A New Dimension of Space-Based Observations of Clouds and Precipitation},
  journal = {Bulletin of the American Meteorological Society},
  volume  = {83},
  number  = {12},
  pages   = {1771--1790},
  year    = {2002},
  doi     = {10.1175/BAMS-83-12-1771}
}

@article{elsaesserMultisensorAdvancedClimatology2017,
  author  = {Elsaesser, Gregory S. and O'Dell, Christopher W. and Lebsock, Matthew D. and Bennartz, Ralf and Greenwald, Thomas J. and Wentz, Frank J.},
  title   = {The Multisensor Advanced Climatology of Liquid Water Path ({MAC-LWP})},
  journal = {Journal of Climate},
  volume  = {30},
  number  = {24},
  pages   = {10193--10210},
  year    = {2017},
  doi     = {10.1175/JCLI-D-16-0902.1}
}

@article{mearsConstructionUncertaintyEstimation2018,
  author  = {Mears, Carl A. and Smith, Deborah K. and Ricciardulli, Lucrezia and Wang, Junhong and Huelsing, Hannah and Wentz, Frank J.},
  title   = {Construction and Uncertainty Estimation of a Satellite-Derived Total Precipitable Water Data Record Over the World's Oceans},
  journal = {Earth and Space Science},
  volume  = {5},
  number  = {5},
  pages   = {197--210},
  year    = {2018},
  doi     = {10.1002/2018EA000363}
}

@article{durreEnhancedComprehensiveExtraction2018,
  author  = {Durre, Imke and Yin, Xungang and Vose, Russell S. and Applequist, Scott and Arnfield, Jeff},
  title   = {Enhancing the Data Coverage in the {Integrated Global Radiosonde Archive}},
  journal = {Journal of Atmospheric and Oceanic Technology},
  volume  = {35},
  number  = {9},
  pages   = {1753--1770},
  year    = {2018},
  doi     = {10.1175/JTECH-D-17-0223.1}
}

@article{hersbachERA5GlobalReanalysis2020,
  author  = {Hersbach, Hans and Bell, Bill and Berrisford, Paul and Hirahara, Shoji and Hor{\'a}nyi, Andr{\'a}s and Mu{\~n}oz-Sabater, Joaqu{\'\i}n and Nicolas, Julien and Peubey, Carole and Radu, Raluca and Schepers, Dinand and others},
  title   = {The {ERA5} global reanalysis},
  journal = {Quarterly Journal of the Royal Meteorological Society},
  volume  = {146},
  number  = {730},
  pages   = {1999--2049},
  year    = {2020},
  doi     = {10.1002/qj.3803}
}

@article{hobbsEnergyConservationAnalysis2016,
  author  = {Hobbs, Will and Palmer, Matthew D. and Monselesan, Didier},
  title   = {An Energy Conservation Analysis of Ocean Drift in the {CMIP5} Global Coupled Models},
  journal = {Journal of Climate},
  volume  = {29},
  number  = {5},
  pages   = {1639--1653},
  year    = {2016},
  doi     = {10.1175/JCLI-D-15-0477.1}
}

@article{irvingMassEnergyConservation2021,
  author  = {Irving, Damien and Hobbs, Will and Church, John and Zika, Jan},
  title   = {A Mass and Energy Conservation Analysis of Drift in the {CMIP6} Ensemble},
  journal = {Journal of Climate},
  volume  = {34},
  number  = {8},
  pages   = {3157--3170},
  year    = {2021},
  doi     = {10.1175/JCLI-D-20-0281.1}
}

@article{loebSatelliteOceanData2021,
  author  = {Loeb, Norman G. and Johnson, Gregory C. and Thorsen, Tyler J. and Lyman, John M. and Rose, Fred G. and Kato, Seiji},
  title   = {Satellite and Ocean Data Reveal Marked Increase in {Earth's} Heating Rate},
  journal = {Geophysical Research Letters},
  volume  = {48},
  number  = {13},
  pages   = {e2021GL093047},
  year    = {2021},
  doi     = {10.1029/2021GL093047}
}

@article{zhangLinearityOutgoingLongwave2020,
  author  = {Zhang, Yi and Jeevanjee, Nadir and Fueglistaler, Stephan},
  title   = {Linearity of Outgoing Longwave Radiation: From an Atmospheric Column to Global Climate Models},
  journal = {Geophysical Research Letters},
  volume  = {47},
  number  = {17},
  pages   = {e2020GL089235},
  year    = {2020},
  doi     = {10.1029/2020GL089235}
}

@article{donohoeRelationshipITCZLocation2013,
  author  = {Donohoe, Aaron and Marshall, John and Ferreira, David and McGee, David},
  title   = {The Relationship between {ITCZ} Location and Cross-Equatorial Atmospheric Heat Transport: From the Seasonal Cycle to the {Last Glacial Maximum}},
  journal = {Journal of Climate},
  volume  = {26},
  number  = {11},
  pages   = {3597--3618},
  year    = {2013},
  doi     = {10.1175/JCLI-D-12-00467.1}
}

@misc{huffmanGPMIMERGFinal2023,
  author       = {Huffman, George J. and Stocker, Erich F. and Bolvin, David T. and Nelkin, Eric J. and Tan, Jackson},
  title        = {{GPM IMERG} Final Precipitation {L3} 1 day 0.1 degree x 0.1 degree {V07}},
  howpublished = {Goddard Earth Sciences Data and Information Services Center (GES DISC), Greenbelt, MD},
  year         = {2023},
  doi          = {10.5067/GPM/IMERGDF/DAY/07}
}

@misc{platnickMODISAtmosphereL32017,
  author       = {Platnick, Steven and King, Michael D. and Hubanks, Paul A.},
  title        = {{MODIS} Atmosphere {L3} Monthly Product ({MYD08\_M3}, Collection 6.1)},
  howpublished = {NASA MODIS Adaptive Processing System, Goddard Space Flight Center},
  year         = {2017},
  doi          = {10.5067/MODIS/MYD08_M3.061}
}

@article{durackStandardizedDataSets2018,
  title = {Toward Standardized Data Sets for Climate Model Experimentation},
  author = {Durack, P. J. and Taylor, K. E. and Eyring, V. and Ames, S. K. and Hoang, T. and Nadeau, D. and Doutriaux, C. and Stockhause, M. and Gleckler, P. J.},
  year = 2018,
  journal = {Eos},
  volume = {99},
  doi = {10.1029/2018EO101751}
}

@article{rossBenchmarkingMachineLearning2023,
  author = {Ross, Andrew and Li, Ziwei and Perezhogin, Pavel and Fernandez-Granda, Carlos and Zanna, Laure},
  title = {Benchmarking of Machine Learning Ocean Subgrid Parameterizations in an Idealized Model},
  journal = {Journal of Advances in Modeling Earth Systems},
  volume = {15}, number = {1}, pages = {e2022MS003258}, year = {2023},
  doi = {10.1029/2022MS003258}
}

@article{zannaDataDrivenEquationDiscovery2020,
  author = {Zanna, Laure and Bolton, Thomas},
  title = {Data-Driven Equation Discovery of Ocean Mesoscale Closures},
  journal = {Geophysical Research Letters},
  volume = {47}, number = {17}, pages = {e2020GL088376}, year = {2020},
  doi = {10.1029/2020GL088376}
}

@article{guillauminStochasticDeepLearning2021,
  author = {Guillaumin, Arthur P. and Zanna, Laure},
  title = {Stochastic-Deep Learning Parameterization of Ocean Momentum Forcing},
  journal = {Journal of Advances in Modeling Earth Systems},
  volume = {13}, number = {9}, pages = {e2021MS002534}, year = {2021},
  doi = {10.1029/2021MS002534}
}

@article{perezhoginGenerativeDataDriven2023,
  author = {Perezhogin, Pavel and Zanna, Laure and Fernandez-Granda, Carlos},
  title = {Generative Data-Driven Approaches for Stochastic Subgrid Parameterizations in an Idealized Ocean Model},
  journal = {Journal of Advances in Modeling Earth Systems},
  volume = {15}, number = {10}, pages = {e2023MS003681}, year = {2023},
  doi = {10.1029/2023MS003681}
}

@article{saneParameterizingVerticalMixing2023,
  author = {Sane, Aakash and Reichl, Brandon G. and Adcroft, Alistair and Zanna, Laure},
  title = {Parameterizing Vertical Mixing Coefficients in the Ocean Surface Boundary Layer Using Neural Networks},
  journal = {Journal of Advances in Modeling Earth Systems},
  volume = {15}, number = {10}, pages = {e2023MS003890}, year = {2023},
  doi = {10.1029/2023MS003890}
}

@article{dheeshjithSamudraAIGlobal2025,
  author = {Dheeshjith, Surya and Subel, Adam and Adcroft, Alistair and Busecke, Julius and Fernandez-Granda, Carlos and Gupta, Shubham and Zanna, Laure},
  title = {Samudra: An {AI} Global Ocean Emulator for Climate},
  journal = {Geophysical Research Letters},
  volume = {52}, number = {10}, pages = {e2024GL114318}, year = {2025},
  doi = {10.1029/2024GL114318}
}

@article{duncanSamudrACEFastAccurate2026,
  author = {Duncan, James P. C. and Wu, Elynn and Dheeshjith, Surya and Subel, Adam and Arcomano, Troy and Clark, Spencer K. and Henn, Brian and Kwa, Anna and McGibbon, Jeremy and Perkins, W. Andre and Gregory, William and Fernandez-Granda, Carlos and Busecke, Julius and Watt-Meyer, Oliver and Hurlin, William J. and Adcroft, Alistair and Zanna, Laure and Bretherton, Christopher S.},
  title = {{SamudrACE}: Fast and Accurate Coupled Climate Modeling With {3D} Ocean and Atmosphere Emulators},
  journal = {Geophysical Research Letters},
  volume = {53}, number = {11}, pages = {e2025GL119340}, year = {2026},
  doi = {10.1029/2025GL119340}
}

@article{eyringPushingFrontiersClimate2024,
  author = {Eyring, Veronika and Collins, William D. and Gentine, Pierre and Barnes, Elizabeth A. and Barreiro, Marcelo and Beucler, Tom and Bocquet, Marc and Bretherton, Christopher S. and Christensen, Hannah M. and Dagon, Katherine and others},
  title = {Pushing the Frontiers in Climate Modelling and Analysis With Machine Learning},
  journal = {Nature Climate Change},
  volume = {14}, number = {9}, pages = {916--928}, year = {2024},
  doi = {10.1038/s41558-024-02095-y}
}

\end{document}